\documentclass[prb,aps,english,twocolumn,notitlepage,superscriptaddress]{revtex4-2}

\usepackage{graphicx}
\usepackage{dcolumn}
\usepackage{rotating}
\usepackage{lipsum}
\usepackage{xcolor}
\usepackage{bm}
\usepackage[T1]{fontenc}  
\usepackage{multirow,amsmath,amssymb}
 \usepackage{babel} 
 \usepackage{txfonts}
 \usepackage{epsfig,epsf,psfrag} 
\usepackage{graphicx} 
\usepackage{pslatex}
\usepackage{epic,eepic} 
\usepackage{color,pstcol}
\usepackage{pstricks} 
\usepackage{fancyhdr} 
\usepackage{tabularx}
\usepackage{physics}
\usepackage{url}
\usepackage{listings}
\usepackage{color} 
\usepackage{caption}
\usepackage{ulem}
\usepackage{balance}
\usepackage{subcaption}
\usepackage{float}
\usepackage[utf8]{inputenc}
\usepackage{comment}

\usepackage{hyperref}
\hypersetup{
 colorlinks=true,
 linkcolor=blue,
 anchorcolor = blue,
 citecolor = blue,
 filecolor = blue,
 urlcolor = blue
}
\usepackage{dsfont}

\graphicspath{{Images/}}

\def \be {\begin{equation}} 
\def \ee {\end{equation}}

\begin{document}

\preprint{APS/123-QED}

\title{Probing the dichotomy between Yu-Shiba-Rusinov and Majorana bound states via conductance, quantum noise and $\Delta_T$ noise}

\title{Probing the dichotomy between Yu-Shiba-Rusinov and Majorana bound states via conductance, quantum noise and $\Delta_T$ noise}
\author{Sachiraj Mishra}%
\email{sachiraj29mishra@gmail.com}

\author{Colin Benjamin}%
\email{colin.nano@gmail.com}
\affiliation{School of Physical Sciences, National Institute of Science Education and Research, HBNI, Jatni-752050, India}
\affiliation{Homi Bhabha National Institute, Training School Complex, AnushaktiNagar, Mumbai, 400094, India }

\begin{abstract}
{We explore the potential of charge and spin conductance as well as charge and spin quantum noise and $\Delta_T$ noise as probes to analyze and contrast Yu-Shiba-Rusinov (YSR) states from Majorana bound states (MBS) in a one-dimensional metal/spin-flipper/metal/insulator/superconductor junction. YSR states, induced by magnetic impurities acting as spin-flippers within the superconducting gap, are distinct from MBS, which can also arise in systems with magnetic impurities, such as magnetic adatoms on superconductors, often leading to false positives in MBS detection. Replacing a trivial $s$-wave superconductor with a topological superconductor featuring triplet pairing, e.g., chiral-$p$ or spinless $p$-wave superconducting nanowire, we analyze and establish clear distinctions between YSR states and MBS. This work provides an unique signature for YSR states, demonstrating that charge or spin conductance as well as both charge and spin quantum noise, along their $\Delta_T$ noise counterparts are effective in identifying YSR states and distinguishing them from MBS.}
\end{abstract}

\maketitle

\section{Introduction}
{Quantum noise serves as an useful tool to explore exotic phenomena in mesoscopic systems, offering profound insights into the underlying physics \cite{noise, qnoise, thermalnoise, kobayashi2021shot, martin2005course}. Defined as the current-current correlation between contacts arising from fluctuations of current around its mean value, quantum noise can be characterized through both charge and spin transport in these systems. Noise due to charge current is referred to as charge quantum noise, while that due to spin transport is known as spin quantum noise. Quantum noise can be further divided into two components: quantum thermal noise, arising at finite temperatures, and quantum shot noise, due to scattering of particles. Research in quantum noise has an exciting past, with significant attention being devoted to quantum shot noise because of its ability to probe intriguing aspects of mesoscopic physics \cite{noise, qnoise, thermalnoise, kobayashi2021shot, martin2005course, shotnoise, PhysRevLett.48.1144, nakamura2020direct, PhysRevLett.79.2526, de1998direct, reznikov1999observation, dolev2008observation}. Early studies, see Ref. \cite{PhysRevB.49.16070}, reported a doubling of shot noise in a normal metal-superconductor junction \cite{PhysRevB.25.4515} at zero Kelvin; this work was later extended in Ref. \cite{PhysRevB.53.16390}, to analyze quantum noise in superconducting junctions at finite temperatures. Experimental techniques have since leveraged this method to measure the charge of Cooper pairs in metal-superconductor junctions \cite{PhysRevLett.84.3398, kozhevnikov2000shot, jehl2000detection} and to investigate pairing symmetries in both trivial and topological superconductors, aiding in the detection of Majorana fermions \cite{PhysRevB.106.125402}. Recent advancements further highlight its potential to distinguish topological (chiral/helical) from trivial edge mode transport \cite{PhysRevB.108.115301}. For a deeper exploration of quantum noise and its numerous applications in mesoscopic physics, readers may refer to Refs. \cite{noise, qnoise, thermalnoise, kobayashi2021shot, martin2005course}.}
 {A recent area of intense research in mesoscopic physics is $\Delta_T$ noise, which is the shot noise like contribution to quantum noise at finite temperature bias but with zero charge current transport. $\Delta_T$ noise has been theoretically predicted ~\cite{PhysRevLett.125.086801, popoff2022scattering, PhysRevLett.127.136801, PhysRevB.107.075409, PhysRevB.107.245301} as well as experimentally observed ~\cite{shein2022electronic, lumbroso2018electronic, sivre2019electronic, PhysRevLett.125.106801, melcer2022absent}, and there has been a recent 
focus on $\Delta_T$ noise in superconducting hybrid junctions \cite{mishra2024andreevreflectionmediateddeltat, pierattelli2024deltatnoisemultiterminalhybrid}. Additionally, spin current holds significant relevance in spintronics \cite{PhysRevLett.120.037201}, particularly within superconducting junctions. In this paper, we propose to probe YSR states and distinguish them from MBS by exploring various quantities such as (i) {charge/}spin conductance, (ii) charge/spin quantum noise both at zero temperature bias, (iii) charge/spin quantum noise at finite temperature bias and finally (iv) charge/spin $\Delta_T$ noise in a normal metal/spin-flipper/metal/insulator/superconductor (N-sf-N-I-S) junction.}

\par



YSR states arise from the interaction between the magnetic impurity's spin and the electron-like or hole-like quasiparticles spin due to Andreev reflection.
This discovery, made independently by Yu, Shiba, and Rusinov \cite{luh1965bound, shiba1968classical, rusinov1969superconductivity}, laid the foundation for our understanding of YSR states. Experimental verification of YSR states has been achieved in recent years using scanning tunneling spectroscopy and atomic scale shot noise spectroscopy ~\cite{PhysRevLett.115.087001, PhysRevLett.128.247001}. {In recent years, this YSR state has been discovered in a setup where a magnetic insulator like CrBr$_3$ interacts with the Cooper pair of the $s$-wave superconductor like NbSe$_2$ \cite{kezilebieke2020topological}. YSR states are notable for their zero-bias charge conductance peaks (ZBCPs), which are a hallmark of YSR states. However, these YSR induced ZBCPs are not quantized, and arise due to distinct reasons.} {If the $s$-wave superconductor is replaced with a topological superconductor (either chiral-$p$ \cite{leggett2021symmetry, jiao2020chiral, PhysRevB.78.195125, PhysRevB.90.085438, RevModPhys.88.035005, PhysRevB.106.125402} or spinless $p$-wave superconducting nanowire \cite{PhysRevB.91.214513, pal2023honing}), it gives rise to MBS. Similar to YSR states, MBS also exhibits ZBCP. However, the physical origin and nature of these ZBCPs are totally different and thus needs to be distinguished from each other. A YSR state should be distinguished from a MBS which too can occur due to interaction with magnetic impurities,
e.g., magnetic adatoms on superconductors \cite{PhysRevB.88.155420, PhysRevB.93.140503, pawlak2016probing}. The primary goal of our study is to find an unique signature for YSR states and compare these with MBS via the charge/spin conductance, charge/spin quantum noise and $\Delta_T$ noise. The reason being, both YSR and MBS show zero-bias charge conductance peaks (ZBCPs) at zero temperature, thus rendering charge conductance as an ineffective discriminator. However, with a spin-flipper, we show that even charge conductance can be an effective discriminator. This is crucial, particularly in the current era, where the detection of MBS holds immense significance due to their potential applications in topological quantum computing \cite{beenakker2013search, sarma2015majorana, alicea2012new}. Ensuring that ZBCPs are not misinterpreted as signatures of YSR is critical, as false positives for MBS can lead to misleading conclusions and hinder progress in the field \cite{pal2023honing, pal2018yu, PhysRevB.110.045432}.}

{Historically, ZBCPs have been extensively employed as a tool for detecting MBS. However, this approach has proven to be unreliable in many instances. Several experimental claims of MBS detection based on ZBCPs have subsequently been retracted \cite{retraction2, gazibegovic2017retracted, yin2015observation, Nayak_2021}. The fundamental issue lies in the fact that ZBCPs are not exclusive to systems hosting MBS. Instead, they can also emerge in systems where MBS are absent, leading to false-positives. For example, in systems where the spin of a magnetic impurity interacts with the Cooper pairs in a superconductor, YSR states can form \cite{pal2018yu}, producing ZBCPs that appear almost identical to those arising from MBS. This overlap underscores the importance of developing additional tools to distinguish between the two types of bound states conclusively \cite{PhysRevLett.119.136803}.}

{
It is important to emphasize that the quantized zero-bias charge conductance peak (ZBCP) associated with Majorana bound states (MBS) is strictly valid at zero temperature. As the temperature increases, this quantization is gradually lost \cite{PhysRevLett.119.136803, PhysRevB.96.184520}. Experimental observations in Ref. \cite{PhysRevLett.119.136803} first demonstrated the temperature scaling behavior of charge conductance, showing that it follows a Lorentzian line shape. This finding was later reinforced and theoretically elaborated in Ref. \cite{PhysRevB.96.184520}.
In our work, we extend this analysis by investigating the scaling behavior of charge conductance for both YSR ($G_{ch}^{YSR}$) and MBS ($G_{ch}^{MBS}$) states, where both of them exhibit a Lorentzian line shape and obey $G_{ch}^{YSR} < \frac{G_{ch}^{MBS}}{2}$, regardless of temperature. Our results indicate that the YSR and MBS states can be distinguished at both zero temperature as well as at finite temperatures. We also explore measures that can effectively differentiate YSR states from MBS states at zero and arbitrary temperatures, ensuring a more robust characterization of these exotic quantum states in experimental settings.}

{Complementary to charge conductance, we analyze spin conductance in our setups both at zero temperature as well as finite temperatures, considering either trivial ($s$-wave) superconductors or topological superconductors, e.g., chiral-
$p$-wave and spinless-
$p$-wave nanowires. Our findings suggest that the spin conductance serves as an excellent measure to probe YSR states and distinguish them from MBS.
Specifically, we observe that for YSR states, the spin conductance exhibits a Lorentzian line shape, similar to charge conductance at both zero as well as finite temperatures. In contrast, for MBS, the spin conductance shows an inverted Lorentzian line shape regardless of temperature, providing a clear qualitative distinction between the YSR states and MBS.
Moreover, a crucial quantitative difference arises in the behavior of spin conductance at zero bias. For YSR states, spin conductance remains finite, reflecting the localized magnetic impurity-induced bound state characteristics. On the other hand, for MBS, spin conductance completely vanishes at zero bias and at zero temperature, which is a signature of topologically protected nature of Majorana states. We also note an essential difference between YSR states and MBS for zero bias and at finite temperatures. Interestingly, we observe that the spin conductance at zero bias decreases for the YSR state as the temperature increases, whereas for MBS, it increases with increasing temperature. This stark contrast in zero-bias behavior both at zero and finite temperatures further reinforces the utility of spin conductance as a diagnostic tool for differentiating YSR and MBS states in hybrid superconducting systems.}

{Further, we identify several unique features in charge and spin quantum noise at both zero and finite temperature biases, as well as in the charge and spin $\Delta_T$
noise, which serve as clear distinguishing signatures of YSR states in contrast to MBS. Specifically, for YSR states, we find that charge quantum noise remains lower compared to that of MBS at zero {voltage} bias for parameters where YSR states occur, regardless of temperature, i.e., $Q_{11}^{ch} (YSR) < \frac{Q_{11}^{ch} (MBS)}{2}$ {at zero voltage bias}. This can help detect YSR states and distinguish them from MBS effectively. Similarly, when we analyze the spin quantum noise, a reversal occurs, i.e., at zero {voltage} bias, the spin quantum noise for YSR states is {finite and much} larger than that of MBS, {$Q_{11}^{sp} (YSR) \neq 0$, while $Q_{11}^{sp} (MBS) \to 0$ at zero voltage bias} regardless of temperature. This striking difference provides a robust criterion for distinguishing YSR states from MBS in hybrid superconducting systems, via quantum noise.}

{Further, at finite temperature bias and at zero voltage bias, charge and spin quantum noise exhibit peaks or dips {around} parameter values where YSR peaks occur. Similar behavior is also shown by charge and spin $\Delta_T$ noise. This characteristic response is a hallmark of YSR states and arises due to the interaction between Cooper pairs and the spin of the spin flipper. In contrast, for MBS, the behavior changes drastically, neither charge/spin quantum noise nor charge/spin $\Delta_T$ noise exhibit any peaks or dips, at or near parameter values where YSR states occur.}

This paper is structured as follows: {Section~\ref{theory} provides an overview of charge and spin transport, which focuses on the calculation of spin-polarized scattering amplitudes in a N-sf-N-I-S junction, where we consider $s$-wave, chiral $p$ and spinless $p$-wave superconducting nanowire. Subsequently, we discuss and calculate charge/spin conductance, quantum noise and spin-polarized $\Delta_T$ noise at a finite temperature gradient and zero applied voltage bias. Section~\ref{results} delves into the YSR states and MBS along with their respective signatures via charge/spin conductance both at zero and finite temperature, charge/spin quantum noise and $\Delta_T$ noise and analyze all the probes via plots and tables. In Sec. \ref{Analysis}, we explain all our results intuitively and also compare our work with the existing literature. Section~\ref{conclusion} concludes this paper with a discussion on the experimental realization of our work. The derivation of charge (spin) current in a N-sf-N-I-S junction is given in Appendix~\ref{App_I}, while Appendix~\ref{App_Qn} presents the calculation of spin-polarized quantum noise in a N-sf-N-I-S junction. We provide the plots of normal and Andreev reflection probabilities for both YSR states and MBS and provide their zero energy values in Tables \ref{Table170}-\ref{Table173} in Appendix \ref{App:prob}. The Mathematica code to calculate charge (spin) quantum noise and $\Delta_T$ noise is uploaded to Github \cite{github}.}

\section{Theory}
\label{theory}

\begin{figure}[H]
\centering
\includegraphics[width=1.10\linewidth]{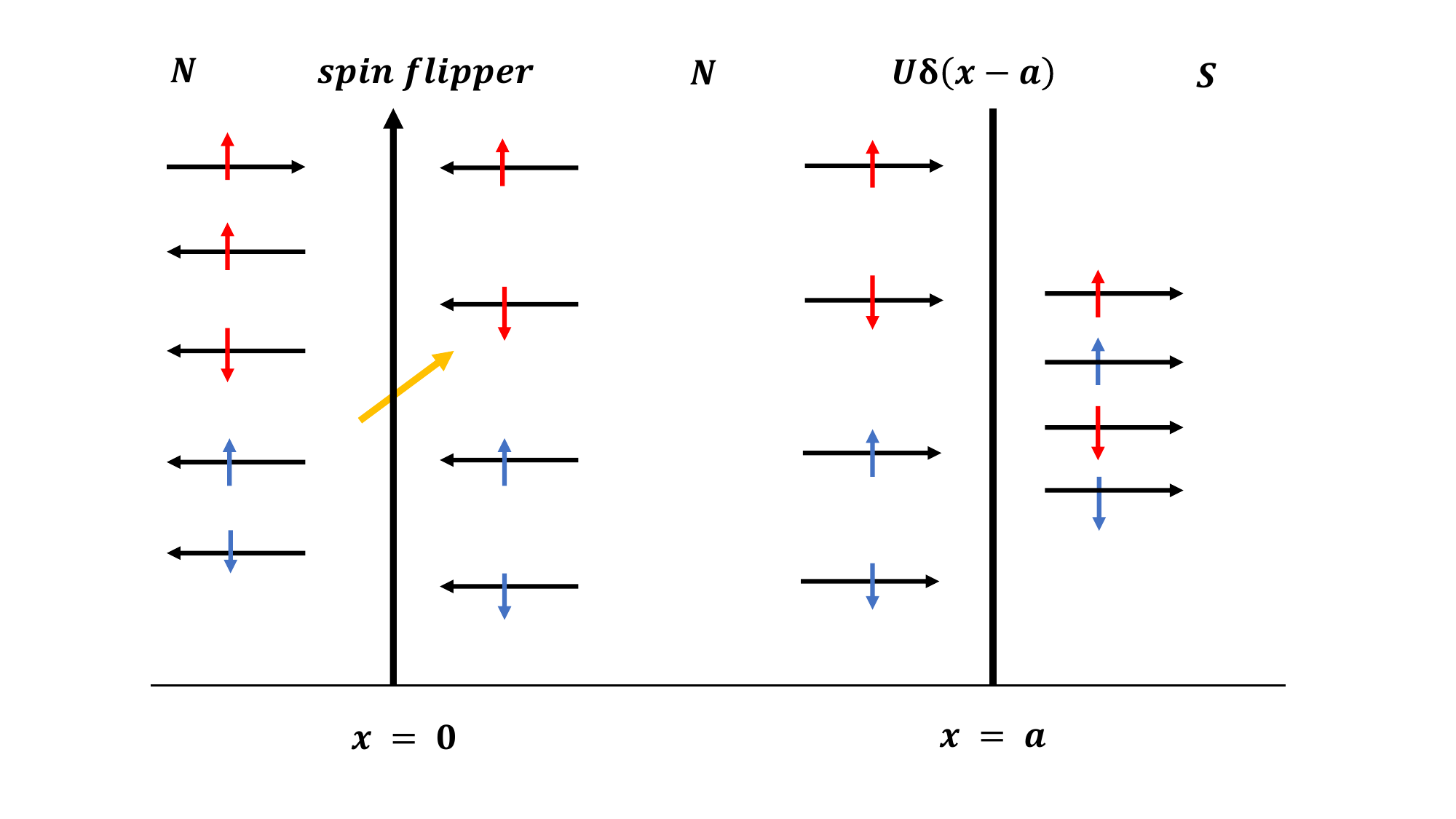}
\caption{Schematic representation of 1D N-sf-N-I-S junction. Here, superconductor can be either $s$-wave, or chiral $p$-wave or spinless $p$-wave nanowire. A magnetic impurity shown by yellow arrow acts as a spin-flipper at $x=0$, while an insulator representing the interface N-S at $x = a$. Red $\uparrow$ arrows depict electron with up spin, red $\downarrow$ arrows depict electron with down spin, blue $\uparrow$ arrows depict hole with up spin, while blue $\downarrow$ depict hole with down spin.}
\label{fig:1}
\end{figure}

{In this section, we discuss the scattering processes in N-sf-N-I-S junction with $s$-wave, chiral $p$-wave and spinless $p$-wave nanowire. Employing the Bogoliubov-de Gennes (BDG) formalism, the Hamiltonian for any generic N-sf-N-I-S junction (see, Fig. \ref{fig:1}) is given as }
{
\begin{equation}
    \mathcal{H}=\left(
	\begin{array}{cc}
	(H_0 (k) + 	H_{sf}) \hat{I} & \hat{\Delta}(\textbf{k}) \Theta(x-a)  \\
		-\hat{\Delta}^*(-\textbf{k}) \Theta(x-a)  & -(H_0^* (-k) + 	H_{sf}^*) \hat{I}
	\end{array}
	\right),
	\label{eq1}
\end{equation}
}

{where $H_0(k) = \hbar^2 k^2/2m^* + U \delta (x - a) - E_F$ and $H_{sf} = -J_0 \delta(x) \Vec{\sigma}. \Vec{\Sigma}$, where $k$ being the wave vector, $E_F$ denoting the Fermi energy, $m^*$ representing the effective mass of electron/hole like quasiparticle, $U$ denoting the strength of the delta potential representing the insulating barrier, $\Theta(x)$ being the Heaviside theta function and $\hat{\Delta}(\mathbf{k})$ being the superconducting gap (Ref.~\cite{PhysRevB.106.125402, RevModPhys.88.035005, PhysRevB.78.195125}), with $J_0$ representing the relative strength of the exchange interaction between the quasiparticle electron/hole spin $\vec{\sigma}$ and the magnetic impurity spin $\vec{\Sigma}$. For singlet pairing, i.e., for $s$-wave superconductor, $\hat{\Delta}(\mathbf{k})$ is $i \Delta_0 \chi_y$ and for triplet pairing, i.e., for either chiral $p$ or spinless $p$-wave nanowire, $\hat{\Delta}(\mathbf{k})$ is $i (\Vec{d}(\textbf{k}). \Vec{\chi}) \chi_y$, where $\Vec{\chi} = \sum_i \chi_i \hat{i}$, where $\chi_i$ are the Pauli matrices for $i  \in \{x, y, z \}$. For both chiral $p$ and spinless $p$-wave nanowire, $\Vec{d}(\textbf{k}) = \Delta_0 \hat{z}$.} The exchange interaction in $H_{sf}$ is expressed as: 
\begin{eqnarray}
    \vec{\sigma} \cdot \vec{\Sigma} = \sigma_z \cdot \Sigma_z + \frac{1}{2} (\sigma^- \Sigma^+ + \sigma^+ \Sigma^-),
    \label{eq2}
\end{eqnarray}

here, the raising and lowering electron spin's (spin-flipper's spin) operator are represented as $\sigma^{\pm}=\sigma_x \pm i \sigma_y$ ($\Sigma^{\pm} = \Sigma_x \pm i \Sigma_y$). Additionally, $\sigma_x, \sigma_y, \sigma_z$ represent the $x, y, z$ components of the electron's spin operator, while $\Sigma_x, \Sigma_y, \Sigma_z$ represent the corresponding components of spin-flipper's spin operator.

{Below, we discuss the BDG Hamiltonian and scattering processes in different superconductors such as $s$-wave, chiral $p$-wave and spinless $p$-wave nanowire.}

\subsection{Wavefunctions}
\label{theory1}
\subsubsection{$s$-wave}



From Eq. (\ref{eq1}), the Hamiltonian for a N-sf-N-I-S junction, with $s$-wave superconductor ~\cite{pal2018yu}, is represented as follows:
\begin{equation}
	\mathcal{H}=\left(
	\begin{array}{cc}
	(H_0 (k) +	H_{sf}) \hat{I} & i \Delta_0 \Theta(x-a) \hat{\sigma}_y \\
	-i	\Delta_0^{*} \Theta(x-a) \hat{\sigma}_y^{*} & -(H_0^* (-k) +	H_{sf}^*) \hat{I}
	\end{array}
	\right),
	\label{eq3}
\end{equation}


\begin{widetext}

The wavefunctions in different regions of N-sf-N-I-S junction for a spin-up electron incident from left normal metal are given as,
\begin{eqnarray}
\psi_{N_1}(x) &=&  (e^{i k_e x } + r^{\uparrow \uparrow} e^{-i k_e x })  \phi^{S}_{m} \check{\varphi}_1 
 + r^{\downarrow \uparrow} e^{-i k_e x} \phi^{S}_{m+1} \check{\varphi}_2 + r^{\uparrow \uparrow}_{a} e^{i k_h x} \phi^{S}_{m+1} \check{\varphi}_3 + r^{\downarrow \uparrow}_a e^{i k_h x} \phi^{S}_{m} \check{\varphi}_4 , ~ \text{for} \hspace{0.08cm} x < 0, \nonumber \\
\psi_{N_2}(x) &=& t^{\uparrow \uparrow} e^{i k_e x } \phi^{S}_{m} \check{\varphi}_1 + t^{\downarrow \uparrow} e^{i k_e x} \phi^{S}_{m+1} \check{\varphi}_2 + f^{\uparrow \uparrow} e^{-i k_e (x-a) } \phi^{S}_{m} \check{\varphi}_1  + f^{\downarrow \uparrow} e^{-i k_e (x-a)} \phi^{S}_{m+1} \check{\varphi}_2 \nonumber \\ 
&& + g^{\uparrow \uparrow} e^{i k_h (x-a)} \phi^{S}_{m+1} \check{\varphi}_3  +g^{\downarrow \uparrow} e^{i k_h (x-a)} \phi^{S}_{m} \check{\varphi}_4  + h^{\uparrow \uparrow} e^{-i k_h x} \phi^{S}_{m+1} \check{\varphi}_3  + h^{\downarrow \uparrow} e^{-i k_h x} \phi^{S}_{m} \check{\varphi}_4,  ~\text{ for} \hspace{0.09cm} 0 < x < a, \nonumber \\
\psi_{S}(x) &=& c^{\uparrow \uparrow}  e^{i q_e x } \phi^{S}_{m} \check{\varphi}^S_1 
+ c^{\downarrow \uparrow} e^{i q_e x} \phi^{S}_{m+1} \check{\varphi}^S_2 + d^{\uparrow \uparrow} e^{-i q_h x} \phi^{S}_{m+1} \check{\varphi}^S_3 + d^{\downarrow \uparrow} e^{-i q_h x} \phi^{S}_{m} \check{\varphi}^S_4  ~\text{ for} \hspace{0.09cm} x > a.
\label{eq4}
\end{eqnarray}



\begin{align}
\text{with~~}
\check{\varphi}_{1} &= \left( \begin{array}{c}
1\\
0\\
0\\
0
\end{array} \right), \check{\varphi}_{2} = 
 \left( \begin{array}{c}
0\\
1\\
0\\
0
\end{array} \right) , \check{\varphi}_3 =\left(  \begin{array}{c}
0\\
0\\
1\\
0
\end{array} \right),  \check{\varphi}_4 =\left(  \begin{array}{c}
0\\
0\\
0\\
1
\end{array} \right),~
\check{\varphi}^S_1 =  \left( \begin{array}{c}
u\\
0\\
0\\
v
\end{array} \right),~ \check{\varphi}^S_2 =  \left( \begin{array}{c}
0\\
u\\
-v\\
0
\end{array} \right),~ \check{\varphi}^S_3 =  \left( \begin{array}{c}
0\\
-v\\
u\\
0
\end{array} \right),~ \check{\varphi}^S_4 =  \left( \begin{array}{c}
v\\
0\\
0\\
u
\end{array} \right),
\label{eq5}
\end{align}
\end{widetext}
{where $\check{\varphi}_i$ are the spinors in the normal metal and $\check{\varphi}_i^S$ are the spinors associated with superconductor for $i \in \{1, 2, 3, 4 \}$.} $\phi^s_m$ denotes the eigenfunction of {the $z$-component of spin-flipper} $\Sigma_z$ , i.e., $\Sigma_z \phi^s_m = m \phi^s_m$, where $m$ denotes the spin magnetic moment. $k_{e,h}$ represent the wave-vectors in the normal metal for electrons and holes, given by $k_{e,h}= \sqrt{ \frac{2 m^*}{ \hbar^2} ( E_F \pm E )}$, where $E$ represents the excitation energy of the electron {with respect to Fermi energy $E_F$.}  {The Andreev reflection with and without spin-flip are characterized by the amplitudes $r^{\downarrow \uparrow}_a = s^{he,\downarrow \uparrow}_{11}$ and $r^{\uparrow \uparrow}_a = s^{he,\uparrow \uparrow}_{11}$. In contrast, the normal reflection with and without spin-flip are denoted by the amplitudes $r^{\downarrow \uparrow} = s^{ee,\downarrow \uparrow}_{11}$ and $r^{\uparrow \uparrow} = s^{ee,\uparrow \uparrow}_{11}$. $s_{\alpha \beta}^{\gamma \eta, \sigma' \sigma}$ is the scattering amplitude for a particle of type $\eta \in \{e, h \}$ with spin $\sigma = \{\uparrow, \downarrow \}$ to scatter from terminal $\beta \in \{1,2 \}$ as a particle $\gamma \in \{e, h\}$ with spin $\sigma' \in \{\uparrow, \downarrow \}$ to terminal $\alpha \in \{ 1,2\}$. 
Furthermore, the transmission amplitudes with (without) spin-flip are given as: $c^{\downarrow \uparrow} = s^{ee, \downarrow \uparrow}_{21}$ (or, $c^{\uparrow \uparrow} = s^{ee,\uparrow \uparrow}_{21}$), which represents the transmission amplitude of an up-spin electron transmitted as a down-spin (or, up-spin) electron, and $d^{\uparrow \uparrow} = s^{he,\uparrow \uparrow}_{21}$ (or, $d^{\downarrow \uparrow} = s^{he,\downarrow \uparrow}_{21}$  ) represents the transmission amplitude of an up-spin electron transmitted as a up-spin (or, down-spin) hole, and the respective scattering probabilities are given as $\mathcal{A}^{\uparrow \uparrow} = (k_h/k_e) |r^{\uparrow \uparrow}_a|^2$, $\mathcal{A}^{\downarrow \uparrow} = (k_h/k_e) |r^{\downarrow \uparrow}_a|^2$, $\mathcal{B}^{\uparrow \uparrow}= |r^{\uparrow \uparrow}|^2$, $\mathcal{B}^{\downarrow \uparrow}= |r^{\downarrow \uparrow}|^2$, $C^{\uparrow \uparrow} =(q_e/k_e) (|u|^2 -|v|^2) |c^{\uparrow \uparrow}|^2$, $C^{\downarrow \uparrow} =(q_e/k_e) (|u|^2 -|v|^2) |c^{\downarrow \uparrow}|^2$ and $D^{\uparrow \uparrow}=(q_h/k_e) (|u|^2 -|v|^2) |d^{\uparrow \uparrow}|^2$, $D^{\downarrow \uparrow}=(q_h/k_e) (|u|^2 -|v|^2) |d^{\downarrow \uparrow}|^2$. The coefficients $(k_h/k_e)$, and $(q_{e,h}/k_e)$ are introduced to ensure the conservation of the probability current, as explained in Ref. \cite{PhysRevB.25.4515}. }
Moreover, wave-vectors for electron-like (hole-like) quasiparticles in the superconductor are $q_{e,h} = \sqrt{ \frac{2 m^*}{ \hbar^2} ( E_F \pm \sqrt{E^2-\Delta_0^2} ) }$, while the coherence factors are $u(v) = \left[ \frac{1}{2} \left\{ 1 \pm \frac{ \sqrt[]{ E^2 - \Delta^2_0 }}{E} \right\} \right]^{1/2} $ in Eq. (\ref{eq4}). {We consider the Andreev approximation, i.e., $E \ll E_F$ and the wavevectors are equal to $k_{e(h)}$ and $q_{e(h)}$ are equal to Fermi wave vector $k_F = \sqrt{\frac{2 m^* E_F}{\hbar^2}}$. We specifically consider the Fermi energy {\( E_F = 50\) in units of $k_BT$}, which is significantly larger than any excitation energy \( E \) at the temperatures examined in this work. The temperatures we consider are \( T = 0.1K, 0.5K, \) and \( 1.0K \), where the characteristic excitation energy at finite temperature is of the order of \( k_B T \). Given that \( k_B T \ll E_F \), the \textit{Andreev approximation} remains valid throughout our analysis. Additionally, we assume the superconducting gap at zero temperature to be \( \Delta_0 = 1.76 k_B T_c = 16.20\) {in units of $k_BT$}, corresponding to a critical temperature of \( T_c = 9.2K \) \cite{RevModPhys.35.1}.
}

\subsubsection{chiral $p$-wave}
{The BDG Hamiltonian for the chiral $p$-wave superconductor \cite{PhysRevB.78.195125, PhysRevB.90.085438, RevModPhys.88.035005, PhysRevB.106.125402} is given as:}
{
\begin{equation}
	\mathcal{H}=\left(
	\begin{array}{cc}
		(H_0 (\textbf{k}) + 	H_{sf}) \hat{I} & \Delta_0 \Theta(x-a) \hat{\sigma}_x \\
		\Delta_0^{\dagger} \Theta(x-a) \hat{\sigma}_x & -(H_0^* (\textbf{-k}) + 	H_{sf}^*) \hat{I}
	\end{array}
	\right),
	\label{eq6}
\end{equation}}

{the wavefunction in the superconducting region is similar to $s$-wave case as in Eq. (\ref{eq4}) with only the definitions of $\check{\varphi}^S_1, \check{\varphi}^S_2, \check{\varphi}^S_3, \check{\varphi}^S_4$ changed to $\check{\varphi}^S_1 = \begin{pmatrix}
    u & 0 & 0 & v
\end{pmatrix}^T$, $\check{\varphi}^S_2 = \begin{pmatrix}
    0 & u & v & 0
\end{pmatrix}^T$, $\check{\varphi}^S_3 = \begin{pmatrix}
    0 & -v & u & 0
\end{pmatrix}^T$ and $\check{\varphi}^S_4 = \begin{pmatrix}
    -v & 0 & 0 & u
\end{pmatrix}^T$, where $T$ stands for the transpose of the spinor. Similar to $s$-wave superconductor, here too, we assume $\Delta_0$ to be 16.20 $k_B K$ \cite{leggett2021symmetry, jiao2020chiral} and $E_F = 50k_B K$ in order to ensure Andreev approximation, i.e., $E \ll E_F$.}

\subsubsection{spinless $p$-wave nanowire}

{The BDG Hamiltonian for spinless $p$-wave nanowire is same as Eq. (\ref{eq6}), i.e.,}

{
\begin{equation}
	\mathcal{H}=\left(
	\begin{array}{cc}
		(H_0 (\textbf{k}) + 	H_{sf}) \hat{I} & \Delta_0 \Theta(x-a) \hat{\sigma}_x \\
		\Delta_0^{\dagger} \Theta(x-a) \hat{\sigma}_x & -(H_0^* (\textbf{-k}) + 	H_{sf}^*) \hat{I}
	\end{array}
	\right),
	\label{eq7}
\end{equation}}

{the wavefunctions in the superconducting region of N-sf-N-I-S junction with a spinless $p$-wave nanowire for a spin-up incident electron is given as \cite{PhysRevB.91.214513, pal2023honing},}

{
\begin{align}
\begin{split}
\psi_{S}(x) &= c^{\uparrow \uparrow}  e^{i k_- x } \phi^{S}_{m} \check{\varphi}^S_1 
+ c^{\downarrow \uparrow} e^{i k_- x} \phi^{S}_{m+1} \check{\varphi}^S_2 \\
&+ d^{\uparrow \uparrow} e^{i k_+ x} \phi^{S}_{m+1} \check{\varphi}^S_3 + d^{\downarrow \uparrow} e^{i k_+ x} \phi^{S}_{m} \check{\varphi}^S_4,
\label{eq8}
\end{split}
\end{align}}


{where, $\check{\varphi}^S_1 = \begin{pmatrix}
    \eta_{-} & 0 & 0 & 1
\end{pmatrix}^T$, $\check{\varphi}^S_2 = \begin{pmatrix}
    0 & \eta_{-} & 1 & 0
\end{pmatrix}^T$, $\check{\varphi}^S_3 = \begin{pmatrix}
    0 & \eta_{+} & 1 & 0
\end{pmatrix}^T$ and $\check{\varphi}^S_4 = \begin{pmatrix}
    \eta_{+} & 0 & 0 & 1
\end{pmatrix}^T$, where $\eta_{\pm} = \frac{k_F (E + \hbar^2 k_{\pm}^2/2m - E_F)}{\Delta_0 k_{\pm}}$, where $k_{\pm}$ are the solutions, which satisfy the equation: $E^2 = (\frac{\hbar^2 k^2}{2m} - E_F )^2 + (\Delta_0 k/k_F)^2$ \cite{PhysRevB.91.214513, pal2023honing}. In this superconductor, we consider $\Delta_0 = 16.20 k_B K$ and $E_F = 50 k_B K$ following Refs. \cite{PhysRevB.91.214513, pal2023honing}. The spinless \( p \)-wave nanowire is proposed to be realized in a semiconductor-superconductor heterostructure \cite{PhysRevB.91.214513}, where the superconductor can be an \( s \)-wave type, such as Niobium, with a critical temperature of \( T_C = 9.2K \). Consequently, for the spinless \( p \)-wave nanowire as well, we assume the superconducting gap to be \( \Delta_0 = 16.20 k_B T\). Additionally, we specifically consider \( E_F = 50 k_B T\) to ensure the validity of the Andreev approximation at the temperatures examined in this work.
}

\subsection{Spin-flip scattering}

{The electron's spin operator denoted as $\Vec{\sigma}$ and spin-flipper's spin operator denoted as $\Vec{\Sigma}$ operating on the spin-up electron spinor~\cite{de1984spin, pal2019spin, pal2018yu} and the spin-flipper eigen function gives,}
{
\begin{eqnarray}
	\Vec{\sigma}. \Vec{\Sigma} \left( \begin{array}{c}
		1\\
		0\\
		0\\
		0
	\end{array} \right) \phi^s_m = \frac{m}{2} \left( \begin{array}{c}
		1\\
		0\\
		0\\
		0
	\end{array} \right) \phi^s_{m} + \frac{\tau}{2} \left( \begin{array}{c}
		0\\
		1\\
		0\\
		0
	\end{array} \right) \phi^s_{m+1},
	\label{eq11}
\end{eqnarray}}
{and $\Vec{\sigma}. \Vec{\Sigma}$ acting on the down-spin electron spinor and the spin-flipper eigen function gives,}
{
\begin{eqnarray}
	\Vec{\sigma}. \Vec{\Sigma} \left( \begin{array}{c}
		0\\
		1\\
		0\\
		0
	\end{array} \right) \phi^s_{m} = -\frac{m}{2} \left( \begin{array}{c}
		0\\
		1\\
		0\\
		0
	\end{array} \right) \phi^s_{m} + \frac{\tau_1}{2} \left( \begin{array}{c}
		1\\
		0\\
		0\\
		0
	\end{array} \right) \phi^s_{m-1}.
	\label{eq12}
\end{eqnarray}}
{Furthermore, $\Vec{\sigma}. \Vec{\Sigma}$ acting on the spin-up hole spinor and the spin-flipper eigen function gives,}
{
\begin{eqnarray}
	\Vec{\sigma}. \Vec{\Sigma} \left( \begin{array}{c}
		0\\
		0\\
		1\\
		0
	\end{array} \right) \phi^s_{m} = -\frac{m}{2} \left( \begin{array}{c}
		0\\
		0\\
		1\\
		0
	\end{array} \right) \phi^s_{m} + \frac{\tau_1}{2} \left( \begin{array}{c}
		0\\
		0\\
		0\\
		1
	\end{array} \right) \phi^s_{m-1},
	\label{eq13}
\end{eqnarray}}
{$\Vec{\sigma}. \Vec{\Sigma}$ acting on the spin-down hole spinor and the spin-flipper eigen function gives,}
{
\begin{eqnarray}
	\Vec{\sigma}. \Vec{\Sigma} \left( \begin{array}{c}
		0\\
		0\\
		0\\
		1
	\end{array} \right) \phi^s_m = \frac{m}{2} \left( \begin{array}{c}
		0\\
		0\\
		0\\
		1
	\end{array} \right) \phi^s_{m} + \frac{\tau}{2} \left( \begin{array}{c}
		0\\
		0\\
		1\\
		0
	\end{array} \right) \phi^s_{m+1}.
	\label{eq14}
\end{eqnarray}}
where $\tau=\sqrt[]{(\Sigma-m)(\Sigma+m+1)}$, $\tau_1=\sqrt[]{(\Sigma+m)(\Sigma-m+1)}$ represent the probabilities of spin-flip for electrons with up-spin and down-spin incident on the left normal metal, where $\Sigma$ denotes the spin of the spin-flipper, with the value of $m$ ranging from $-\Sigma$, $-\Sigma + 1$,...., $\Sigma-1$, $\Sigma$.

\subsection{Boundary conditions}

\subsubsection{$s$-wave and chiral $p$-wave}

{The boundary conditions in a N-sf-N-I-S junction with $s$-wave and chiral $p$-wave superconductor at the interface $x=0$ are,}
\begin{eqnarray}
\left. \psi_{N_1} \right\vert_{x=0} &=& \left. \psi_{N_2} \right \vert_{x=0}, \nonumber \\
\left. \frac{d \psi_{N_2}}{dx}\right \vert_{x=0} - \left. \frac{d \psi_{N_1}}{dx} \right \vert_{x=0} &=& \left. \frac{-2 m^* J_0 \vec{\sigma} \cdot \vec{\Sigma}}{\hbar^2} \psi_{N_1} \right \vert_{x=0},
\label{eq15}
\end{eqnarray}
and at the interface $x=a$ in a N-sf-N-I-S junction are,
\begin{eqnarray}
\left. \psi_{N_2} \right \vert_{x=a} &=& \left. \psi_{S} \right \vert_{(x=a)} , \nonumber \\
\left. \frac{d \psi_{S}}{dx}\right \vert_{x=a} - \left. \frac{d \psi_{N_2}}{dx} \right \vert_{x=a} &=& \left. \frac{2 m^* U}{\hbar^2} \psi_{N_2} \right \vert_{x=a}.
	\label{eq16}
\end{eqnarray}
Incorporating Eqs.~(\ref{eq4}) into Eqs.~(\ref{eq15}) and (\ref{eq16}), we get scattering amplitudes for spin-up (or spin-down) incident electron. {We consider dimensionless barrier strength $Z= m^* U/ (\hbar^2 k_F)$ and the dimensionless exchange interaction strength $J= m^* J_0/ (\hbar^2 k_F)$ in the calculation for conductance, quantum noise and $\Delta_T$ noise. }

\subsubsection{Spinless $p$-wave nanowire}

{For the spinless $p$-wave superconducting nanowire, the boundary condition at the interface $x = 0$ are \cite{PhysRevB.91.214513, pal2023honing},}

{
\begin{align}
    \left. \psi_{N_1} \right\vert_{x=0} &= \left. \psi_{N_2} \right \vert_{x=0}, \nonumber \\
2i \left. \frac{d}{dx} \xi_z \psi_{N_2}\right \vert_{x=0} - 2i \left. \frac{d }{dx} \xi_z \psi_{N_1} \right \vert_{x=0} &= \left. -2i J \vec{\sigma} \cdot \vec{\Sigma} \xi_z \psi_{N_1} \right \vert_{x=0},
\label{eq17}
\end{align}
}

{and, the boundary conditions at the interface $x = a$ is \cite{PhysRevB.91.214513, pal2023honing},}

{
\begin{align}
    \left. \psi_{N_2} \right\vert_{x=a} &= \left. \psi_{S} \right \vert_{x=a}, \nonumber \\
\left(-2i\frac{d }{dx} \xi_z + \Delta_0 \xi_x\right)\left. \psi_{S} \right \vert_{x=a} + 2i \left. \frac{d }{dx} \xi_z \psi_{N_2} \right \vert_{x=a} &= \left. -2iZ  \xi_z \psi_{S} \right \vert_{x=a},
\label{eq18}
\end{align}
}
{Where, $\xi_x = \chi_x \otimes I$ and $\xi_z = \chi_z \otimes I$ are the block diagonal Pauli matrices, where $I$ is the identity matrix.}

\subsection{Configurations}
\label{sec2D}
In a N-sf-N-I-S junction, an incident spin-up electron can undergo four possible reflection processes at any interface due to spin-flip scattering and Andreev reflection. Firstly, the electron may undergo reflection without spin-flip, where it is reflected as an electron with the same spin. Alternatively, it may experience Andreev reflection, reflecting the electron as a hole with an opposite spin. Moreover, the electron could undergo reflection with a spin-flip, resulting in its reflection as an electron with a flipped spin. Lastly, Andreev reflection with spin-flip may occur, causing the electron to be reflected as a hole with the same spin.


\begin{figure}[h!]
	\begin{center}
		\boxed{
			\begin{aligned}
				1. \big\uparrow^{e^-} & \otimes \big\Uparrow^S \xrightarrow[]{\text{ $\Sigma=m$ }}
				\frac{m}{2} \left( \big\uparrow^{e^-} \otimes \big\Uparrow^S \right) \nonumber \\
				2. \big\uparrow^{e^-} & \otimes \big\Downarrow^S \xrightarrow[]{\text{ $\Sigma \neq m$ }}
				\frac{m}{2} \left( \big\uparrow^{e^-} \otimes \big\Downarrow^S \right) + \frac{\tau}{2} \left( \big\downarrow^{e^-} \otimes \big\Uparrow^S \right)  \nonumber \\
				3. \big\downarrow^{e^-} & \otimes \big\Uparrow^S \xrightarrow[]{\text{ $\Sigma\neq-m$ }}
				\frac{-m}{2} \left( \big\downarrow^{e^-} \otimes \big\Uparrow^S \right) + \frac{\tau_1}{2} \left( \big\uparrow^{e^-} \otimes \big\Downarrow^S \right)  \nonumber \\
				4. \big\downarrow^{e^-} & \otimes \big\Downarrow^S \xrightarrow[]{\text{ $\Sigma = -m$ }}
				\frac{-m}{2} \left( \big\downarrow^{e^-} \otimes \big\Downarrow^S \right)
		\end{aligned} }
	\end{center}
	\caption{This box illustrates the different spin configurations that arise when an electron (with spin $\uparrow$ or $\downarrow$) interacts with a spin-flipper. The spin-flipper's spin state ($\Sigma$) can be either aligned ($\Sigma = \pm m$) or anti-aligned ($\Sigma \neq \pm m$) with the electron's spin. The spin-flip probabilities for up-spin incident electron and down-spin incident electron are denoted as $\tau$ and $\tau_1$, respectively.}
	\label{fig:avgm}
\end{figure}

When the time between electron collisions (electron's elastic scattering time, $\rho_e$) is significantly longer than the spin-flipper's relaxation time ($\rho_{sf}$), i.e., $\rho_e \gg \rho_{sf}$, the spin-flipper rapidly flips its spin before encountering the next incoming electron, as illustrated in Fig.~\ref{fig:avgm}. This rapid relaxation implies that the magnetic moment ($m$) for spin associated with a particular spin of the spin-flipper ($\Sigma$) fluctuates \cite{pal2018yu, pal2023honing}. Four possible scenarios are an up-spin incident electron interacting with spin-flipper with spin $\Sigma=m$ is denoted as spin-configuration 1, or with $\Sigma \neq m$ is denoted as spin-configuration 2. Similarly, an electron with a down-spin incident from the left side of the normal metal interacting with spin-flipper at interface $x=0$ for $\Sigma \neq -m$ is denoted as spin-configuration 3 and scenario with for $\Sigma = -m$ is represented as spin-configuration 4, see Fig. \ref{fig:avgm}.
{We calculate for each configuration the charge/spin conductance, spin polarized quantum noise, and spin polarized $\Delta_T$ noise, then average over the four configurations. Next, we discuss the charge/spin current, charge/spin conductance, spin polarized quantum noise and spin polarized $\Delta_T$ noise in a N-sf-N-I-S junction. {Configurations 1 and 4 yield identical results because spin-flip scattering is absent in both cases, and the amplitudes corresponding to each scattering process are the same up to a phase factor, resulting in identical probabilities. Likewise, configurations 2 and 3 also produce identical results, as both involve spin-flip scattering, and the associated scattering amplitudes are again equivalent up to a phase factor, leading to the same probabilities. Therefore, it is sufficient to analyze only configurations 1 and 2, without explicitly considering configurations 3 and 4. As a result, our four-configuration analysis can be effectively reduced to just two configurations.
}}

\subsection{Current and spin-polarised quantum noise}
\label{qnoise}

For a N-sf-N-I-S junction{,} the {general expression for} average spin-polarised current in contact $i$ {considering spin degeneracy} {valid for any mesoscopic junction} is~\cite{PhysRevB.53.16390, qnoise},
\begin{eqnarray}
\langle I^{\sigma}_i \rangle = \frac{2e}{h} \sum_{\substack{j,l \in \{1,2\}; \\ y,\gamma,\eta \in \{e,h\} \\  \rho, \rho^{\prime} \in \{\uparrow, \downarrow\} }} sgn(y) \int^{\infty}_{-\infty} dE A^{\rho \rho^{\prime}}_{j \gamma;l \eta}(i y,\sigma) \langle a^{\rho \dagger}_{j \gamma} a^{\rho^{\prime}}_{l \eta} \rangle, ~~
\label{eq21}
\end{eqnarray}

with $sgn(y)=+1 (-1)$ for electron (hole). $A^{\rho \rho^{\prime}}_{j \gamma;l \eta}(i y,\sigma) = \delta_{i j} \delta_{i l} \delta_{y \gamma} \delta_{y \eta}  \delta_{\sigma \rho} \delta_{\sigma \rho^{\prime}} - s^{y \gamma,\sigma \rho \dagger}_{i j} s^{y \eta,\sigma \rho^{\prime}}_{i l}$, with $i, j,l \in \{1,2\}$ indices refer to left normal metal and superconductor, and $y, \eta, \gamma$ denote electron or hole (see, Appendix \ref{App_I}) with $ \sigma$, $\rho^{\prime}$, $\rho \in \{\uparrow, \downarrow \}$. $a^{\rho \dagger}_{j \gamma}$ ($a^{\rho^{\prime}}_{l \eta}$) represents the creation (annihilation) operators for particle of type $\gamma$ ($\eta$) at contact $j$ ($l$) with spin $\rho (\rho^{\prime})$. The expectation value  $\langle a^{\rho \dagger}_{j \gamma} a^{\rho^{\prime}}_{l \eta} \rangle$ is $\delta_{j l} \delta_{\gamma \eta} \delta_{\rho \rho^{\prime}} f_{j \gamma} (E)$, see Ref. \cite{noise}, wherein the Fermi function $f_{j \gamma}(E)$ is independent of spin, with $f_{j \gamma}(E) = \left[ 1+ e^{\frac{E+ sgn(\gamma)e V_{j}}{k_B T_{j}}} \right]^{-1}$ being the Fermi function in contact $j$ for particle $\gamma$, with $sgn(\gamma)= +$ for electron and $-$ for hole, $k_B$ is Boltzmann constant. $T_{j}$ represents the temperature and $V_j$ denotes the applied voltage bias at contact $j$. 

{In our setup, we apply voltage bias ($V_1) = V$ in the normal metal and the superconducting terminal is grounded.}
 In the normal metal $N_1$, Fermi function for the electron is {$f_{1e}(E)= \left[ 1 + e^{\frac{E-eV}{k_B T_1}} \right]^{-1}$, and for the hole is $f_{1h}(E)= \left[ 1 + e^{\frac{E+eV}{k_B T_1}} \right]^{-1}$}. In the superconductor ($S$), Fermi function for electron-like quasiparticles is same as Fermi function for hole-like quasiparticles at $V_2=0$, i.e., $f_{2e}(E)=f_{2h}(E) = \left[ 1 + e^{\frac{E}{k_B T_2}} \right]^{-1}$.

The average charge current ~\cite{de1984spin, pal2019spin, pal2018yu} in the left normal metal ($N_1$) of the N-sf-N-I-S junction can be written as,
\begin{equation}
\begin{split}
\langle I^{ch}_1 \rangle 
&  = \frac{2e}{h} \int^{\infty}_{-\infty} dE \,\,\, F^{ch}_I (f_{1e}(E)-f_{2e}(E)),
\label{eq22}
\end{split}
\end{equation}

{where $F^{ch}_{I}= 1 + \mathcal{A}^{\uparrow \uparrow} + \mathcal{A}^{\downarrow \uparrow} - \mathcal{B}^{\uparrow \uparrow} - \mathcal{B}^{\downarrow \uparrow}$, see Appendix \ref{App_I}. Charge conductance in a N-sf-N-I-S junction at zero temperature at finite bias voltage ($V_1=eV$, $V_2=0$) is calculated as $G^{ch} = d\langle I^{ch}_1 \rangle /dV = G_0 \left( 1 + \mathcal{A}^{\uparrow \uparrow} + \mathcal{A}^{\downarrow \uparrow} - \mathcal{B}^{\uparrow \uparrow} - \mathcal{B}^{\downarrow \uparrow} \right)$, where $G_0 = 2e^2/h$. The charge conductance at finite temperatures {considering spin degeneracy} is
\cite{PhysRevB.96.184520, PhysRevLett.119.136803},}

{\begin{equation}
\begin{split}
    G_{ch} = \frac{2e^2}{h} \int_{-\infty}^{\infty} dE F_I^{ch}(E) \left(-\frac{\partial f(E)}{\partial E}\right),
    \end{split}
    \label{eq:175}
\end{equation}}

{where, $f(E) = \left[1 + e^{\frac{E - eV}{k_B T}}\right]$, with $T_1 = T_2 = T$, where the expressions to $F_{I}^{ch}$ is given below Eqs. (\ref{eq22}). In our calculations, we perform the integration over the range \( V - 10 k_B T \) to \( V + 10 k_B T \) for any generic voltage bias \( V \) and equilibrium temperature \( T \). This truncated integration yields the exact result as the full integral from \( -\infty \) to \( \infty \). Importantly, these integration limits ensure that the integral of \( \left(-\frac{\partial f(E)}{\partial E}\right) \) remains exactly 1, which is also the case for the full integral over \( (-\infty, \infty) \). Thus, our chosen limits accurately capture the relevant contributions while simplifying the computation.
 In the integration done at finite temperatures, we take these limits for all {quantities} like spin conductance, charge/spin quantum noise and charge/spin $\Delta_T$ noise.} 

{The average spin current ~\cite{cheng2013quantum, PhysRevB.60.3572} in $N_1$ {considering spin degeneracy} can be written as,}
{
\begin{equation}
\begin{split}
\langle I^{sp}_1 \rangle
&= \frac{2e}{h} \int^{\infty}_{-\infty} dE \,\,\, F^{sp}_I (f_{1e}(E)-f_{2e}(E)) ,
\label{eq23}
\end{split}
\end{equation}}

{where $F^{sp}_{I} = 1 +   \mathcal{A}^{\uparrow \uparrow} - \mathcal{A}^{\downarrow \uparrow} - \mathcal{B}^{\uparrow \uparrow} + \mathcal{B}^{\downarrow \uparrow}$, see Appendix \ref{App_I}. Spin conductance in a N-sf-N-I-S junction at zero temperature and finite bias voltage ($V_1 = V, V_2 = 0$) is $G_{sp} = d\langle I_1^{sp} \rangle/dV = G_0 (1 +   \mathcal{A}^{\uparrow \uparrow} - \mathcal{A}^{\downarrow \uparrow} - \mathcal{B}^{\uparrow \uparrow} + \mathcal{B}^{\downarrow \uparrow} )$. The spin conductance at finite temperature  is 
\cite{cheng2013quantum, PhysRevB.60.3572},}

{\begin{equation}
\begin{split}
    G_{sp} = \frac{2e^2}{h} \int_{-\infty}^{\infty} dE F_I^{sp}(E) \left(-\frac{\partial f(E)}{\partial E}\right),
    \end{split}
    \label{eq:176}
\end{equation}}

{with $f(E) = \left[1 + e^{\frac{E - eV}{k_B T}}\right]$ and $T_1 = T_2 = T$. }
{Eqs. (\ref{eq22}) and (\ref{eq23}) are applicable to all setups such as N-sf-N-I-$s$, N-sf-N-I-chiral $p$ and N-sf-N-I-spinless $p$-wave nanowire. The integration limits remain same as in Eq. (\ref{eq:175}).}

The current-current correlation at the metal contact $N_1$ is considered to be the quantum noise. Spin polarised quantum noise at $N_1$ between charge carriers with spin $\rho$ and $\rho^{\prime}$ at different times $t$ and $T$ is defined as $Q^{\rho \rho^{\prime}}_{11}(t-T) \equiv \langle \Delta I^{\rho}_{1}(t) \Delta I^{\rho^{\prime}}_{1}(T) + \Delta I^{\rho}_{1}(T) \Delta I^{\rho^{\prime}}_{1}(t) \rangle $ with $\Delta I^{\rho}_{1}(t)= I^{\rho}_{1}(t) - \langle I^{\rho}_{1}(t) \rangle$ \cite{qnoise}. The charge quantum noise auto-correlation at zero frequency ~\cite{PhysRevLett.92.106601} for charge current $I^{ch}_{1}$ is written as,
\begin{equation}
 Q^{ch}_{11} = Q^{ \uparrow \uparrow}_{11} + Q^{ \uparrow \downarrow}_{11} + Q^{ \downarrow \uparrow}_{11} + Q^{ \downarrow \downarrow}_{11},
 \label{eq24}
\end{equation}
while the spin quantum noise auto-correlation at zero frequency for spin current $I^{sp}_{1}$ is,
\begin{equation}
    Q^{sp}_{11} = Q^{ \uparrow \uparrow}_{11} - Q^{ \uparrow \downarrow}_{11} - Q^{ \downarrow \uparrow}_{11} + Q^{ \downarrow \downarrow}_{11},
    \label{eq25}
\end{equation}

where, the spin-polarised quantum noise-auto correlation at zero frequency in N-sf-N-I-S junction $Q_{11}^{\sigma \sigma'}$ \cite{PhysRevB.53.16390, martin2005course} for $\sigma, \sigma' \in \{\uparrow, \downarrow \}$ {considering spin degeneracy} is given as

\begin{widetext}

\begin{equation}
Q^{\sigma \sigma^{\prime}}_{11} = \sum_{x', y' \in \{e,h \}} Q_{11}^{\sigma \sigma', x' y'} = \frac{2e^2}{h} \int_{-\infty}^{\infty} \sum_{\rho, \rho^{\prime} \in \{\uparrow, \downarrow\}} \sum_{ \substack{k,l \in \{1, 2\} ,\\
x^{\prime}, y^{\prime}, \Gamma,\eta \in \{e,h\} } } sgn(x^{\prime}) sgn(y^{\prime}) A^{\sigma^{\prime} \sigma}_{k,\Gamma;l,\eta}(1 x^{\prime},\sigma) A^{\rho^{\prime} \rho}_{l,\eta;k,\Gamma}(1 y^{\prime},\sigma^{\prime}) \textit{f}_{k \Gamma}(E) \{ 1 - \textit{f}_{l \eta}(E) \} dE,
\label{eq26}
\end{equation}
where $sgn(x^{\prime})=+1 (-1)$ for electron (hole). {The integration limits remain same as in Eq. (\ref{eq:175}).}

\end{widetext}

{One can utilize Eq. (\ref{eq26}) to calculate the spin-polarized correlations following Ref. \cite{PhysRevB.53.16390} and the charge quantum noise ($Q_{11}^{ch}$) given in Eq. (\ref{eq24}) and the thermal noise-like contribution ($Q_{11}^{ch; th}$) and shot noise-like contribution ($Q_{11}^{ch; sh}$) is derived in Appendix \ref{App_Qn}. Similarly, the spin quantum noise ($Q_{11}^{sp}$) is given in Eq. (\ref{eq25}) and the thermal noise-like contribution ($Q_{11}^{sp; th}$) and shot noise-like contribution ($Q_{11}^{sp; sh}$) are given in Appendix \ref{App_Qn}.}



{Next, we calculate the charge/spin quantum noise and $\Delta_T$ noise.}

\subsection{Spin polarised quantum noise and $\Delta_T$ noise}
\label{dtnoise}

{In this work, we focus on the charge (spin) quantum noise ($Q_{11}^{ch(sp)}$) at zero temperature bias, i.e., $T_1 - T_2 = \Delta T = 0$. We also investigate charge/spin quantum noise and $\Delta_T$ noise at finite temperature bias ($T_1 - T_2 = \Delta T$) and zero voltage bias, which general expression is given in Appendix \ref{App_Qn}. For the calculation of charge/spin quantum noise and $\Delta_T$ noise at finite temperature bias, we consider $T_1 = T + \frac{\Delta T}{2}$ and $T_2 = T - \frac{\Delta T}{2}$.} {The {shot} noise arising from a non-equilibrium temperature gradient in the absence of net charge (or spin) current at zero bias voltage is termed as charge (spin) $\Delta_T$ noise ~\cite{PhysRevLett.125.086801, popoff2022scattering, PhysRevLett.127.136801, lumbroso2018electronic, sivre2019electronic}. We denote the charge $\Delta_T$ noise as $\Delta^{ch}_T$ and spin $\Delta_T$ noise as $\Delta_T^{sp}$. One can calculate $\Delta^{ch}_T$ from $Q_{11}^{ch; sh}$ (see, Appendix \ref{App_I}) at zero average charge current $\langle I_{1}^{ch} \rangle = 0$. Similarly, one can also calculate the spin $\Delta_T$ noise, denoted as $\Delta_T^{sp}$ from $Q_{11}^{sp; sh}$ (see, Appendix \ref{App_I}) at $\langle I_1^{sp} \rangle = 0$.}

{Therefore, the expression for $\Delta_T^{ch (sp)}$ is given as}
{
\begin{equation}
\begin{split}
\Delta^{ch (sp)}_T &= \frac{4e^2}{h} \int^{\infty}_{-\infty} dE F^{ch (sp)}_{sh} (f_{1e} - f_{2e})^2.
\label{eq28}
\end{split}
\end{equation}}

{where the expressions for \( F_{sh}^{ch} \) and \( F_{sh}^{sp} \), as derived in Eqs.~(\ref{eq:B3}) and (\ref{eq:B4}) in Appendix~\ref{App_Qn}, are obtained under the assumption that the average temperature \( T \) is much smaller than the superconducting critical temperature \( T_c = 9.2\,\mathrm{K} \). This condition ensures that the excitation energy ($E$) remains below the superconducting gap \( \Delta_0 \). The expressions for $F_{sh}^{ch(sp)}$ are given as follows:
}

{
\begin{equation}
\begin{split}
    F_{sh}^{ch} &= - F_{sh}^{sp} = 4 \mathcal{A}^{\uparrow \uparrow} \mathcal{B}^{\downarrow \uparrow} + 4 \mathcal{A}^{\downarrow \uparrow} \mathcal{B}^{\uparrow \uparrow} - 8 \text{Re} \left(r_N^{\downarrow \uparrow} r_{Na}^{\uparrow \uparrow *} r_{Na}^{\downarrow \uparrow} r_N^{\uparrow \uparrow *}\right).
\end{split}
\label{eqn}
\end{equation}}

{From the above equation, it is evident that $\Delta_T^{sp} = -\Delta_T^{ch}$. We emphasize that this relation does not hold in general for all parameter regimes; however, in the low-temperature limit where $k_B T \ll \Delta_0$, it always holds due to the vanishing of transmission probabilities in the subgap region.} {The integration limits remain same as in Eq. (\ref{eq:175}). Next, we study the YSR and MBS bound states in detail by analyzing their zero bias properties. To probe zero bias YSR, we first calculate both the charge/spin condutance, charge/spin quantum noise (both at zero temperature bias and finite temperature bias) and finally the charge/spin $\Delta_T$ noise.}

\section{Results and Discussion}
\label{results}

This section first delves into YSR states in a N-sf-N-I-S junction and the results for the {charge/spin} conductance {$G_{ch(sp)}$,  charge/spin quantum noise $S_{ch(sp)}$ at zero and finite temperature, charge/spin $\Delta_T$ noise and quantum noise, which can effectively probe YSR states and distinguish it from MBS. The Mathematica code for the calculations are given in \cite{github}.} 

{In Fig. \ref{fig:2}, we present the YSR states in the N-sf-N-I-S junction. The bound state energies $E^{\pm}$ can be determined by computing the complex poles of the charge conductance $G_{ch}$ \cite{pal2018yu} (see GitHub \cite{github} for the Mathematica code). The real part of these poles (see Fig. \ref{fig:2}) corresponds to the energy levels ($E/\Delta_0$) at which the YSR peaks appear, while the imaginary part represents the width of these peaks. In Fig. \ref{fig:2}(a), we plot the YSR states (real part of the energy levels) as a function of the spin-flipper barrier strength $J$ for $Z = 0.781$. We observe the presence of two energy-bound states, which coalesce at two distinct spin-flip coupling values, $J=4.5$ and $J=-4.9$. We also observe two distinct energy bound states coalescing at two different spin-flipper barrier strengths $J = 4.5$ and -8.9 for $Z = 1.121$, see Fig. \ref{fig:2}(b)

\begin{widetext}

\begin{figure}[H]
\centering
\includegraphics[width=1.0\linewidth]{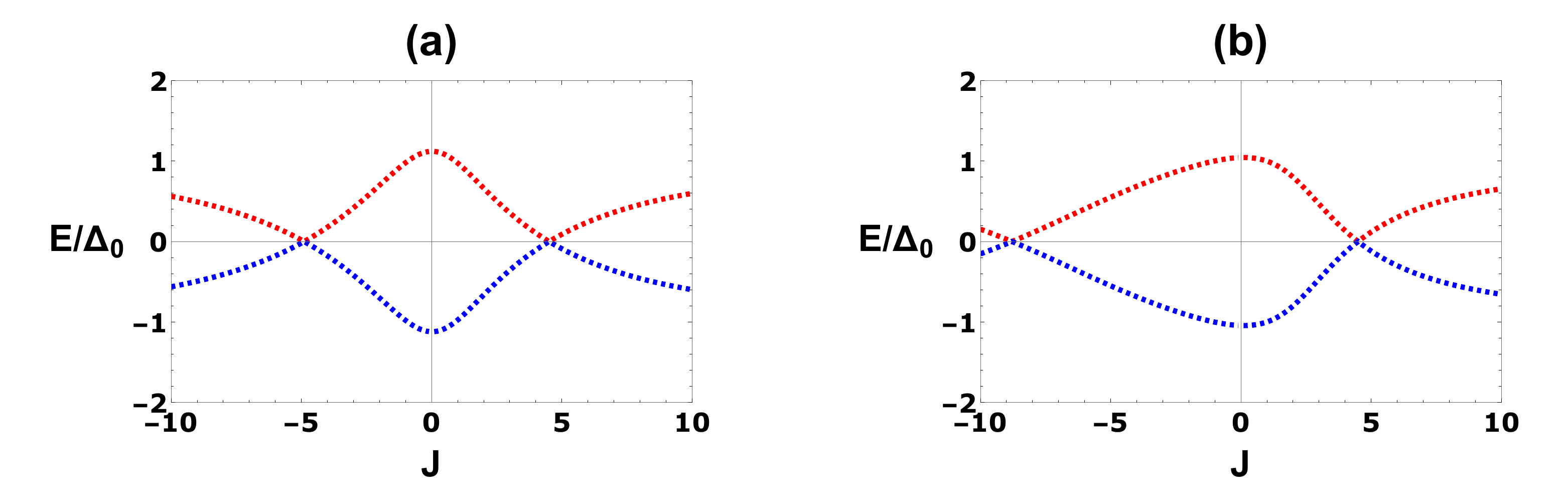}
\caption{YSR states vs. spin-flipper barrier strength ($J$) for N-sf-N-I-N junction with (a) $Z = 0.781$, (b) $Z = 1.121$ and spin-flipper parameters $\Sigma =1/2$, $m=-1/2$ and $k_F a=0.85 \pi$.}
\label{fig:2}
\end{figure}
\end{widetext}


\subsection{Charge and spin conductance}
{In Figs. \ref{fig:19}-\ref{fig:20}, we present the charge conductance ($G_{ch}$) and spin conductance ($G_{sp}$) for an N-sf-N-I-S junction as functions of the spin-flipper barrier strength ($J$) at zero bias ($\frac{eV}{\Delta_0} = 0.0$), considering a fixed barrier strength of $Z = 0.781$ across different average temperatures ($T_1 = T_2 = T = 0.0K, 0.1K, 0.5K$).} 

{As shown in Fig. \ref{fig:19}(a), at $T = 0.0K$, the charge conductance for the YSR state, $G_{ch}^{YSR}$ is around 1 (in units of $2e^2/h$) at coupling $J$ where YSR states emerge, specifically at $J = 4.5$ and $J = -4.9$, where it takes values of 0.93 and 0.83, respectively, see also Table \ref{Table111}. In contrast, for MBS in both chiral $p$-wave and spinless $p$-wave nanowires, the charge conductance, $G_{ch}^{MBS}$, remains quantized at 2 (in units of $2e^2/h$) for all values of $J$, a hallmark of their topological nature, see Figs. \ref{fig:19}(b) and (c) and Table \ref{Table111}. A crucial observation is that charge conductance follows the relation $G_{ch}^{YSR} < \frac{G_{ch}^{MBS}}{2}$, meaning that the charge conductance for YSR states is consistently lower than that of MBS states by more than a factor of 2, making it an effective quantitative discriminator. This significant reduction in charge conductance for YSR states is attributed to spin-flip scattering, whereas MBS states remain unaffected by spin-flip scattering. This robustness of MBS charge conductance against spin-flip scattering further distinguishes them from YSR states. Thus, our findings establish that charge conductance provides a reliable method for differentiating YSR states and MBS, reinforcing its role as an effective experimental probe.}

{Furthermore, this also occurs at finite temperatures. As temperature increases, MBS begin to lose their quantization, as shown in Fig. \ref{fig:19}(b) and (c), yet we still observe that $G_{ch}^{YSR}$ remains less than $G_{ch}^{MBS}$ by more than a factor of 2. For instance, at $T = 0.1K$, $G_{ch}^{YSR}$ is 0.92 and 0.82 at $J = 4.5$ and $J = -4.9$, respectively, in units of {$\frac{2e^2}{h}$}. In contrast, for the MBS in a chiral $p$-wave superconductor, $G_{ch}^{MBS}$ takes values of 1.99 both at $J = 4.5$ and $J = -4.9$, respectively, which are more than twice the corresponding values for YSR states. Similarly, for the MBS state in a spinless $p$-wave nanowire, $G_{ch}^{MBS}$ is 1.99 and 1.99 both at $J = 4.5$ and $J = -4.9$, respectively, again exceeding twice the values observed for YSR states. Therefore, even at finite temperatures, the relation $G_{ch}^{YSR} < \frac{G_{ch}^{MBS}}{2}$ holds, proving to be an effective criterion for distinguishing YSR and MBS states. As the temperature is further increased to $0.5K$, $G_{ch}^{YSR}$ is again less than $ \frac{G_{ch}^{MBS}}{2}$ for both chiral $p$-wave superconductor and spinless $p$-wave nanowire and the inequality $G_{ch}^{YSR} < \frac{G_{ch}^{MBS}}{2}$ persists. The zero bias values of charge conductance are provided in Table \ref{Table111}.}

\begin{widetext}

 \begin{figure}[H]
\centering
\includegraphics[width=1.0\linewidth]{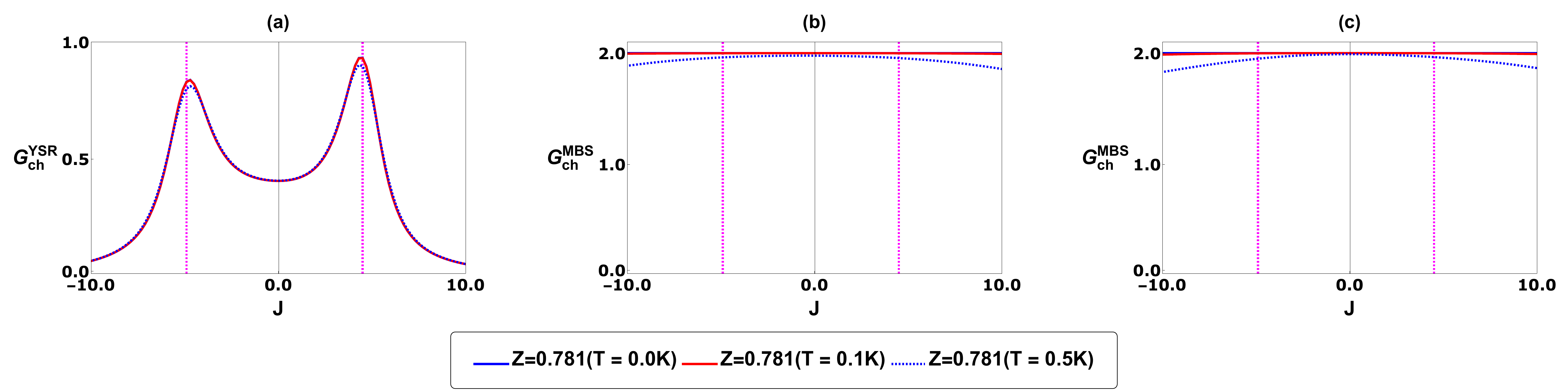}
\caption{Charge conductance in units of {$\frac{2e^2}{h}$} {taking the spin degeneracy into account} in the N-sf-N-I-S setup for (a) $s$-wave superconductor, (b) chiral-$p$ wave superconductor and (c) spinless $p$-wave superconducting nanowire vs. dimensionless spin-flipper strength $J$ at dimensionless barrier strength $Z = 0.781$ for temperatures $T = 0.0K, 0.1K, 0.5K$. Other parameters: $\frac{eV}{\Delta_0} = 0.0, k_F a = 0.85 \pi$. {The pink verticals at $J = 4.5$ and -4.9 denote where YSR peaks occur.}}
\label{fig:19}
\end{figure}

      \begin{table}[H]
    \caption{Charge conductance in units of {$\frac{2e^2}{h}$} at $Z=0.781$ for different values of average temperature $T=0.0K, 0.1K, 0.5K$ {at zero voltage bias $\left(\frac{eV}{\Delta_0} \to 0.0\right)$} for $J=4.5$ and -4.9 {(from Fig. \ref{fig:19})}.}
    \centering
    \begin{tabular}{|c|c|c|c|c|c|c|}
        \hline
        \multirow{2}{*}{$T$} & \multicolumn{2}{c|}{YSR ($s$-wave)} & \multicolumn{2}{c|}{MBS (chiral $p$-wave)} & \multicolumn{2}{c|}{MBS (Spinless $p$-wave nanowire)} \\ \cline{2-7} 
        & $J=4.5$ & $J=-4.9$ & $J=4.5$ & $J=-4.9$ & $J=4.5$ & $J=-4.9$ \\ \hline
        0.0$K$ & 0.93 & 0.83  & 2.00  & 2.00  & 2.00  & 2.00  \\ \hline
        0.1$K$ & 0.92  & 0.82  & 1.99  & 1.99  & 1.99  & 1.99  \\ \hline
        0.5$K$ & 0.90  & 0.80  & 1.95  & 1.96  & 1.96  & 1.94  \\ \hline
    \end{tabular}
\label{Table111}
\end{table}

\end{widetext}

{Similar to charge conductance, we analyze the behavior of spin conductance ($G_{sp}$) as a function of the barrier strength $J$ at different temperatures, as shown in Fig. \ref{fig:20}. At zero temperature, the spin conductance for the YSR state, $G_{sp}^{YSR}$, takes values of 0.95 and 0.96 (in units of {$\frac{2e^2}{h}$}) at $J = 4.5$ and $J = -4.9$, respectively, see Fig. \ref{fig:20}(a) and Table \ref{Table17}. In contrast, for MBS states, the spin conductance, $G_{sp}^{MBS}$, remains exactly zero for all values of $J$. This implies that in the presence of MBS, spin conductance vanishes, and is unaffected by spin-flip scattering, whereas this is not the case for YSR states. The persistence of finite spin conductance for YSR states and its complete absence for MBS provide a definitive criterion to distinguish between these two distinct states, making spin conductance a highly effective experimental probe for their identification.}
{As the temperature increases, the spin conductance associated with YSR states, $G_{sp}^{YSR}$, decreases, whereas the spin conductance for MBS, $G_{sp}^{MBS}$ increases from zero. However, $G_{sp}^{YSR}$ remains considerably larger than $G_{sp}^{MBS}$. For instance, at $T = 0.1K$, $G_{sp}^{YSR}$ is 0.95 and 0.94 for spin-flipper strengths $J = 4.5$ and $J = -4.9$, respectively. In contrast, for the MBS in a chiral $p$-wave superconductor, $G_{sp}^{MBS}$ at $T=0.1K$, is significantly lower, with values of 0.0025 and 0.0012 for the same $J$ values. The spin conductance for MBS in a spinless $p$-wave nanowire is even smaller, measuring 0.0012 and 0.0014, respectively (see Fig. \ref{fig:20} and Table \ref{Table17}). This stark contrast in magnitude highlights a key distinction between YSR states and MBS. As the temperature rises further to $T = 0.5K$, $G_{sp}^{YSR}$ continues to decrease, while $G_{sp}^{MBS}$ further increases. This trend reinforces the conclusion that spin conductance also serves as a reliable means to distinguish between YSR and MBS states: in YSR systems, spin conductance diminishes with increasing temperature, whereas in MBS systems, it exhibits an opposite behavior and grows with temperature. This fundamental difference in response provides a robust criterion for identifying the nature of the underlying states (see Fig. \ref{fig:20} and Table \ref{Table17}). In the low temperature regime $0.0K < T < 1.0K$, $G_{sp}^{YSR} > 2G_{sp}^{MBS}$.}

\begin{widetext}

\begin{figure}[H]
\centering
\includegraphics[width=1.0\linewidth]{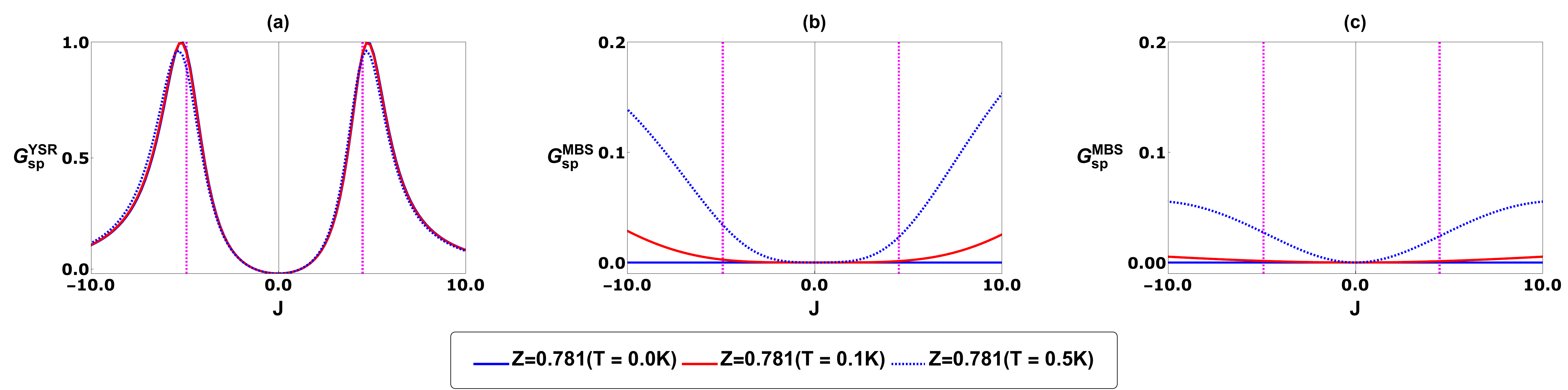}
\caption{Spin conductance in units of {$\frac{2e^2}{h}$} {taking the spin degeneracy into account} in the N-sf-N-I-S setup for (a) $s$-wave superconductor, (b) chiral-$p$ wave superconductor and (c) spinless $p$-wave superconducting nanowire vs. dimensionless spin-flipper strength $J$ at dimensionless barrier strength $Z = 0.781$ for temperatures $T = 0.0K, 0.1K, 0.5K$. Other parameters: $\frac{eV}{\Delta_0} = 0.0, k_F a = 0.85 \pi$. {The pink verticals at $J = 4.5$ and -4.9 denote where YSR peaks occur.}}
\label{fig:20}
\end{figure} 

\begin{table}[H]
    \caption{Spin conductance in units of {$\frac{2e^2}{h}$} at $Z=0.781$ for different values of average temperature $T=0.0K, 0.1K, 0.5K$ {at zero voltage bias $\left(\frac{eV}{\Delta_0} \to 0.0\right)$} for $J=4.5$ and -4.9 {(from Fig. \ref{fig:20})}.}
    \centering
    \begin{tabular}{|c|c|c|c|c|c|c|}
        \hline
        \multirow{2}{*}{$T$} & \multicolumn{2}{c|}{YSR ($s$-wave)} & \multicolumn{2}{c|}{MBS (chiral $p$-wave)} & \multicolumn{2}{c|}{MBS (Spinless $p$-wave nanowire)} \\ \cline{2-7} 
        & $J=4.5$ & $J=-4.9$ & $J=4.5$ & $J=-4.9$ & $J=4.5$ & $J=-4.9$ \\ \hline
        0.0$K$ & 0.95  & 0.96  & 0.00  & 0.00  & 0.00  & 0.00 \\ \hline
        0.1$K$ & 0.95  & 0.94  & 0.0025  & 0.0012  & 0.0012  & 0.0014  \\ \hline
        0.5$K$ & 0.93  & 0.88  & 0.034  & 0.023  & 0.023  & 0.027  \\ \hline
    \end{tabular}
\label{Table17}
\end{table}

\end{widetext}

{In Figs. \ref{fig:4}-\ref{fig:5}, we present the charge (\( G_{ch} \)) conductance for the N-sf-N-I-S junction as a function of voltage bias ($eV/\Delta_0$). The superconductor in this setup can either be an \( s \)-wave type, which supports YSR states, or a \( p \)-wave type, such as chiral-\( p \)-wave, spinless nanowire superconductors, which host MBS. As shown in Figs. \ref{fig:4}(a) and \ref{fig:5}(a), the charge conductance \( G_{ch}^{YSR} \) exhibits a distinct zero-bias peak (\( \frac{eV}{\Delta_0} = 0.0 \)) for barrier strength \( Z = 0.781 \) at \( J = 4.5 \) and -4.9, a hallmark feature of YSR states at zero temperature ($T=0.0K$). Although this conductance peak is not perfectly quantized at \( 2e^2/h \), it bears a striking resemblance qualitatively to the zero-bias conductance peak (ZBCP) observed in \( G_{ch}^{MBS} \) at $T=0.0K$ for MBS in chiral-\( p \)-wave, spinless \( p \)-wave, as seen in Figs. \ref{fig:4}(b), \ref{fig:4}(c) and \ref{fig:5}(b), \ref{fig:5}(c){, which exhibit quantized conductance peaks. However, we observe that the charge conductance for the YSR state is half that of the MBS state, which itself highlights a key distinction between these two states quantitatively.} It is important to emphasize that the charge conductance follows a Lorentzian line shape, where $F_I^{ch}(V) = \frac{P \Gamma^2}{(eV/\Delta_0)^2 + \Gamma^2}$, where $P$ is the peak value of the conductance and $\Gamma$ is the width of the line shape. In Fig. \ref{fig:4} and \ref{fig:5}, we consider those values of $P$ and $\Gamma$, which correctly resemble that of charge conductance $G_{ch}$ at zero temperature. The values of $P$ and $\Gamma$ are given in Table \ref{Table1} for both YSR states and MBS{, which clearly shows distinguishing features between YSR and MBS states quantitatively.} For instance, for the YSR state at $Z=0.781$ and $J=4.5$, the value of $P$ is 0.932, whereas for the MBS state in both chiral-$p$ and spinless-$p$ wave nanowires, $P$ is 2—more than twice that of the YSR state—providing a clear distinction between YSR and MBS states (see, Table \ref{Table1}). Similarly, for $Z=0.781$ and $J=-4.9$, $P$ for the YSR state is 0.830, while for the MBS state, it is quantized with $P=2$, further reinforcing the distinction between these two states as shown in Table \ref{Table1}.}

{At finite temperatures (\( T = 0.1K \) and \( 0.5K \)), the zero-bias peak diminishes further, indicating a loss of quantization in the MBS (see Figs. \ref{fig:4} and \ref{fig:5}). Even at finite temperatures, the charge conductance retains a Lorentzian line shape, with \( F_I^{ch}(E) \) following the same form as at zero temperature (with $eV$ replaced by $E$). Table \ref{Table1} presents the values of peak and width ($P$ and $\Gamma$) for \(Z=0.781\) and \(J=4.5\), obtained. We observe that the value of $P$ for the YSR state is 0.930, whereas for the MBS in the chiral-$p$ case, the peak value is twice those of the YSR state, with $P=1.99$ (see, Table \ref{Table1}). Similarly, for the spinless $p$-wave nanowire, $P$ is 1.99, which is again significantly higher than that of the YSR state, as shown in Table \ref{Table1}. This analysis quantitatively establishes a clear distinction between YSR states and MBS. A similar trend is observed for \(Z=0.781\) and $J=-4.9$ (Table \ref{Table1}), confirming that \(G_{ch}\) can help in identifying YSR states and distinguish it from MBS quantitatively at elevated temperatures.}

\begin{widetext}

\begin{figure}[H]
\centering
\includegraphics[width=1.0\linewidth]{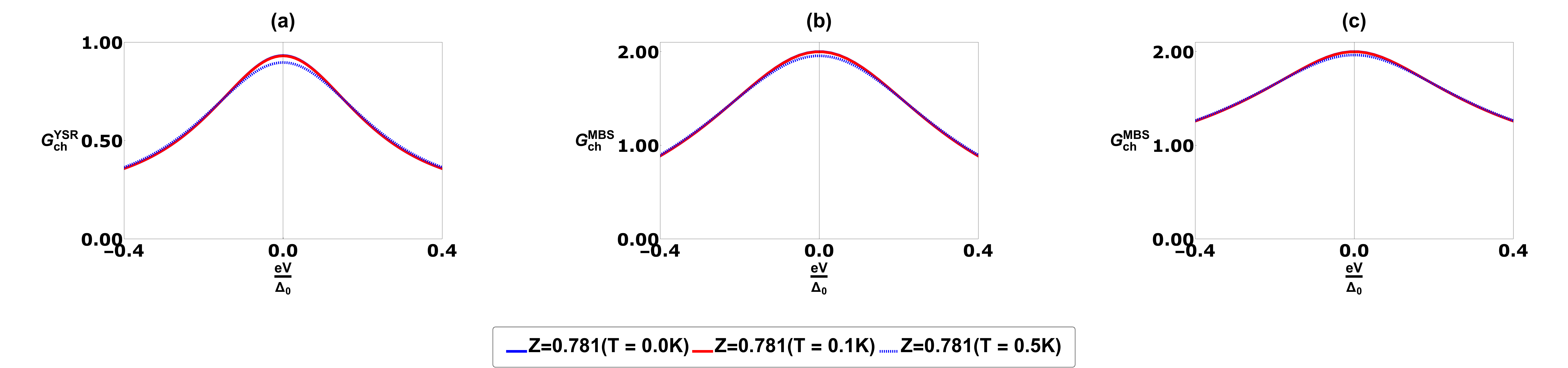}
\caption{Charge conductance in units of {$\frac{2e^2}{h}$} {taking the spin degeneracy into account} in the N-sf-N-I-S setup for (a) $s$-wave superconductor, (b) chiral-$p$ wave superconductor and (c) spinless $p$-wave superconducting nanowire vs. voltage bias applied $\left(\frac{eV}{\Delta_0}\right)$ for $Z = 0.781$ at temperatures $T = 0.0K, 0.1K, 0.5K$. Other parameters: $J = 4.5, k_F a = 0.85 \pi$.}
\label{fig:4}
\end{figure}

       \begin{figure}[H]
\centering
\includegraphics[width=1.0\linewidth]{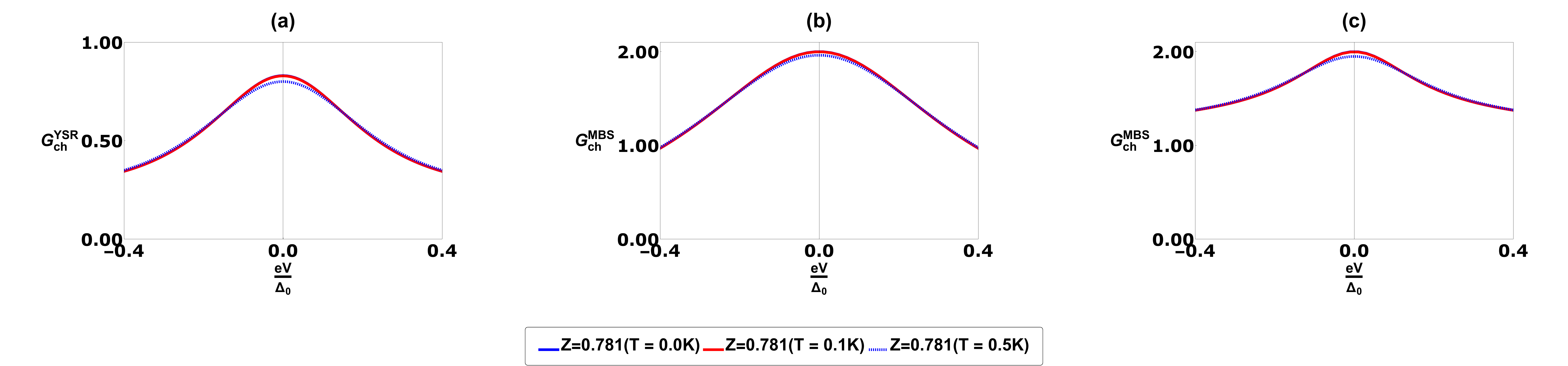}
\caption{Charge conductance in units of {$\frac{2e^2}{h}$} {taking the spin degeneracy into account} in the N-sf-N-I-S setup for (a) $s$-wave superconductor, (b) chiral-$p$ wave superconductor and (c) spinless $p$-wave superconducting nanowire vs. voltage bias applied $\left(\frac{eV}{\Delta_0}\right)$ for $Z = 0.781$ at temperatures $T = 0.0K, 0.1K, 0.5K$. Other parameters: $J = -4.9, k_F a = 0.85 \pi$.}
\label{fig:5}
\end{figure}

       \begin{table}[H]
    \caption{The charge conductance in units of {$\frac{2e^2}{h}$} follows a Lorentzian line shape of $G = \frac{P \Gamma^2}{E^2 + \Gamma^2}$ regardless of temperature at $Z = 0.781$, and below we provide the values of $P$ and $\Gamma$ at $J=4.5$ and -4.9 {at zero voltage bias $\left(\frac{eV}{\Delta_0} \to 0.0\right)$}. Both $G_{ch}^{YSR}$ and $G_{ch}^{MBS}$ decrease with increase in temperature, but the rule $G_{ch}^{YSR}$(peak) < $\frac{G_{ch}^{MBS}\text{(peak)}}{2}$ always holds {(from Figs. \ref{fig:4} and \ref{fig:5})}.}
    \centering
    \begin{tabular}{|c|c|c|c|c|c|c|}
        \hline
        \multirow{2}{*}{$T$} & \multicolumn{2}{c|}{YSR ($s$-wave)} & \multicolumn{2}{c|}{MBS (chiral $p$-wave)} & \multicolumn{2}{c|}{MBS (Spinless $p$-wave nanowire)} \\ \cline{2-7} 
        & $J=4.5$ & $J=-4.9$ & $J=4.5$ & $J=-4.9$ & $J=4.5$ & $J=-4.9$ \\ \hline
        0.0$K$ & $P=0.932, \Gamma = 0.275$ & $P=0.830, \Gamma = 0.285$ & $P=2, \Gamma = 0.36$ & $P=2, \Gamma = 0.39$ & $P=2, \Gamma = 0.416$ & $P=2, \Gamma = 0.38$ \\ \hline
        0.1$K$ & $P=0.930, \Gamma = 0.277$ & $P=0.829, \Gamma = 0.286$ & $P=1.99, \Gamma = 0.37$ & $P=1.99, \Gamma = 0.391$ & $P=1.99, \Gamma = 0.406$ & $P=1.99, \Gamma = 0.372$ \\ \hline
        0.5$K$ & $P=0.897, \Gamma = 0.281$ & $P=0.800, \Gamma = 0.289$ & $P=1.953, \Gamma = 0.393$ & $P=1.960, \Gamma = 0.394$ & $P=1.96, \Gamma = 0.395$ & $P=1.94, \Gamma = 0.243$ \\ \hline
    \end{tabular}
\label{Table1}
\end{table}

\end{widetext}

{The spin conductance \( G_{sp} \) provides an additional and crucial tool for distinguishing between YSR states and MBS. For YSR states, the spin conductance \( G_{sp}^{YSR} \) exhibits a pronounced zero-bias peak for the same barrier strengths, \( Z = 0.781 \) and \( Z = 1.121 \), as shown in Fig. \ref{fig:6}(a) at \( T = 0.0K \). Notably, \( G_{sp}^{YSR} \) follows a Lorentzian line shape, which is consistent with the presence of YSR states.}

{In stark contrast, the behavior of the spin conductance for MBS is fundamentally different. In chiral-\( p \)-wave and spinless \( p \)-wave nanowire systems, the spin conductance \( G_{sp}^{MBS} \) vanishes identically for all values of \( Z \), including \( Z = 0.781 \) and \( Z = 1.121 \), as illustrated in Figs. \ref{fig:6}(b) and \ref{fig:6}(c). This vanishing spin conductance is a direct consequence of the topological nature of MBS and provides a striking contrast to the nonzero spin conductance associated with YSR states. Furthermore, we observe that for MBS, \( G_{sp} \) follows an inverse Lorentzian line shape, which serves as an additional distinguishing feature between YSR states and MBS.}

{We observe that the spin conductance at $\frac{eV}{\Delta_0} = 0.0$ exhibits contrasting temperature dependence for YSR and MBS states. Specifically, for the YSR state, the spin conductance decreases as the temperature increases, whereas for the MBS state, it follows the opposite trend and increases with temperature. This distinct behavior is evident in both Fig. \ref{fig:6} and Table \ref{Table3}. The stark contrast in the thermal response of spin conductance between these two states provides a clear and effective criterion for distinguishing YSR states from MBS states.} 
{This difference arises because YSR states are induced by magnetic impurities in conventional superconductors, while MBS are topologically protected zero-bias states in superconducting systems. Thus, this observation serves as a robust discriminator between YSR and MBS states, offering a valuable experimental signature for their identification.}

\begin{widetext}

\begin{figure}[H]
\centering
\includegraphics[width=1.0\linewidth]{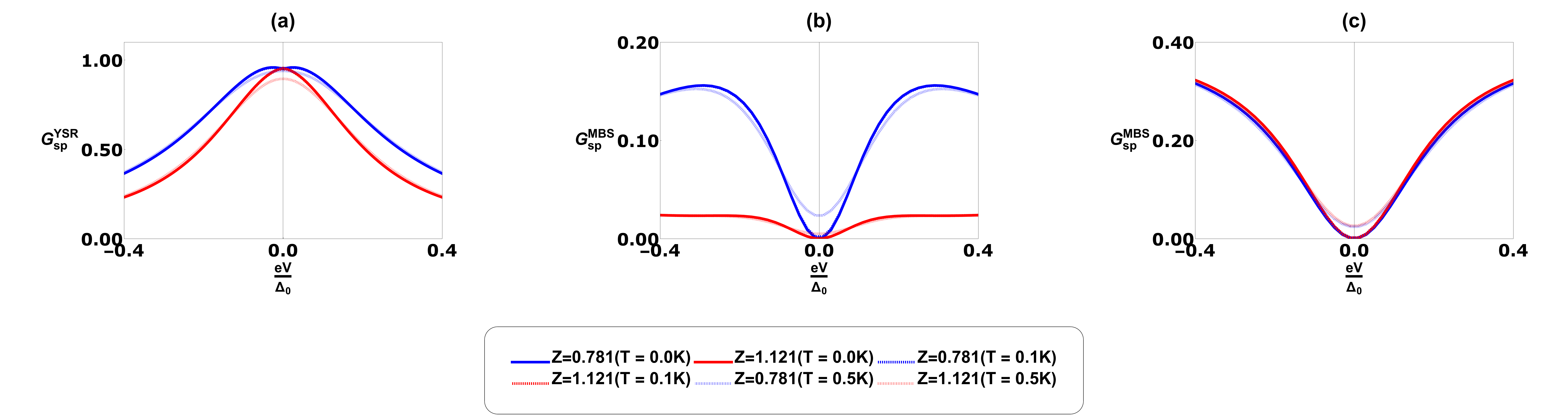}
\caption{Spin conductance in units of {$\frac{2e^2}{h}$} {taking the spin degeneracy into account} in the N-sf-N-I-S setup for (a) $s$-wave superconductor, (b) chiral-$p$ wave superconductor and (c) spinless $p$-wave superconducting nanowire vs. voltage bias applied $\left(\frac{eV}{\Delta_0}\right)$ for $Z = 0.781$ and $Z=1.121$ at temperatures $T = 0.0K, 0.1K, 0.5K$. Other parameters: $J = 4.5, k_F a = 0.85 \pi$. }
\label{fig:6}
\end{figure}

\begin{table}[H]
\centering
\caption{Spin conductance in units of {$\frac{2e^2}{h}$} for both YSR and MBS states at both $Z =0.781$ and 1.121 at $J=4.5$ with {at zero voltage bias $\left(\frac{eV}{\Delta_0} \to 0.0\right)$} $G_{sp}^{YSR} (\text{Peak}) > 2 G_{sp}^{MBS} (\text{Peak})$ regardless of temperature {(from Fig. \ref{fig:6})}.}
\renewcommand{\arraystretch}{1.3} 
\scalebox{0.85}{
\begin{tabular}{|c|c|c|c|c|c|c|}
\hline
\multirow{2}{*}{States} & \multicolumn{2}{c|}{$T=0.0K$} & \multicolumn{2}{c|}{$T=0.1K$} & \multicolumn{2}{c|}{$T=0.5K$} \\ \cline{2-7} 
                        & $Z=0.781$ & $Z=1.121$ & $Z=0.781$ & $Z=1.121$ & $Z=0.781$ & $Z=1.121$ \\ \hline
YSR ($s$-wave) & \multicolumn{1}{m{3cm}|}{\centering Lorentzian \\ $G_{sp} = 0.949$ (eV = 0)} & \multicolumn{1}{m{3cm}|}{\centering Lorentzian \\ $G_{sp} = 0.955$ (eV = 0)} & \multicolumn{1}{m{3cm}|}{\centering Lorentzian \\ $G_{sp} = 0.948$ $(eV = 0)$} & \multicolumn{1}{m{3cm}|}{\centering Lorentzian \\ $G_{sp} = 0.954$ $(eV = 0)$} & \multicolumn{1}{m{3cm}|}{\centering Lorentzian \\ $G_{sp} = 0.939$ $(eV = 0)$} & \multicolumn{1}{m{3cm}|}{\centering Lorentzian \\ $G_{sp} = 0.895$ $(eV = 0)$} \\ \hline
MBS (chiral-$p$ wave) & \multicolumn{1}{m{3cm}|}{\centering Inverted Lorentzian \\ $G_{sp} = 0$ $(eV = 0)$} & \multicolumn{1}{m{3cm}|}{\centering Inverted Lorentzian \\ $G_{sp} = 0$ $(eV = 0)$} & \multicolumn{1}{m{3cm}|}{\centering Inverted Lorentzian \\ $G_{sp} = 0.0023$ $(eV = 0)$} & \multicolumn{1}{m{3cm}|}{\centering Inverted Lorentzian \\ $G_{sp} = 0.0012$ $(eV = 0)$} & \multicolumn{1}{m{3cm}|}{\centering Inverted Lorentzian \\ $G_{sp} = 0.0238$ $(eV = 0)$} & \multicolumn{1}{m{3cm}|}{\centering Inverted Lorentzian \\ $G_{sp} = 0.0051$ $(eV = 0)$} \\ \hline
MBS (spinless-$p$ wave) & \multicolumn{1}{m{3cm}|}{\centering Inverted Lorentzian \\ $G_{sp} = 0$ $(eV = 0)$} & \multicolumn{1}{m{3cm}|}{\centering Inverted Lorentzian \\ $G_{sp} = 0$ $(eV = 0)$} & \multicolumn{1}{m{3cm}|}{\centering Inverted Lorentzian \\ $G_{sp} = 0.02$ $(eV = 0)$} & \multicolumn{1}{m{3cm}|}{\centering Inverted Lorentzian \\ $G_{sp} = 0.002$ $(eV = 0)$} & \multicolumn{1}{m{3cm}|}{\centering Inverted Lorentzian \\ $G_{sp} = 0.02$ $(eV = 0)$} & \multicolumn{1}{m{3cm}|}{\centering Inverted Lorentzian \\ $G_{sp} = 0.02$ $(eV = 0)$} \\ \hline
\end{tabular}}
\label{Table3}
\end{table}
\end{widetext}

{In this analysis, we observed that spin conductance could distinguish YSR states from MBS qualitatively, while charge conductance also could do so, and conversely, charge/spin quantum noise at zero temperature bias could also distinguish YSR states from MBS. Further, we search for a more comprehensive approach that goes beyond conventional charge and spin quantum noise at zero temperature bias to unambiguously identify and differentiate between YSR and MBS states by utilizing charge/spin quantum noise and $\Delta_T$ noise at finite temperature bias. First, it is experimentally feasible and practical to implement ~\cite{lumbroso2018electronic, sivre2019electronic, PhysRevLett.125.106801, melcer2022absent}. Second, it enables a clear distinction between YSR states and MBS in both charge and spin channels. The quantum noise and \( \Delta_T \) noise approach provides a unified framework for differentiating between YSR states and MBS, irrespective of whether charge or spin transport is considered, making it a promising tool for experimental and theoretical investigations.}

\subsection{Dependence of charge and spin quantum noise at zero temperature bias on applied voltage bias}

{In Figs. \ref{fig:21} and \ref{fig:23}, we present the charge and spin quantum noise ($Q_{11}^{ch/sp}$) at zero temperature bias, i.e., when $T_1 = T_2 = T$, as a function of voltage bias ($eV/\Delta_0$), considering temperatures $T=0.1K$ and $0.5K$. As shown in Fig. \ref{fig:21} at zero bias ($\frac{eV}{\Delta_0} = 0.0$), $Z=0.781$, $J=4.5$, and $T=0.1K$, the charge quantum noise ($Q_{11}^{ch}$) for the YSR state is 1.89 (in units of {$\frac{4e^2}{h} k_B T$}). In contrast, for the MBS in a chiral $p$-wave superconductor, it is significantly higher at 4 (in units of {$\frac{4e^2}{h} k_B T$}), which is more than twice the value for the YSR state, see Table \ref{Table16}. A similar trend is observed for the MBS state in a spinless $p$-wave superconductor, where $Q_{11}^{ch}$ reaches 4 (in units of {$\frac{4e^2}{h} k_B T$}), again exceeding twice the value of the YSR case, see Table \ref{Table16}. This clear quantitative difference follows the inequality $Q_{11}^{ch} (YSR) < \frac{Q_{11}^{ch} (MBS)}{2}$, establishing an effective criterion for distinguishing between YSR and MBS states. Furthermore, at a higher temperature of $T=0.5K$, we observe that this inequality remains valid, as seen in Fig. \ref{fig:21} and Table \ref{Table16}. Very similar results to that shown in Fig. \ref{fig:21} and Table \ref{Table16} are seen for $Z=1.121$ with $J = 4.5$ and -8.9 (where YSR states occur) at the same temperatures $T = 0.1K$ and 0.5$K$, and this trend persists: at $\frac{eV}{\Delta_0} = 0.0$, the charge quantum noise in the MBS case is consistently more than twice that of YSR case, reinforcing the robustness of this distinguishing feature. The significant enhancement of charge quantum noise in MBS states compared to YSR states provides a strong and measurable signature that can be utilized to differentiate between these two quantum states.}

\begin{widetext}

\begin{figure}[H]
\centering
\includegraphics[width=1.0\linewidth]{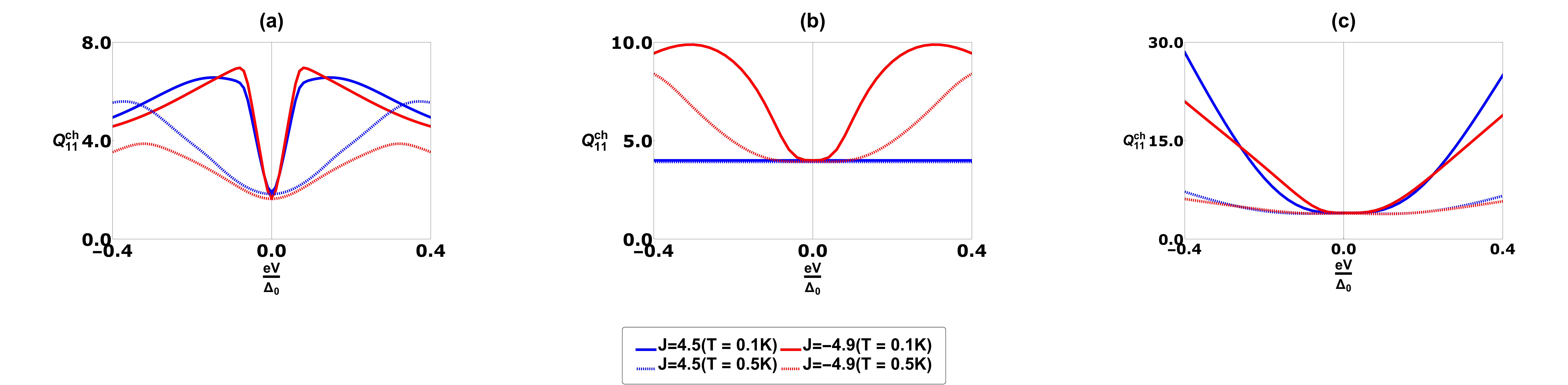}
\caption{Charge quantum noise in units of {$\frac{4e^2}{h} k_B T$} of the N-sf-N-I-S setup for (a) $s$-wave superconductor, (b) chiral-$p$ wave superconductor and (c) spinless $p$-wave superconducting nanowire vs. voltage bias applied $\left(\frac{eV}{\Delta_0}\right)$ for $Z = 0.781$ at temperatures $T = 0.1K, 0.5K$ with $k_F a = 0.85 \pi$. }
\label{fig:21}
\end{figure}

\begin{table}[H]
\caption{Charge quantum noise in units of {$\frac{4e^2}{h} k_B T$} {at zero voltage bias $\left(\frac{eV}{\Delta_0} \to 0.0\right)$} for $Z = 0.781$ at $J=4.5$ and $J=-4.9$, where YSR states occur {(from Fig. \ref{fig:21})}.}
\centering
\begin{tabular}{|l|ll|ll|}
\hline
\multirow{2}{*}{States} & \multicolumn{2}{l|}{$T=0.1K$}    & \multicolumn{2}{l|}{$T=0.5K$}    \\ \cline{2-5} 
                  & \multicolumn{1}{l|}{$J=4.5$} & $J=-4.9$ & \multicolumn{1}{l|}{$J=4.5$} & $J=-4.9$ \\ \hline
         YSR ($s$-wave)         & \multicolumn{1}{l|}{1.89} & 1.68 & \multicolumn{1}{l|}{1.82} & 1.63 \\ \hline
          MBS (chiral $p$-wave)        & \multicolumn{1}{l|}{3.99} & 3.99 & \multicolumn{1}{l|}{3.91} & 3.93 \\ \hline
            MBS (spinless $p$-wave)      & \multicolumn{1}{l|}{3.99} & 3.93 & \multicolumn{1}{l|}{3.95} & 3.93 \\ \hline
\end{tabular}
\label{Table16}
\end{table}
\end{widetext}

{When we extend this analysis to spin quantum noise ($Q_{11}^{sp}$), see Fig. \ref{fig:23}, {We see that the spin quantum noise changes sign and turns positive for YSR states, whereas for MBS states, it is always negative. This observation distinguishes YSR state from MBS.} Also quantitatively, we observe the reverse phenomenon compared to charge quantum noise. At zero bias ($\frac{eV}{\Delta_0} = 0.0$) and at $J = 4.5$, the spin quantum noise for the YSR state is 1.37 (in units of {$\frac{4e^2}{h} k_B T$}), whereas for the MBS state in a chiral $p$-wave superconductor, it {vanishes}. Similarly, in the case of an MBS state in a spinless $p$-wave nanowire, $Q_{11}^{sp}$ {vanishes}. This establishes the inequality {$Q_{11}^{sp} (YSR) \neq 0$ at $\frac{eV}{\Delta_0} \to 0$} as evident from Fig. \ref{fig:23} and Table \ref{Table18}. Importantly, even at a higher temperature of $T=0.5K$, this inequality remains valid, as seen in Fig. \ref{fig:23} and Table \ref{Table18}. Very similar results to that shown in Fig. \ref{fig:23} and Table \ref{Table18} are seen at $Z = 1.121$ for $J= 4.5$ and -8.9 at different temperatures $T = 0.1K$ and $0.5K$. This consistent observation suggests that large spin quantum noise {for} YSR states and {vanishing} spin quantum noise {for} MBS states serve as a crucial signature distinguishing {these states}.}

\begin{widetext}

       \begin{figure}[H]
\centering
\includegraphics[width=1.0\linewidth]{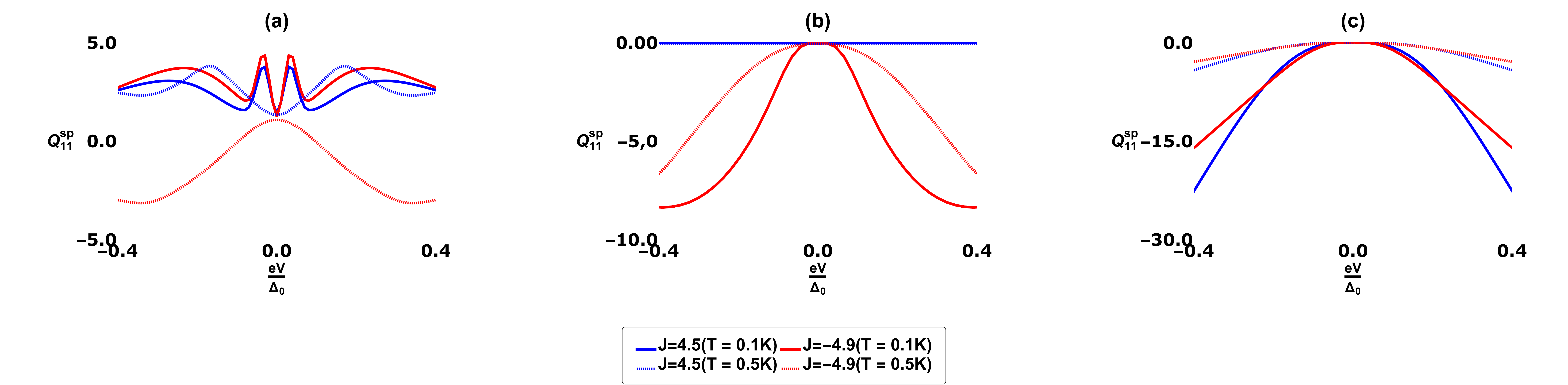}
\caption{Spin quantum noise in units of {$\frac{4e^2}{h} k_B T$} of the N-sf-N-I-S setup for (a) $s$-wave superconductor, (b) chiral-$p$ wave superconductor and (c) spinless $p$-wave superconducting nanowire vs. voltage bias applied $\left(\frac{eV}{\Delta_0}\right)$ for $Z = 0.781$ at temperatures $T = 0.1K, 0.5K$ with $k_F a = 0.85 \pi$.  }
\label{fig:23}
\end{figure}

\begin{table}[H]
\caption{Spin quantum noise in units of {$\frac{4e^2}{h} k_B T$} {at zero voltage bias $\left(\frac{eV}{\Delta_0} \to 0.0\right)$} for $Z = 0.781$ at $J=4.5$ and $J=-4.9$, where YSR states occur {(from Fig. \ref{fig:23})}.}
\centering
\begin{tabular}{|l|ll|ll|}
\hline
\multirow{2}{*}{States} & \multicolumn{2}{l|}{$T=0.1K$}    & \multicolumn{2}{l|}{$T=0.5K$}    \\ \cline{2-5} 
                  & \multicolumn{1}{l|}{$J=4.5$} & $J=-4.9$ & \multicolumn{1}{l|}{$J=4.5$} & $J=-4.9$ \\ \hline
         YSR ($s$-wave)         & \multicolumn{1}{l|}{1.37} & 1.22 & \multicolumn{1}{l|}{1.30} & 1.05 \\ \hline
          MBS (chiral $p$-wave)        & \multicolumn{1}{l|}{{vanishing}} & {vanishing} & \multicolumn{1}{l|}{{vanishing}} & {vanishing} \\ \hline
            MBS (spinless $p$-wave)      & \multicolumn{1}{l|}{{vanishing}} & {vanishing} & \multicolumn{1}{l|}{{vanishing}} & {vanishing} \\ \hline
\end{tabular}
\label{Table18}
\end{table}

\end{widetext}

\subsection{Dependence of charge and spin quantum noise at finite temperature bias ($T_1 \neq T_2$) on spin flip strength}
{
To investigate the behavior of charge and spin quantum noise under a finite temperature bias, we consider a system with two reservoirs at different temperatures, \(T_1\) and \(T_2\), where \(T_1 > T_2\). The temperature difference is defined as \(\Delta T = T_1 - T_2\), and we assume that this difference is small relative to the average temperature \(T = \frac{T_1 + T_2}{2}\), ensuring the condition \(\Delta T \ll T\). To maintain a controlled thermal bias, we express the individual temperatures as:  $T_1 = T + \frac{\Delta T}{2}, T_2 = T - \frac{\Delta T}{2}$.  
This formulation allows us to systematically study quantum noise in the presence of a small thermal gradient while keeping the applied voltage bias at zero (\(eV = 0\)).}

{Figs. \ref{fig:11}–\ref{fig:13} summarize our key findings on charge and spin quantum noise. For our analysis, we consider three different average temperatures: \(T = 0.1K\), \(0.5K\), and \(1.0K\), each with a temperature bias of \(\Delta T = 0.1 T\).}  

{Our results reveal that the charge quantum noise, denoted as \(Q_{11}^{ch}\), exhibits distinct peaks at \(J = -4.9\) and \(J = 4.5\) for a barrier strength of \(Z = 0.781\). These peaks coincide precisely with the locations of YSR states at the lowest temperature considered (\(T = 0.1K\)), as depicted in Fig. \ref{fig:11}(a) and confirmed in Table \ref{Table8}. This observation indicates a strong correlation between the occurrence of charge noise peaks and the presence of YSR states.  
In contrast, when considering MBS, \(Q_{11}^{ch}\) does not exhibit any peaks or dips at these values of \(J\), highlighting a fundamental distinction between YSR states and MBS. The lack of such peak/dip in the charge quantum noise for MBS states serves as a key signature for differentiating them from YSR states.  
At higher temperatures (\(T = 0.5K\) and \(T = 1.0K\)), \(Q_{11}^{ch}\) continues to show distinguishable features for YSR states, including prominent peaks and dips as a function of \(J\), reinforcing their characteristic noise response (see Figs. \ref{fig:11}(b) and \ref{fig:11}(c), along with Table \ref{Table8}). This temperature-dependent evolution suggests that, although thermal effects may modify the noise profile, the fundamental distinction between YSR and MBS states remains evident in the charge noise characteristics.  
A very similar pattern is observed at different barrier strength of \(Z = 1.121\), where \(Q_{11}^{ch}\) exhibits YSR peaks at \(J = 4.5\) and \(J = -8.9\) across all temperature regimes considered. This further supports the robustness of charge quantum noise as a diagnostic tool for detecting YSR states under thermal bias. Once again, for MBS, no such peaks or dips appear, reinforcing their distinct noise behavior.}

\begin{widetext}

       \begin{figure}[H]
\centering
\includegraphics[width=1.0\linewidth]{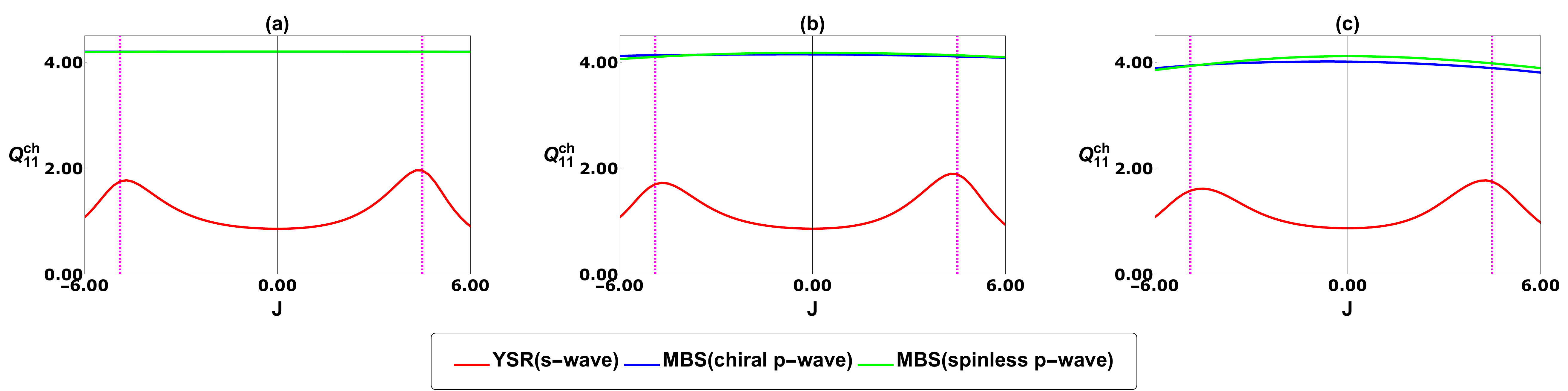}
\caption{Charge quantum noise in units of {$\frac{4e^2}{h} k_B T$} at (a) 0.1$K$, (b) 0.5$K$ and (c) 1.0$K$ vs. the spin-flipper strength ($J$). Other parameters: $Z = 0.781, k_F a = 0.85 \pi$, $\frac{eV}{\Delta_0} = 0.0$ and $\frac{\Delta T}{T} = 0.1$. {The pink verticals at $J = 4.5$ and -4.9 denote where YSR peaks occur.}}
\label{fig:11}
\end{figure}

              \begin{table}[H]
\caption{{Behavior of charge quantum noise} for $Z=0.781$ at $J=4.5$ and -4.9 {(pink verticals)}, where YSR peaks occur. ($\times$) denotes no peak or dip {(from Fig. \ref{fig:11})}.}
\centering
\begin{tabular}{|l|ll|ll|ll|}
\hline
\multirow{2}{*}{$T$} & \multicolumn{2}{l|}{YSR ($s-$wave)}    & \multicolumn{2}{l|}{MBS (chiral $p-$wave)}    & \multicolumn{2}{l|}{MBS (spinless $p-$wave)}    \\ \cline{2-7} 
                  & \multicolumn{1}{l|}{$J=4.5$} & $J=-4.9$ & \multicolumn{1}{l|}{$J=4.5$} & $J=-4.9$ & \multicolumn{1}{l|}{$J=4.5$} & $J=-4.9$ \\ \hline
                 0.1$K$ & \multicolumn{1}{l|}{Peak} & Peak & \multicolumn{1}{l|}{$\times$} & $\times$ & \multicolumn{1}{l|}{$\times$} & $\times$ \\ \hline
                 0.5$K$ & \multicolumn{1}{l|}{Peak} & Peak & \multicolumn{1}{l|}{$\times$} & $\times$ & \multicolumn{1}{l|}{$\times$} & $\times$ \\ \hline
                 1.0$K$ & \multicolumn{1}{l|}{Peak} & Peak & \multicolumn{1}{l|}{$\times$} & $\times$ & \multicolumn{1}{l|}{$\times$} & $\times$ \\ \hline
\end{tabular}
\label{Table8}
\end{table}
\end{widetext}

{Additional insights are obtained by analyzing spin quantum noise as a function of the exchange coupling strength \(J\). Fig. \ref{fig:13} presents the results of our study on spin quantum noise, denoted as \(Q_{11}^{sp}\).  
At the lowest temperature considered (\(T = 0.1K\)), \(Q_{11}^{sp}\) exhibits pronounced peaks at \(J = 4.5\) and near \(J = -4.9\), aligning closely with the locations of YSR states (see Fig. \ref{fig:13}(a) and Table \ref{Table10}). In contrast, for MBS, no such peaks or dips appear, reinforcing the clear difference in their quantum noise signatures.  
This distinguishing behavior persists at higher temperatures (\(T = 0.5K\) and \(T = 1.0K\)), where \(Q_{11}^{sp}\) continues to exhibit a peak near \(J = 4.5\) along with a small dip around \(J = -4.9\) (see Fig. \ref{fig:13}(b) and Table \ref{Table10}). Notably, these features remain absent for MBS, further confirming that spin quantum noise, like charge quantum noise, serves as a reliable probe for differentiating YSR states and MBS. {It is also noteworthy that the $Q_{11}^{sp}$ changes sign for YSR states, whereas for MBS it never changes sign, which further provides compelling justification towards distinction between YSR states and MBS.}

\begin{widetext}

\begin{figure}[H]
\centering
\includegraphics[width=1.0\linewidth]{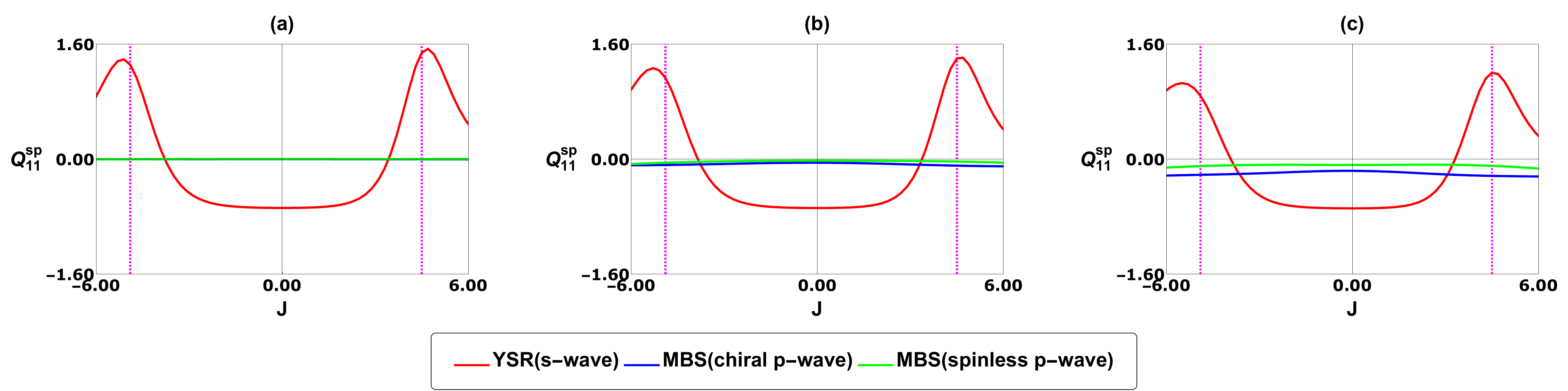}
\caption{Spin quantum noise in units of {$\frac{4e^2}{h} k_B T$} at (a) 0.1$K$, (b) 0.5$K$ and (c) 1.0$K$ vs. the spin-flipper strength ($J$). Other parameters: $Z = 0.781, k_F a = 0.85 \pi$, $\frac{eV}{\Delta_0} = 0.0$ and $\frac{\Delta T}{T} = 0.1$. {The pink verticals at $J = 4.5$ and -4.9 denote where YSR peaks occur.}}
\label{fig:13}
\end{figure}

          \begin{table}[H]
\caption{{Behavior of spin quantum noise} for $Z=0.781$ at $J=4.5$ and -4.9 {(pink verticals)}, where YSR peaks occur. ($\times$) denotes no peak or dip {(from Fig. \ref{fig:13})}.}
\centering
\begin{tabular}{|l|ll|ll|ll|}
\hline
\multirow{2}{*}{$T$} & \multicolumn{2}{l|}{YSR ($s-$wave)}    & \multicolumn{2}{l|}{MBS (chiral $p-$wave)}    & \multicolumn{2}{l|}{MBS (spinless $p-$wave)}    \\ \cline{2-7} 
                  & \multicolumn{1}{l|}{$J=4.5$} & $J=-4.9$ & \multicolumn{1}{l|}{$J=4.5$} & $J=-4.9$ & \multicolumn{1}{l|}{$J=4.5$} & $J=-4.9$ \\ \hline
                 0.1$K$ & \multicolumn{1}{l|}{Peak} & Peak & \multicolumn{1}{l|}{$\times$} & $\times$ & \multicolumn{1}{l|}{$\times$} & $\times$ \\ \hline
                 0.5$K$ & \multicolumn{1}{l|}{Peak} & Peak & \multicolumn{1}{l|}{$\times$} & $\times$ & \multicolumn{1}{l|}{$\times$} & $\times$ \\ \hline
                 1.0$K$ & \multicolumn{1}{l|}{Peak} & Peak & \multicolumn{1}{l|}{$\times$} & $\times$ & \multicolumn{1}{l|}{$\times$} & $\times$ \\ \hline
\end{tabular}
\label{Table10}
\end{table}
\end{widetext}

{In summary, our findings demonstrate that both charge and spin quantum noise provide robust signatures for distinguishing YSR states from MBS under a finite temperature bias. Charge quantum noise exhibits clear peaks at specific values of \(J\) where YSR states occur, while MBS remain devoid of such features. Similarly, spin quantum noise further reinforces this distinction by exhibiting characteristic peaks and dips for YSR states, which are absent in the case of MBS.  
These observations hold across different barrier strengths (\(Z = 0.781\) and \(Z = 1.121\)) and temperature regimes, underscoring the effectiveness of quantum noise analysis as a diagnostic tool. The results highlight the potential of quantum noise measurement. {Here, charge and spin quantum noise are plotted at zero charge and spin current, respectively, and they are approximately of the same magnitude as their respective thermal noise component effectively, known as charge/spin $\Delta_T$ thermal noise. This is because the charge/spin shot noise-like contribution measured at zero charge/spin current, i.e., charge/spin $\Delta_T$ noise, is approximately 1000 times smaller than the thermal noise-like component, see Sec. \ref{results} D.}}

\subsection{Charge and spin $\Delta_T$ noise}

{
We present our results on charge and spin \(\Delta_T\) noise in Figs. \ref{fig:15}–\ref{fig:18}, investigating their behavior under a finite temperature bias. These results are obtained using Eq. (\ref{eq28}) by averaging over both spin-flip and no spin-flip contributions. Consistent with our previous analysis of charge and spin quantum noise, we consider three different average temperatures: \(T = 0.1K\), \(0.5K\), and \(1.0K\), each with a temperature bias of $\frac{\Delta T}{T} = 0.1$, while maintaining zero applied voltage bias (\(eV = 0\)) throughout the analysis.}  

{Our findings reveal that the charge \(\Delta_T\) noise, denoted as \(\Delta_T^{ch}\), exhibits pronounced dips around \(J = 4.5\) and \(J = -4.9\) for barrier strength: \(Z = 0.781\) at lowest temperature considered (\(T = 0.1K\)). In contrast, for MBS, no such peaks or dips appear, highlighting a fundamental distinction between YSR states and MBS (see Fig. \ref{fig:15}(a) and Table \ref{Table12}). As the temperature increases to \(T = 0.5K\) and \(T = 1.0K\), these characteristic dips in \(\Delta_T^{ch}\) persist for YSR states, while MBS do not show any such features (see Figs. \ref{fig:15}(b) and (c) and Table \ref{Table12}). This temperature-dependent behavior further reinforces the ability of charge \(\Delta_T\) noise to effectively differentiate between YSR states and MBS.}  

\begin{widetext}

\begin{figure}[H]
\centering
\includegraphics[width=1.0\linewidth]{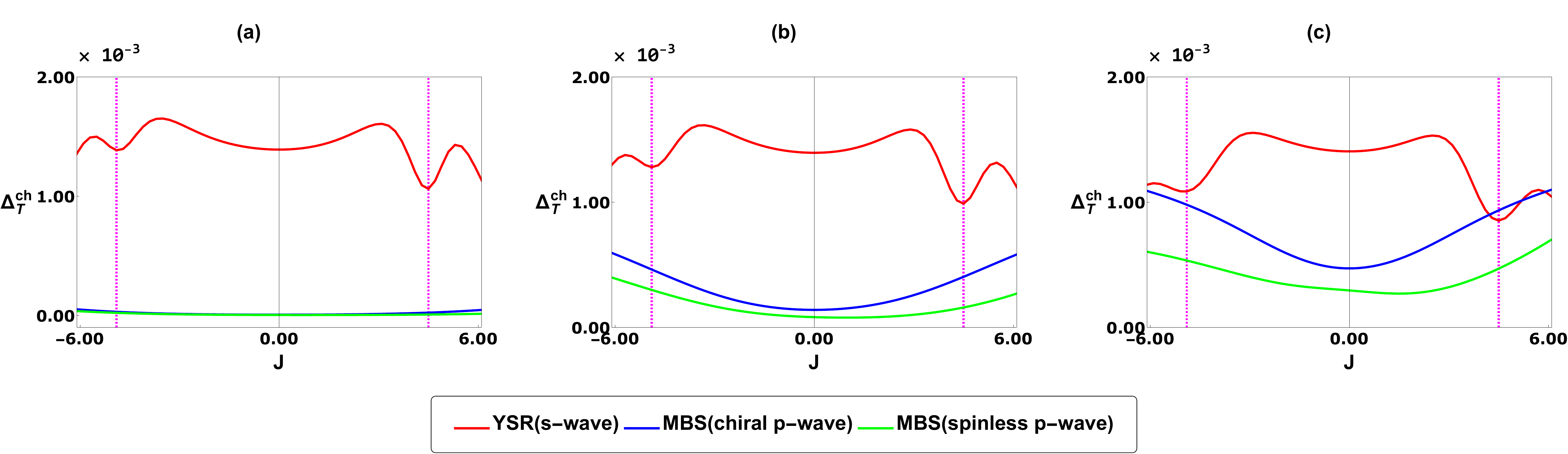}
\caption{Charge $\Delta_T$ noise in units of {$\frac{4e^2}{h} k_B T$} at (a) 0.1$K$, (b) 0.5$K$ and (c) 1.0$K$ vs. the spin-flipper strength ($J$). Other parameters: $Z = 0.781, k_F a = 0.85 \pi$, $\frac{eV}{\Delta_0} = 0.0$ and $\frac{\Delta T}{T} = 0.1$. {The pink verticals at $J = 4.5$ and -4.9 denote where YSR peaks occur.}}
\label{fig:15}
\end{figure}

        \begin{table}[H]
\caption{{Behavior of charge $\Delta_T$ noise} for $Z=0.781$ at $J=4.5$ and -4.9 {(pink verticals)}, where YSR peaks occur. ($\times$) denotes no peak or dip {(from Fig. \ref{fig:15})}.}
\centering
\begin{tabular}{|l|ll|ll|ll|}
\hline
\multirow{2}{*}{$T$} & \multicolumn{2}{l|}{YSR ($s-$wave)}    & \multicolumn{2}{l|}{MBS (chiral $p-$wave)}    & \multicolumn{2}{l|}{MBS (spinless $p-$wave)}    \\ \cline{2-7} 
                  & \multicolumn{1}{l|}{$J=4.5$} & $J=-4.9$ & \multicolumn{1}{l|}{$J=4.5$} & $J=-4.9$ & \multicolumn{1}{l|}{$J=4.5$} & $J=-4.9$ \\ \hline
                 0.1$K$ & \multicolumn{1}{l|}{Dip} & Dip & \multicolumn{1}{l|}{$\times$} & $\times$ & \multicolumn{1}{l|}{$\times$} & $\times$ \\ \hline
                 0.5$K$ & \multicolumn{1}{l|}{Dip} & Dip & \multicolumn{1}{l|}{$\times$} & $\times$ & \multicolumn{1}{l|}{$\times$} & $\times$ \\ \hline
                 1.0$K$ & \multicolumn{1}{l|}{Dip} & Dip & \multicolumn{1}{l|}{$\times$} & $\times$ & \multicolumn{1}{l|}{$\times$} & $\times$ \\ \hline
\end{tabular}
\label{Table12}
\end{table}
\end{widetext}

{A similar trend is observed when considering a different barrier strength of \(Z = 1.121\). Here, \(\Delta_T^{ch}\) exhibits a dip {around} \(J = 4.5\) and a peak around \(J = -8.9\) for YSR states at \(T = 0.1K\) (see Fig. \ref{fig:16}(a) and Table \ref{Table13}). However, for MBS, no such peaks or dips emerge, reinforcing their distinct noise characteristics. As the temperature increases to \(T = 0.5K\) and \(T = 1.0K\), \(\Delta_T^{ch}\) continues to display dips and peaks at \(J = 4.5\) and \(J = -8.9\), respectively, for YSR states, while for MBS, the absence of any such signatures remains consistent (see Figs. \ref{fig:16}(b) and (c) and Table \ref{Table13}). This temperature-independent contrast between YSR states and MBS further validates the robustness of charge \(\Delta_T\) noise as a distinguishing tool.}

\begin{widetext} 

 \begin{figure}[H]
\centering
\includegraphics[width=1.0\linewidth]{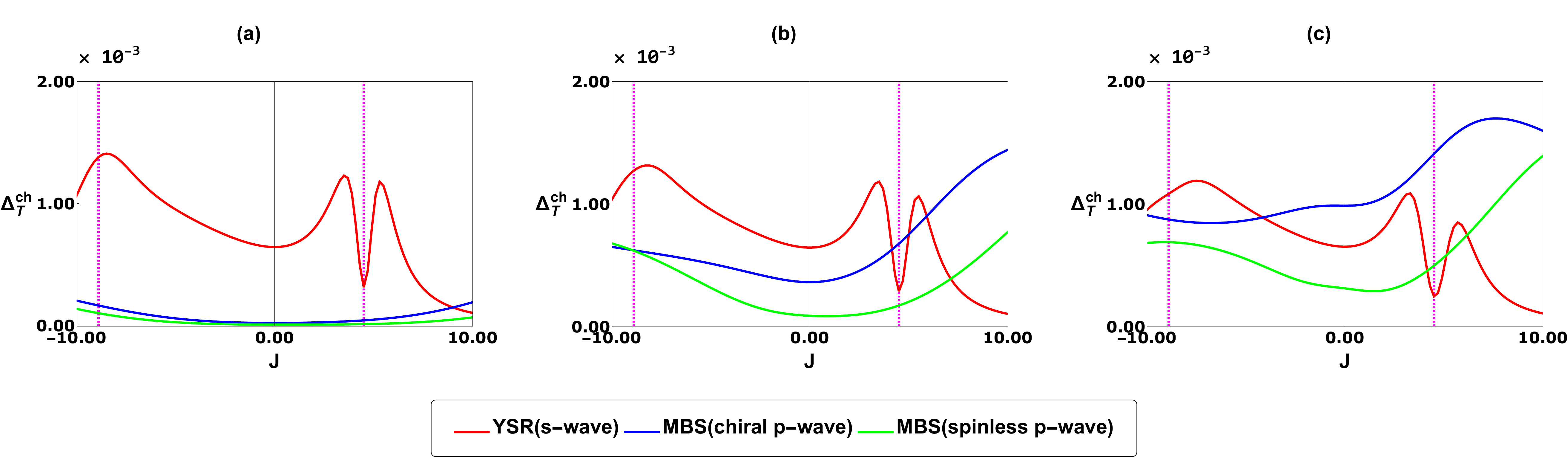}
\caption{Charge $\Delta_T$ noise in units of {$\frac{4e^2}{h}  k_B T$} at (a) 0.1$K$, (b) 0.5$K$ and (c) 1.0$K$ vs. the spin-flipper strength ($J$). Other parameters: $Z = 1.121, k_F a = 0.85 \pi$, $\frac{eV}{\Delta_0} = 0.0$ and $\frac{\Delta T}{T} = 0.1$. {The pink verticals at $J = 4.5$ and -8.9 denote where YSR peaks occur.}  }
\label{fig:16}
\end{figure}      

       \begin{table}[H]
\caption{{Behavior of charge $\Delta_T$ noise} for $Z=1.121$ at $J=4.5$ and -8.9 {(pink verticals)}, where YSR peaks occur. ($\times$) denotes no peak or dip {(from Fig. \ref{fig:16})}.}
\centering
\begin{tabular}{|l|ll|ll|ll|}
\hline
\multirow{2}{*}{$T$} & \multicolumn{2}{l|}{YSR ($s-$wave)}    & \multicolumn{2}{l|}{MBS (chiral $p-$wave)}    & \multicolumn{2}{l|}{MBS (spinless $p-$wave)}    \\ \cline{2-7} 
                  & \multicolumn{1}{l|}{$J=4.5$} & $J=-8.9$ & \multicolumn{1}{l|}{$J=4.5$} & $J=-8.9$ & \multicolumn{1}{l|}{$J=4.5$} & $J=-8.9$ \\ \hline
                 0.1$K$ & \multicolumn{1}{l|}{Dip} & Peak & \multicolumn{1}{l|}{$\times$} & $\times$ & \multicolumn{1}{l|}{$\times$} & $\times$ \\ \hline
                 0.5$K$ & \multicolumn{1}{l|}{Dip} & Peak & \multicolumn{1}{l|}{$\times$} & $\times$ & \multicolumn{1}{l|}{$\times$} & $\times$ \\ \hline
                 1.0$K$ & \multicolumn{1}{l|}{Dip} & Peak & \multicolumn{1}{l|}{$\times$} & $\times$ & \multicolumn{1}{l|}{$\times$} & $\times$ \\ \hline
\end{tabular}
\label{Table13}
\end{table}
\end{widetext}

{{Extending our analysis to spin \(\Delta_T\) noise provides additional valuable insights (see, Fig. \ref{figspin}). For YSR states, spin \(\Delta_T\) noise, denoted as \(\Delta_T^{sp}\), exhibits {peaks near \(J = 4.5\)} and \(J = -4.9\) for \(Z = 0.781\). However, for MBS states originating from a chiral-\(p\) wave superconductor, \(\Delta_T^{sp}\) {does not show any such peak}, indicating a fundamental difference between these states and YSR states (see, Fig. \ref{figspin}(a) and Table \ref{Tablespin}). Importantly, this behavior strongly distinguishes YSR states from MBS states, as the latter do not exhibit any such dips or peaks.}  
{Additionally, for MBS states arising from a spinless-\(p\) wave nanowire, spin \(\Delta_T\) noise does not exhibit any peaks or dips, unlike the YSR states.}}  

{{This distinguishing behavior persists even at elevated temperatures (\(T = 0.5K\) and \(T = 1.0K\)). Specifically, for YSR states, \(\Delta_T^{sp}\) continues to exhibit {peaks} around \(J = 4.5\) and \(J = -4.9\), whereas for MBS states, no such peaks or dips appear at any temperature (see, Figs. \ref{figspin}(b)). As temperature further increases to $T = 1.0K$, we continue to observe {peak} around $J = 4.5$ and $J = -4.9$ for YSR state, see Fig. \ref{figspin}(c). }}
\begin{widetext}

 \begin{figure}[H]
\centering
\includegraphics[width=1.0\linewidth]{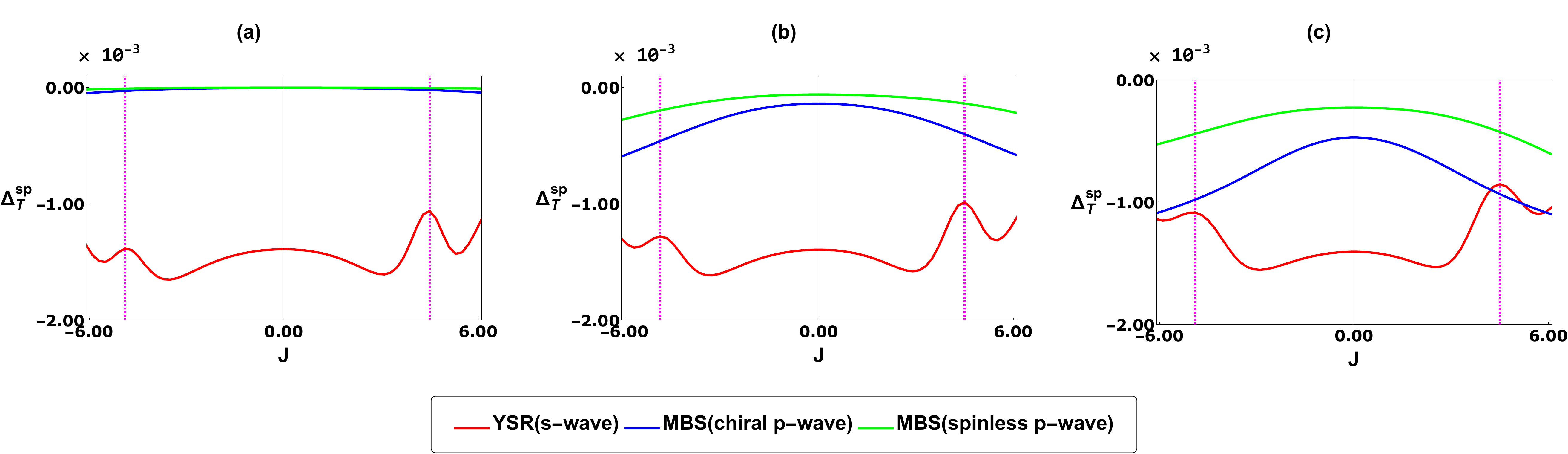}
\caption{{Spin $\Delta_T$ noise in units of {$\frac{4e^2}{h}  k_B T$} at (a) 0.1$K$, (b) 0.5$K$ and (c) 1.0$K$ vs. the spin-flipper strength ($J$). Other parameters: $Z = 0.781, k_F a = 0.85 \pi$, $\frac{eV}{\Delta_0} = 0.0$ and $\frac{\Delta T}{T} = 0.1$. {The pink verticals at $J = 4.5$ and -4.9 denote where YSR peaks occur.}}}
\label{figspin}
\end{figure}   

\begin{table}[H]
\caption{{Behavior of spin $\Delta_T$ noise for $Z=0.781$ at $J=4.5$ and $-4.9$ (pink verticals), where YSR peaks occur. ($\times$) denotes no peak or dip (from Fig.~\ref{figspin}).}}
\centering
\begin{tabular}{|l|ll|ll|ll|}
\hline
\multirow{2}{*}{$T$} & \multicolumn{2}{l|}{{YSR ($s$-wave)}}    & \multicolumn{2}{l|}{{MBS (chiral $p$-wave)}}    & \multicolumn{2}{l|}{{MBS (spinless $p$-wave)}}    \\ \cline{2-7} 
                  & \multicolumn{1}{l|}{{$J=4.5$}} & {$J=-4.9$} & \multicolumn{1}{l|}{{$J=4.5$}} & {$J=-4.9$} & \multicolumn{1}{l|}{{$J=4.5$}} & {$J=-4.9$} \\ \hline
                 {0.1$K$} & \multicolumn{1}{l|}{{Peak}} & {Peak} & \multicolumn{1}{l|}{{$\times$}} & {$\times$} & \multicolumn{1}{l|}{{$\times$}} & {$\times$} \\ \hline
                 {0.5$K$} & \multicolumn{1}{l|}{{Peak}} & {Peak} & \multicolumn{1}{l|}{{$\times$}} & {$\times$} & \multicolumn{1}{l|}{{$\times$}} & {$\times$} \\ \hline
                 {1.0$K$} & \multicolumn{1}{l|}{{Peak}} & {Peak} & \multicolumn{1}{l|}{{$\times$}} & {$\times$} & \multicolumn{1}{l|}{{$\times$}} & {$\times$} \\ \hline
\end{tabular}
\label{Tablespin}
\end{table}
\end{widetext}

{Similarly, at $Z = 1.121$, $\Delta_T^{sp}$ exhibits a peak around $J = 4.5$ and a dip at $J = -8.9$ at $T = 0.1K$, see Fig. \ref{fig:18}(a) and Table \ref{Table15}. However, for MBS states no such peak or dip are seen, which clearly distinguishes YSR states from MBS effectively. Similarly, as temperature increases to $T = 0.5K$ or 1.0K, the peaks at $J = 4.5$ and a dip at $J = - 8.9$ are observed for YSR states, whereas for MBS, no such peaks or dips are seen, see Figs. \ref{fig:18} (b) and (c). This suggests that spin \(\Delta_T\) noise is an effective probe for distinguishing between YSR and MBS states even at higher temperature regimes.}

\begin{widetext}

 \begin{figure}[H]
\centering
\includegraphics[width=1.0\linewidth]{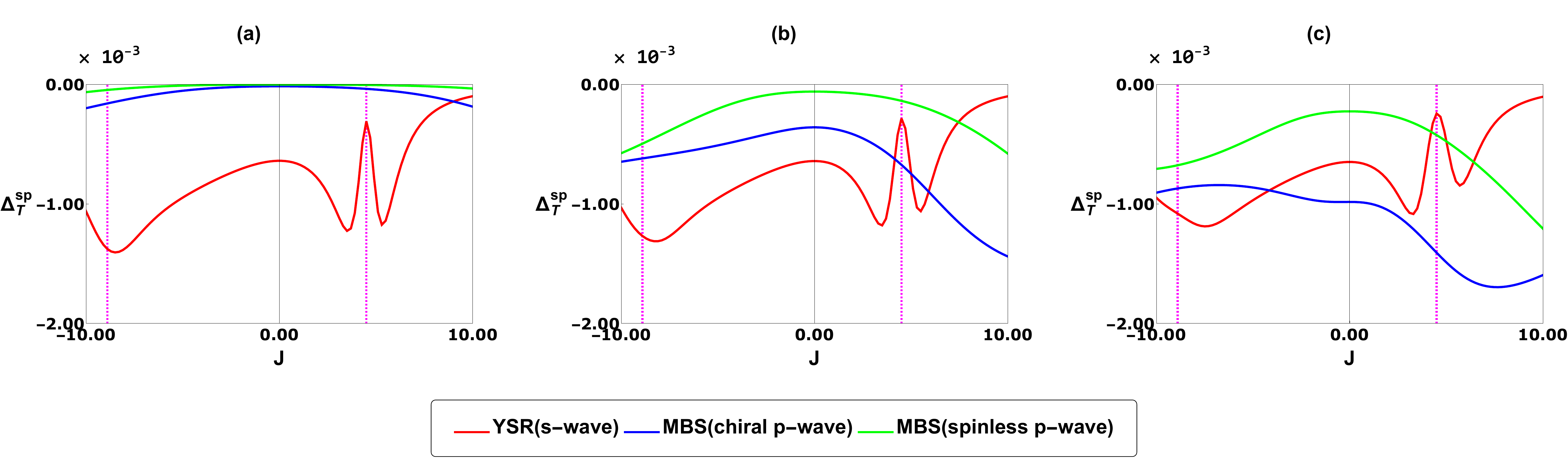}
\caption{Spin $\Delta_T$ noise in units of {$\frac{4e^2}{h}  k_B T$} at (a) 0.1$K$, (b) 0.5$K$ and (c) 1.0$K$ vs. the spin-flipper strength ($J$). Other parameters: $Z = 1.121, k_F a = 0.85 \pi$, $\frac{eV}{\Delta_0} = 0.0$ and $\frac{\Delta T}{T} = 0.1$. {The pink verticals at $J = 4.5$ and -8.9 denote where YSR peaks occur.}}
\label{fig:18}
\end{figure}   

       \begin{table}[H]
\caption{{Behavior of spin $\Delta_T$ noise} for $Z=1.121$ at $J=4.5$ and -8.9 {(pink verticals)}, where YSR peaks occur. ($\times$) denotes no peak or dip {(from Fig. \ref{fig:18})}.}
\centering
\begin{tabular}{|l|ll|ll|ll|}
\hline
\multirow{2}{*}{$T$} & \multicolumn{2}{l|}{YSR ($s-$wave)}    & \multicolumn{2}{l|}{MBS (chiral $p-$wave)}    & \multicolumn{2}{l|}{MBS (spinless $p-$wave)}    \\ \cline{2-7} 
                  & \multicolumn{1}{l|}{$J=4.5$} & $J=-8.9$ & \multicolumn{1}{l|}{$J=4.5$} & $J=-8.9$ & \multicolumn{1}{l|}{$J=4.5$} & $J=-8.9$ \\ \hline
                 0.1$K$ & \multicolumn{1}{l|}{Peak} & Dip & \multicolumn{1}{l|}{$\times$} & $\times$ & \multicolumn{1}{l|}{$\times$} & $\times$ \\ \hline
                 0.5$K$ & \multicolumn{1}{l|}{Peak} & Dip & \multicolumn{1}{l|}{$\times$} & $\times$ & \multicolumn{1}{l|}{$\times$} & $\times$ \\ \hline
                 1.0$K$ & \multicolumn{1}{l|}{Peak} & Dip & \multicolumn{1}{l|}{$\times$} & $\times$ & \multicolumn{1}{l|}{$\times$} & $\times$ \\ \hline
\end{tabular}
\label{Table15}
\end{table}

\end{widetext}

{In conclusion, our results demonstrate that both charge and spin \(\Delta_T\) noise provide powerful and consistent signatures for differentiating YSR states from MBS states under a finite temperature bias. Charge \(\Delta_T\) noise exhibits distinct dips and peaks at specific values of \(J\) for YSR states, whereas MBS states do not exhibit any such features, reinforcing their fundamentally different noise characteristics. Likewise, spin \(\Delta_T\) noise further strengthens this distinction, with YSR states showing clear peaks and dips at characteristic \(J\) values, while MBS states exhibit an entirely different noise behavior depending on their origin (chiral-\(p\) wave or spinless-\(p\) wave).}  

{These findings highlight the robustness of quantum noise analysis as a diagnostic tool for identifying YSR states and distinguishing them from different types of MBS states in superconducting systems. The ability of charge and spin \(\Delta_T\) noise to retain their distinguishing features across different barrier strengths and temperature regimes further enhances their potential for experimental applications in the study of exotic quantum states in mesoscopic superconducting systems.  }

{
In summary, while charge/spin conductance provides initial insights about the distinction between YSR and MBS states both qualitatively and quantitatively, whereas charge/spin quantum noise, as well as charge/spin $\Delta_T$ noise analyses offer additional and effective probes for distinguishing YSR and MBS states.}

\section{Analysis}
\label{Analysis}

{
In this work, we have explored multiple probing techniques capable of detecting YSR states and distinguishing them from MBS. These probes include charge conductance, spin conductance, charge and spin quantum noise, as well as charge and spin $\Delta_T$ noise. In this section, we provide an intuitive explanation of our results and further compare our work with the existing literature.} 

\subsection{Intuitive explanantion}

{To begin with, we analyze charge and spin conductance for both YSR states and MBS. As illustrated in Fig. \ref{fig:2}, we observe that the two YSR state energies merge at specific values of the spin-flipper strength $J$. For barrier strength of $Z = 0.781$, these states coalesce at $J = 4.5$ and $J = -4.9$ (see, Fig. \ref{fig:2}(a)), while for $Z = 1.121$, the coalescing occurs at $J = 4.5$ and $J = -8.9$ (see, Fig. \ref{fig:2}(b)). The merging of bound states at particular energies results in the emergence of YSR peaks, as demonstrated in Figs. \ref{fig:4}(a) and \ref{fig:5}(a). {However, the precise value
of $k_F a$ does not affect the main findings, charge/spin conductance, charge/spin quantum noise and $\Delta_T$ noise, as long as it allows a pair of zero-energy YSR
states to form.}}

{As shown in Section \ref{theory} E, Eq. (\ref{eq:175}), charge conductance ($G_{ch}$) is determined by a combination of Andreev and normal reflection probabilities. Specifically, it is proportional to $
(1 + \mathcal{A}^{\uparrow \uparrow} + \mathcal{A}^{\downarrow \uparrow} - \mathcal{B}^{\uparrow \uparrow} - \mathcal{B}^{\downarrow \uparrow})$.
Similarly, spin conductance ($G_{sp}$), shown in Eq. (\ref{eq:176}) and is proportional to $
(1 + \mathcal{A}^{\uparrow \uparrow} - \mathcal{A}^{\downarrow \uparrow} - \mathcal{B}^{\uparrow \uparrow} + \mathcal{B}^{\downarrow \uparrow})$.
Here, $\mathcal{A}^{\sigma' \sigma}$ represents the Andreev reflection probability, and $\mathcal{B}^{\sigma' \sigma}$ represents the normal reflection probability for an initial spin state $\sigma \in \{\uparrow, \downarrow\}$ transitioning to a final spin state $\sigma' \in \{\uparrow, \downarrow\}$. We compute charge and spin conductance for each configuration and then average over the four possible configurations, which are described in detail in Section \ref{theory} D. Further, the values of $\mathcal{A}^{\uparrow \uparrow}, \mathcal{A}^{\downarrow \uparrow}, \mathcal{B}^{\uparrow \uparrow}$ and $\mathcal{B}^{\downarrow \uparrow}$ at $\frac{eV}{\Delta_0} = 0.0$ are given in Tables \ref{Table170}-\ref{Table173} in Appendix \ref{App:prob}.}  

{At zero temperature ($T = 0.0K$), the zero-bias peaks of $G_{ch}$ for YSR states ($G_{ch}^{YSR}$) are not strictly quantized at $2e^2/h$, yet they exhibit a strong resemblance to Majorana bound states (MBS), see Figs. \ref{fig:4} and \ref{fig:5}. In contrast, the charge conductance for MBS ($G_{ch}^{MBS}$) is precisely quantized at $2e^2/h$ at zero bias ($\frac{eV}{\Delta_0} = 0.0$), see Figs. \ref{fig:4} and \ref{fig:5}. This perfect quantization results from a fully transparent junction, ensuring perfect Andreev reflection for MBS. For MBS, $\mathcal{A}^{\downarrow \uparrow} = 1$ for all configurations at zero energy, while $\mathcal{A}^{\uparrow \uparrow} = 0$, $\mathcal{B}^{\uparrow \uparrow} = 0$, and $\mathcal{B}^{\downarrow \uparrow} = 0$ at $\frac{eV}{\Delta_0} = 0.0$. Consequently, $G_{ch}^{MBS}$ remains strictly quantized even after averaging over configurations. However, for YSR states, perfect Andreev reflection does not occur, although $G_{ch}^{YSR}$ reaches a maximum at specific values such as $J = 4.5$ and $-4.9$ for $Z = 0.781$, leading to observable charge conductance peaks. These peaks arise due to dominant Andreev reflection, which although imperfect, meaning $\mathcal{A}^{\uparrow \uparrow}$ and $\mathcal{A}^{\downarrow \uparrow} < 1$, particularly in configuration-2, where spin-flip scattering occurs. Similarly, the normal reflection probabilities $\mathcal{B}^{\uparrow \uparrow}$ and $\mathcal{B}^{\downarrow \uparrow}$ are finite and remain below 1, but large ($\simeq 0.8$) in configuration-2 at zero energy, wherein we have spin-flip scattering.}  
{In configuration-1, where spin-flip scattering does not occur, $\mathcal{A}^{\uparrow \uparrow} = 0$ and $\mathcal{B}^{\downarrow \uparrow} = 0$ always hold. At $J = 4.5$ and $J = -4.9$ for $Z = 0.781$, the reflection probabilities in configuration-1 are $\mathcal{A}^{\downarrow \uparrow} = 0.12$, $\mathcal{A}^{\uparrow \uparrow} = 0$, $\mathcal{B}^{\uparrow \uparrow} = 0.88$, and $\mathcal{B}^{\downarrow \uparrow} = 0$ at zero enery. Similarly, in configuration-2, $\mathcal{A}^{\downarrow \uparrow} = 0.04$, $\mathcal{A}^{\uparrow \uparrow} = 0.78$, $\mathcal{B}^{\uparrow \uparrow} = 0.009$, and $\mathcal{B}^{\downarrow \uparrow} = 0.17$ at zero energy. Evaluating charge conductance for each configuration and averaging over them gives 0.93$\frac{2e^2}{h}$. However, in configuration-2, the Andreev reflection probability $\mathcal{A}^{\uparrow \uparrow} = 0.78$ is significantly large, which results in a conductance peak. If the value of $J$ deviates from $J = 4.5$ or $J = -4.9$, Andreev reflection is suppressed, and the conductance peak disappears. As a result, peaks emerge only at these values of $J$, see Fig. \ref{fig:19}(a). In contrast, for MBS, $G_{ch}^{MBS}$ remains perfectly quantized regardless of the value of $J$ and also regardless of configuration, due to perfect Andreev reflection confirming its topological origin, as illustrated in Figs. \ref{fig:19}(b) and \ref{fig:19}(c).}
{At finite temperatures, the calculation of charge conductance requires contributions from zero energy as well as finite energies (see, Eq.~(\ref{eq:175})). At finite energies, there is no perfect Andreev reflection, even for MBS.
At non-zero temperatures, $G_{ch}^{YSR}$ still exhibits peak at $J = 4.5$ and $J = -4.9$, while the quantization of $G_{ch}^{MBS}$ gradually deteriorates, see Fig. \ref{fig:19}. The charge conductance for MBS loses perfect quantization and the conductance starts varying as a function of $J$ at finite temperatures. A crucial distinction between YSR states and MBS is that the charge conductance for YSR states always remains less than that for MBS regardless of temperatures, obeying the condition $G_{ch}^{YSR} < \frac{G_{ch}^{MBS}}{2}$ at zero voltage bias ($\frac{eV}{\Delta_0} = 0.0$). This quantitative difference serves as a good discriminator between YSR states and MBS. Moreover, the charge conductance at zero bias for both YSR states and MBS decrease with an increase in temperature, yet the aforementioned condition remains valid.}

{Furthermore, the spin conductance exhibits a more distinctive behavior that allows a clearer distinction between YSR and MBS states. In YSR states, the spin conductance ($G_{sp}^{YSR}$) follows a Lorentzian line shape, whereas for MBS ($G_{sp}^{MBS}$), it exhibits an inverted Lorentzian line shape, see Fig. \ref{fig:6} and Table \ref{Table3}. Specifically, $G_{sp}^{YSR}$ shows a peak at zero bias ($\frac{eV}{\Delta_0} = 0.0$), while $G_{sp}^{MBS}$ vanishes at zero bias. The vanishing of $G_{sp}^{MBS}$ again arises due to perfect Andreev reflection ($\mathcal{A}^{\downarrow \uparrow} = 1$ for both configuration-1 and 2), whereas for $G_{sp}^{YSR}$, despite enhanced maximum Andreev reflection occurring ($\mathcal{A}^{\uparrow \uparrow}$  = 0.78 for configuration-2), it is insufficient to generate a dip similar to MBS.}
{At higher temperatures (after considering contribution from finite energies, see Eq. (\ref{eq:176})), the peak for $G_{sp}^{YSR}$ at $\frac{eV}{\Delta_0} = 0.0$ remains visible, while $G_{sp}^{MBS}$ shows a dip at zero bias, maintaining its inverted Lorentzian shape, see Fig. \ref{fig:6} and Table \ref{Table3}. When plotting $G_{sp}^{YSR}$ and $G_{sp}^{MBS}$, peaks appear at $J = 4.5$ and $J = -4.9$ for $Z = 0.781$, whereas no such peaks are observed in $G_{sp}^{MBS}$, see Fig. \ref{fig:20} and Table \ref{Table17}. In fact, at zero temperature, $G_{sp}^{MBS}$ vanishes entirely for all values of $J$ due to perfect Andreev reflection ($\mathcal{A}^{\downarrow \uparrow} = 1, \mathcal{A}^{\uparrow \uparrow} = 0$ in both configurations 1 and 2). As temperature increases, peaks in $G_{sp}^{YSR}$ persist at the aforementioned values of $J$, whereas no corresponding peaks are seen in $G_{sp}^{MBS}$ (see, Fig. \ref{fig:20} and Table \ref{Table17}).} 

{Now, coming to the charge quantum noise at finite temperature bias and zero voltage bias, which consists of both a thermal noise-like contribution and a shot noise-like contribution, we observe distinct peaks at $J = 4.5$ and $J = -4.9$ for the YSR state, see, Fig. \ref{fig:11}. However, no such peaks are present for MBS, as shown in Fig. \ref{fig:11}. The calculation of total quantum noise also requires the contribution from zero energy as well as finite energies too, see Eq. (\ref{eq:B3}) in Appendix \ref{App_Qn}. The charge quantum noise exhibits qualitative behavior similar to charge conductance. This occurs because, when computing charge quantum noise at finite temperature bias and zero voltage bias, the thermal noise-like component, i.e., $\Delta_T$ thermal noise significantly outweighs the shot noise-like component, and the $\Delta_T$ thermal noise conveys the same information as conductance. The dominance of the thermal noise-like contribution is approximately 1000 times greater than that of shot noise. Consequently, peaks in charge quantum noise appear at $J = 4.5$ and $J = -4.9$ for YSR states, while MBS do not exhibit such peaks. This distinction is crucial for differentiating between YSR states and MBS.}

{When analyzing charge \(\Delta_T\) noise, which represents the shot noise-like contribution to charge quantum noise at finite temperature bias and zero voltage bias, we observe distinct dips at \( J = 4.5 \) and \( J = -4.9 \) for YSR states, see, Fig. \ref{fig:15}  and Table \ref{Table12}. This behavior is in direct contrast to charge quantum noise, which follows the same trend as charge conductance under finite temperature bias and zero voltage bias. The charge \(\Delta_T\) noise is proportional to the expression {$(\mathcal{A}^{\uparrow \uparrow} \mathcal{B}^{\downarrow \uparrow} + \mathcal{A}^{\downarrow \uparrow} \mathcal{B}^{\uparrow \uparrow}) (f_1 - f_2)^2 $}, which requires contribution from both zero as well as finite energies. {The additional terms in Eq.~(\ref{eqn})---specifically those involving only the product of amplitudes, such as \( r_N^{\downarrow \uparrow} r_{Na}^{\uparrow \uparrow^*} r_N^{\uparrow \uparrow^*} r_{Na}^{\downarrow \uparrow} \)---make a negligible contribution to the charge noise \( \Delta_T \). In configuration-1, the amplitudes \( r_{Na}^{\uparrow \uparrow} \) and \( r_N^{\downarrow \uparrow} \) are identically zero, while in configuration-2, the amplitudes \( r_{Na}^{\downarrow \uparrow} \) and \( r_N^{\uparrow \uparrow} \) remain very small near zero energy. As a result, this product of amplitudes remains negligible across both configurations.
} We compute the charge \(\Delta_T\) noise for each configuration and then take the average. {In configuration-1, \(\mathcal{A}^{\uparrow \uparrow} = 0\) and $\mathcal{A}^{\downarrow \uparrow} = 0.12$, whereas $\mathcal{B}^{\uparrow \uparrow} = 0.88$ and $\mathcal{B}^{\downarrow \uparrow} = 0$, while in configuration-2, \(\mathcal{A}^{\uparrow \uparrow} \approx 0.78\), $\mathcal{A}^{\downarrow \uparrow} = 0.04$ at zero energy and $\mathcal{B}^{\uparrow \uparrow} = 0.009$ and $\mathcal{B}^{\downarrow \uparrow} = 0.17$.} The significantly larger value of \(\mathcal{A}^{\uparrow \uparrow}\) in configuration-2 is primarily responsible for the dip in charge \(\Delta_T\) noise. As \(\mathcal{A}^{\uparrow \uparrow}\) increases, the charge \(\Delta_T\) noise tends to diminish. In the case of YSR states, enhanced Andreev reflection occurs at \( J = 4.5 \) and \( J = -4.9 \) when the barrier strength is \( Z = 0.781 \). As a result, the charge noise \( \Delta_T \) is expected to decrease in this regime. At any other barrier strengths, these dips do not appear. Consequently, we observe a dip at these specific barrier strength values for YSR states (see, Fig.~\ref{fig:15} and Table~\ref{Table12}). In contrast, for MBS, Andreev reflection remains maximum and perfect across all barrier strengths, meaning that no dips are observed at particular values, unlike in YSR states. A similar trend is observed for \( J = 4.5 \) at \( Z = 1.121 \) (see, Fig.~ and Table~XII of the revised manuscript). However, for \( J = -8.9 \), a peak rather than a dip is observed (see, Fig.~\ref{fig:16} and Table~\ref{Table13}). {This peak, instead of a dip, arises because at finite temperatures, contributions from finite-energy states become significant, and these contributions can reverse the expected dip into a peak.} Notably, for MBS, neither peaks nor dips are present under these conditions.}    

{We observe that the spin quantum noise {at zero voltage bias} changes sign in the presence of YSR states, whereas no such sign change occurs for MBS, as shown in Fig.~\ref{fig:13}. The evaluation of spin quantum noise involves contributions from both zero-energy and finite-energy states, as outlined in Eq.~(\ref{eq:B4}) in Appendix~\ref{App_Qn}. From Eq.~(\ref{eq:B4}), the spin quantum noise is dependent primarily on the terms $\mathcal{A}^{\uparrow \uparrow^2} + \mathcal{A}^{\downarrow \uparrow^2} + \left(1 - \mathcal{B}^{\uparrow \uparrow} + \mathcal{B}^{\downarrow \uparrow}\right)^2 + 2 \mathcal{A}^{\uparrow \uparrow} - 2 \mathcal{A}^{\downarrow \uparrow} - 2 \mathcal{A} \mathcal{B}$. The remaining terms in Eq.~(\ref{eq:B4}) that involve only products of amplitudes contribute negligibly to the total spin quantum noise. This is because, in configuration-1, the amplitudes \( r_{N}^{\downarrow \uparrow} \) and \( r_{Na}^{\uparrow \uparrow} \) vanish, while in configuration-2, the amplitudes \( r_{N}^{\uparrow \uparrow} \) and \( r_{Na}^{\downarrow \uparrow} \) remain very small near zero energy. Consequently, the contributions from these amplitude-product terms to the total spin \( \Delta_T \) noise are negligible in both configurations. The dominant contribution near the values of $J$ where YSR states appear arises from the terms $\mathcal{A}^{\uparrow \uparrow^2} + \mathcal{A}^{\downarrow \uparrow^2} + \left(1 - \mathcal{B}^{\uparrow \uparrow} + \mathcal{B}^{\downarrow \uparrow}\right)^2 + 2 \mathcal{A}^{\uparrow \uparrow}$, which outweigh the remaining terms due to enhanced value of $\mathcal{A}^{\uparrow \uparrow}$, i.e., Andreev reflection with spin-flip scattering (see, Figs. \ref{fig:201}(a) and \ref{fig:203}(a)} around zero energy, resulting in a positive spin quantum noise. In contrast, for MBS, the quantity $\mathcal{A}^{\downarrow \uparrow}$ remains fixed at unity around zero energy (see, Figs. \ref{fig:200}-\ref{fig:203}) for both the configurations, which causes the spin quantum noise to remain negative regardless of the value of the spin-flip barrier strength.
}

{The shot noise-like contribution to the spin quantum noise at finite temperature bias and zero voltage bias is spin $\Delta_T$ noise. However, the spin \(\Delta_T\) noise exhibits {peaks around both \( J = 4.5 \) and \( J = -4.9 \) for \( Z = 0.781 \) at \( T = 0.1K \), \( T = 0.5K \) and $T = 1.0K$, see Fig. \ref{figspin}}. The expression for spin \(\Delta_T\) noise is proportional to {$(
- 4 \mathcal{A}^{\downarrow \uparrow} \mathcal{B}^{\uparrow \uparrow} - 4 \mathcal{A}^{\uparrow \uparrow} \mathcal{B}^{\downarrow \uparrow}) (f_1 - f_2)^2$
(see, Eq.~(\ref{eqn})) and requires contribution from both zero energy and finite energies and this has same magnitude as charge $\Delta_T$ noise. Thus, the spin \(\Delta_T\) noise has a negative sign, but the magnitude is same as charge $\Delta_T$ noise.} {Similarly, in Fig. \ref{fig:18}, we observe a peak at $J = 4.5$ and a dip at $J = -8.9$ for $Z = 1.121$, which is opposite of the behavior of charge $\Delta_T$ noise shown in Fig. \ref{fig:16}.}

\subsection{Comparison with existing literature}
{Several works \cite{PhysRevB.96.075161, PhysRevB.96.184520} have extensively used charge conductance as the primary probe to differentiate between trivial and topological ZBCPs. However, none of the existing literature has explored alternative transport probes to distinguish trivial zero-bias modes from topological ones. This gap in research is precisely what our work addresses. We present, for the first time, a comprehensive study incorporating also spin conductance, charge/spin quantum noise, and charge/spin \(\Delta_T\) noise as novel diagnostic tools in the presence of both YSR states and MBS, while providing a fresh perspective on charge conductance itself. This represents a significant advancement in the field, providing deeper insight into the nature of these exotic quantum states beyond what charge conductance alone can reveal.}

{A study presented in Refs. \cite{PhysRevB.96.075161, PhysRevB.96.184520} attempted to distinguish the zero-bias peaks arising from topological Majorana bound states and non-topological Andreev bound states (ABS) motivated by an experiment done in Ref. \cite{deng2016majorana}. Specifically in Ref. \cite{PhysRevB.96.184520}, a scaling relation was derived that described the dependence of charge conductance on temperature and barrier strength. The charge conductance for MBS was shown to follow a Lorentzian line shape as a function of these parameters. However, it was established that this scaling relation holds only at low temperatures and high barrier strengths, and not at high temperatures and low barrier strengths. When the same analysis was applied to non-topological ABS, it was found that their charge conductance exhibited a similar dependence on temperature and barrier strength, making it difficult to qualitatively distinguish between the two bound states. Similarly, in Ref. \cite{PhysRevB.96.075161} also, an attempt towards the trivial and topological zero bias modes are attempted, but failed to provide a clear distinction between them. Therefore, the authors further suggest experimental techniques to effectively differentiate between topological and trivial zero-energy conductance peaks.  
One possible approach to making this distinction is by varying the quantum dot confinement potential. This variation can significantly affect the conductance in the trivial zero-bias mode case, whereas the conductance of the topological zero-bias mode remains unchanged \cite{PhysRevB.96.075161}.}

{Our work demonstrates that while charge conductance for both YSR and MBS states follows a Lorentzian line shape, the fundamental distinction lies in the fact that $G_{ch}^{YSR} < \frac{G_{ch}^{MBS}}{2}$ at all temperatures. On the other hand, for YSR states, the zero-bias charge conductance remains always lower than that of MBS states across the temperature range considered. In Refs. \cite{PhysRevB.96.184520, PhysRevB.96.075161}, it is reported that the zero-bias charge conductance for an MBS is always lower than that of an ABS, regardless of temperature.} 

{Beyond charge conductance, our study looks at other probes like the spin conductance, charge and spin quantum noise {at zero charge and spin current respectively}, and charge and spin $\Delta_T$ noise. These probes prove to be quite effective in distinguishing between YSR and MBS states. One particularly striking distinction arises in spin conductance. For YSR states, spin conductance exhibits a Lorentzian line shape as a function of voltage bias at any given temperature. In contrast, for MBS states, the spin conductance follows an inverted Lorentzian line shape. This crucial difference provides a strong and clear method for distinguishing the two bound states. Notably, Ref. \cite{PhysRevB.96.184520} did not consider spin conductance in its analysis and only focused on charge conductance. Furthermore, when analyzing charge quantum noise, we find that for YSR states (\( Q_{11}^{ch} (YSR) \)) and MBS (\( Q_{11}^{ch} (MBS) \)), the relationship  $Q_{11}^{ch} (YSR) < \frac{Q_{11}^{ch} (MBS)}{2}$
holds true. However, for spin quantum noise,  
{$Q_{11}^{sp} (YSR) \neq 0$, while $Q_{11}^{sp} (MBS) \to 0$ at $\frac{eV}{\Delta_0} \to 0$, i.e., $Q_{11}^{sp}$(MBS) vanishes at zero voltage bias, while $Q_{11}^{sp}$ (YSR) remains finite.}  
This contrast in quantum noise behavior serves as a reliable probe to differentiate between YSR states and MBS.  
Additionally, both charge and spin quantum noise for YSR states exhibit peaks at specific barrier strengths corresponding to the presence of YSR states. In contrast, for MBS, neither charge nor spin quantum noise displays such peaks, reinforcing their distinct transport signatures.  
A similar distinction is observed in charge and spin \(\Delta_T\) noise. For YSR states, charge \(\Delta_T\) noise (\(\Delta_T^{ch} (YSR)\)) and spin \(\Delta_T\) noise (\(\Delta_T^{sp} (YSR)\)) exhibit dips {around} the barrier strengths where YSR states occur. However, for MBS, neither charge nor spin \(\Delta_T\) noise shows such dips at these barrier strengths. This further confirms that charge and spin \(\Delta_T\) noise can serve as effective probes for distinguishing between YSR states and MBS.}

{All these works on distinguishing trivial from topological ZBCPs only make use of charge conductance. In conclusion, our study is the first comprehensive investigation that explores the entire range of transport probes including charge/spin conductance as well as charge/spin quantum noise and $\Delta_T$ noise to distinguish between Yu-Shiba-Rusinov (YSR) states and Majorana bound states (MBS).}

\section{Experimental realization and Conclusion}
\label{conclusion}

{Recently, researchers have successfully measured $\Delta_T$ noise arising from finite temperature differences in experimental setups \cite{lumbroso2018electronic, sivre2019electronic, PhysRevLett.125.106801}. In our study, conducted on a metal-spin-flipper-metal-insulator-superconductor junction, we analyze a range of physical observables: charge/spin conductance, charge/spin quantum noise and $\Delta_T$ noise under finite temperature gradients and zero applied voltage bias.}  

{We consider the pairing symmetry of the superconductor in two cases: singlet ($s$-wave) or triplet ($p$-wave). The $s$-wave superconductor stands out because the interaction between Cooper pairs and the spin of a magnetic impurity can induce Yu-Shiba-Rusinov (YSR) states. On the other hand, $p$-wave superconductors, whether chiral $p$-wave, spinless $p$-wave nanowires, are topological in nature and host Majorana bound states (MBS) regardless of the presence of a spin flipper. Our analysis demonstrates that {quantitative analysis of charge conductance can distinguish MBS from YSR states similar to what was seen earlier in Ref. \cite{PhysRevB.96.184520}.} However, we observe that the spin conductance, charge/spin quantum noise and $\Delta_T$ noise serve as more efficient and robust tools to identify YSR states and distinguish them from MBS.} 

{For the $s$-wave superconductor, materials such as Niobium (Nb) \cite{RevModPhys.35.1} can be used. Similarly, potential candidates for chiral $p$-wave superconductors, include Sr$_2$RuO$_4$ \cite{leggett2021symmetry} or heavy fermion superconductors like UTe$_2$ \cite{jiao2020chiral}. The spinless \( p \)-wave nanowire can be realized in a quantum wire based on a semiconductor-superconductor heterostructure, where the superconductor can be an \( s \)-wave type superconductor such as Niobium, and the semiconductor needs to exhibit strong spin-orbit coupling, for example, Indium Arsenide (InAs) \cite{PhysRevB.91.214513, PhysRevLett.105.077001, PhysRevLett.105.177002, PhysRevB.81.125318}.
In our setup, the spin flipper serves as a magnetic impurity, conceptually similar to an Anderson impurity but distinct from a Kondo impurity \cite{de1984spin, pal2018yu, dargys2012boundary}. One practical realization could be a quantum dot containing spin-paired electrons and an additional unpaired electron, which acts as a magnetic impurity or spin flipper \cite{PhysRevLett.106.206801}.}

{In this work, we arrive at several key conclusions that can guide future studies. We demonstrate that spin conductance and charge/spin quantum noise at arbitrary temperatures are reliable probes to distinguish between YSR and MBS states quantitatively and qualitatively. Further, we explore $\Delta_T$ noise under finite temperature gradients with zero applied voltage bias, where the average charge and spin currents always vanish and see they too can be effective probes.}

{We find that charge and spin quantum noise {measured at zero charge and spin current} are very effective indicator of YSR states. These exhibit distinct peaks at specific barrier strengths ($Z = 0.781$ and $Z = 1.121$ in our setup) at particular spin-flipper barrier strengths. Through the analysis of charge and spin quantum noise, we arrive at a key conclusions that increasing the temperature plays a major role, making it possible to distinguish YSR and MBS states via its scaling with average temperature.}  

{Furthermore, $\Delta_T$ noise analysis provides additional insights. Charge $\Delta_T$ noise exhibits peaks or dips at the barrier strengths where YSR states occur, whereas for MBS states, it does not show any such differentiating behavior. Similarly, spin $\Delta_T$ noise also exhibits peaks or dips at the barrier strengths for YSR states but does not show any such characteristic for MBS states, offering another strong discriminator.} 

{Experimentally, to isolate the thermal noise-like contribution from shot noise, one can use the following approach. First, to measure the thermal noise contribution, one should set the temperatures of the normal metal and the superconductor to be equal, i.e., $T_1 = T_2 = T$ at zero voltage bias. In this equilibrium setup, the quantum noise measured will correspond entirely to the thermal noise-like contribution. Next, to measure the combined effect of both thermal and shot noise, a temperature bias is introduced such that $T_1 = T + \Delta T/2$ and $T_2 = T - \Delta T/2$, but at zero voltage bias. The total quantum noise measured under this condition will be a sum of both thermal and shot noise components. Finally, to isolate the shot noise contribution, the thermal noise calculated from the equilibrium measurements (where $T_1 = T_2 = T$) is subtracted from the total quantum noise measured with the applied temperature bias. The remaining noise will then correspond to the shot noise-like contribution, see Refs. \cite{lumbroso2018electronic, sivre2019electronic}. The shot noise-like contribution at finite temperature bias and at zero voltage bias is the $\Delta_T$ noise. These approaches offer crucial insights into spin-dependent transport phenomena and provide powerful experimental means to distinguish between different bound states in hybrid superconducting systems.}

{The experimental measurement of charge conductance is straightforward, as it can be directly obtained from the charge current. The measurement of spin conductance is inherently more complex than charge conductance and requires specialized techniques to accurately probe spin transport. One key method involves creating an imbalance in the population of spin-up and spin-down particles, which can be achieved through various means. A widely used approach in superconducting materials is the application of an in-plane magnetic field, which lifts spin degeneracy and induces a net spin current by favoring one spin orientation over the other \cite{kuzmanovic2020evidence}. This imbalance further generates a measurable spin current. Once the spin current is determined, spin conductance can be calculated in a manner analogous to charge conductance. Furthermore, the fluctuations in spin current or spin magnetization, which lead to spin quantum noise, can be probed using high-precision optical spectroscopy techniques such as spin-noise spectroscopy \cite{zapasskii2013spin}. In the beginning, one can measure the total spin quantum noise at equilibrium condition ($T_1 = T_2 = T$) at zero voltage bias, which will yield quantum thermal noise and then later again measure spin quantum measure by applying temperature bias, but at zero voltage bias. To finally measure, the shot noise-like contribution, one can subtract the total spin quantum noise measured under non-equilibrium condition from that measured at equilibrium condition. This particular shot noise-like contribution to the spin quantum noise at finite temperature bias and zero voltage bias is referred to as spin $\Delta_T$ noise. {We observe that both the charge and spin $\Delta_T$ noise are 1000 times smaller than the thermal noise-like contribution to the total charge or spin quantum noise. {Charge $\Delta_T$ noise for a setup with molecular junction made up of Hydrogen molecules inserted between two sharp electrodes made up of normal metal such as Gold has been already experimentally detected, see Refs. \cite{lumbroso2018electronic, shein2022electronic}, wherein the charge $\Delta_T$ noise contribution was successfully isolated from the thermal noise like contribution.}. Therefore, to measure the charge and spin $\Delta_T$ noise experimentally, one requires a measurement precision of 1 part per thousand to effectively separate the $\Delta_T$ noise from the thermal noise-like contribution.}}

{Our study significantly advances the understanding of $\Delta_T$ noise as a powerful tool in mesoscopic transport research. It demonstrates that $\Delta_T$ noise, coupled with quantum noise, can effectively probe YSR states and distinguish them from MBS. This work addresses a longstanding challenge in the field, where previous experimental efforts to identify such probes have faced limitations. By providing a detailed theoretical framework and actionable insights, we lay the groundwork for experimental verification and further advancements.}

{We believe this work can inspire future research in the domains of topological quantum computation and spintronics. The ability to accurately distinguish between YSR and MBS states is a crucial step towards realizing Majorana bound states and exotic phases of matter. Our findings open new avenues for exploring novel transport phenomena in mesoscopic systems, driving the state-of-the-art in quantum technologies.}\\

\section*{Appendix}

{The appendix is divided into three parts. The first section, Appendix \ref{App_I}, covers the derivation of current due to charge and spin transport in normal metal ($N_1$) for a N-sf-N-I-S junction. Subsequently, in Appendix \ref{App_Qn}, we calculate the expressions for quantum noise, which are spin-polarized. In Appendix \ref{App:prob}, we provide the plots of normal and Andreev reflection probabilities for both YSR states and MBS and provide their zero energy values in Tables \ref{Table170}-\ref{Table173}.} 
 

\appendix
\label{appendix}

\begin{widetext}

\section[Spin-polarised average current]{Spin-polarised average current}
\label{App_I}

The expression for the spin-polarized average current ($\langle I^{\sigma}_1 \rangle$) in the normal metal ($N_1$) is given by:

\begin{eqnarray}
\langle I^{\sigma}_1 \rangle = \sum_{\substack{k,l \in \{1,2\}; \\ \alpha,\Gamma,\eta \in \{e,h\} \\  \rho, \rho^{\prime} \in \{\uparrow, \downarrow\} }} \frac{2e}{h}  \text{sgn}(\alpha) \int^{\infty}_{-\infty} dE A_{k \Gamma;l \eta}(1 \alpha,\sigma) \langle a^{\rho \dagger}_{k \Gamma} a^{\rho^{\prime}}_{l \eta} \rangle,
\label{A_I}
\end{eqnarray}

where $ \text{sgn}(\alpha)=+1 (-1)$ for electron (hole). The term $A_{k \Gamma;l \eta}(1 \alpha,\sigma) = \delta_{1 k} \delta_{1 l} \delta_{\alpha \Gamma} \delta_{\alpha \eta}  \delta_{\sigma \rho} \delta_{\sigma \rho^{\prime}} - s^{\alpha \Gamma,\sigma \rho \dagger}_{1 k} s^{\alpha \eta,\sigma \rho^{\prime}}_{1 l}$ represents the matrix element, where $k,l \in \{ 1, 2 \}$ indices label the normal metal ($N_1$) and superconductor ($S$) contacts respectively, and $\alpha, \Gamma$, $\eta$ denote electron or hole. The indices in Eq. (\ref{A_I}), $\sigma$, $\rho^{\prime}$, and $\rho$ denote the spin of the particle (electron or hole), specifically up-spin ($\uparrow$) or down-spin ($\downarrow$). The operators $a^{\rho \dagger}_{k \Gamma}$ and $a^{\rho^{\prime}}_{l \eta}$ are the creation and annihilation operators, respectively, for particle $\Gamma$ in contact $k$ with spin $\rho$, and for particle $\eta$ in contact $l$ with spin $\rho^{\prime}$. The expectation value of the product of these operators simplifies to $\langle a^{\rho \dagger}_{k \Gamma} a_{\rho^{\prime} l \eta} \rangle = \delta_{k l} \delta_{\Gamma \eta} \delta_{\rho \rho^{\prime}} f_{k \Gamma}$, where Fermi function is denoted as $f_{k \Gamma} =  \left[ 1+ e^{\frac{E+ \text{sgn}(\Gamma) V_{k}}{k_B T_{k}}} \right]^{-1}$ in contact $k$ (normal metal or superconductor) and for particles $\Gamma$ (electron or hole), and $\text{sgn}(\Gamma)=+1 (-1)$ for electron (hole).

In our setup with a finite bias voltage ($V_1=eV, V_2=0$), the Fermi function for electrons in the normal metal is given by $f_{1e} = \left( 1 + e^{\frac{E-eV}{k_B T_1}} \right)^{-1}$. In the superconductor at $V_2=0$, Fermi functions for electron-like quasiparticles are the same as hole-like quasiparticles and represented as $f_{2e} = f_{2h} = \left( 1 + e^{\frac{E}{k_B T_2}} \right)^{-1}$ \cite{PhysRevB.53.16390, noise}.
Utilizing the properties $f_{1h}(E)= 1- f_{1e}(-E)$, $\mathcal{A}^{\uparrow \uparrow}(-E)=\mathcal{A}^{\uparrow \uparrow}$, $\mathcal{A}^{\downarrow \uparrow}(-E)=\mathcal{A}^{\downarrow \uparrow}$, $\mathcal{B}^{\uparrow \uparrow}(-E)=\mathcal{B}^{\uparrow \uparrow}$, and $\mathcal{B}^{\downarrow \uparrow}(-E)=\mathcal{B}^{\downarrow \uparrow}$ \cite{PhysRevB.25.4515, lambert1991generalized, noise}, the average charge current in $N_1$ (see, Refs.~\cite{PhysRevB.25.4515, lambert1991generalized}) can be simplified as,
\begin{equation}
\langle I^{ch}_1 \rangle  = \frac{2e}{h} \int^{\infty}_{-\infty}\left( 1 + \mathcal{A}^{\uparrow \uparrow} + \mathcal{A}^{\downarrow \uparrow} - \mathcal{B}^{\uparrow \uparrow} - \mathcal{B}^{\downarrow \uparrow}\right) (f_{1e} -f_{2e}) dE = \frac{2e}{h} \int^{\infty}_{-\infty} F^{ch}_I (f_{1e}-f_{2e}) dE,
\label{A_II}
\end{equation}
where $F^{ch}_{I}= 1 + \mathcal{A}^{\uparrow \uparrow} + \mathcal{A}^{\downarrow \uparrow} - \mathcal{B}^{\uparrow \uparrow} - \mathcal{B}^{\downarrow \uparrow} $. The mean spin current in the left normal metal $N_1$ (see, Refs.~\cite{PhysRevB.25.4515, lambert1991generalized, cheng2013quantum, PhysRevB.60.3572}) as follows,

\begin{equation}
\langle I^{sp}_1 \rangle  = \frac{2e}{h} \int^{\infty}_{-\infty}\left( 1 + \mathcal{A}^{\uparrow \uparrow} - \mathcal{A}^{\downarrow \uparrow} - \mathcal{B}^{\uparrow \uparrow} + \mathcal{B}^{\downarrow \uparrow} \right) (f_{1e} -f_{2e}) dE = \frac{2e}{h} \int^{\infty}_{-\infty} F^{sp}_I (f_{1e}-f_{2e}) dE,
\label{A_II}
\end{equation}
where $F^{sp}_{I}= 1 + \mathcal{A}^{\uparrow \uparrow} - \mathcal{A}^{\downarrow \uparrow} - \mathcal{B}^{\uparrow \uparrow} + \mathcal{B}^{\downarrow \uparrow}$.

Next, we calculate the spin-polarised quantum noise in a N-sf-N-I-S junction.

\section{Spin-polarised quantum noise}
\label{App_Qn}

Spin-polarised quantum noise is defined as the correlation between current in contact $p$ and current in contact $q$ with spin $\sigma$ and $\sigma^{\prime}$  ~\cite{noise, thermalnoise}, such as  $Q^{\sigma \sigma^{\prime}}_{p q} (t-T) = \frac{1}{2 \pi} \langle \Delta I^{\sigma}_{p}(t) \Delta I^{\sigma^{\prime}}_{q}(T) + \Delta I^{\sigma^{\prime}}_{q}(T) \Delta I^{\sigma}_{p}(t) \rangle$, with $\Delta I^{\sigma}_{p} = I^{\sigma}_{p} - \langle I^{\sigma}_{p} \rangle$, where $I^{\sigma}_{p}$ is the spin-polarised current in lead $p$ with spin $\sigma \in \{\uparrow, \downarrow \}$. Quantum noise power can be obtained by taking the Fourier transform of the quantum noise, expressed as $ 2 \pi \delta(\omega + \bar{\omega}) Q^{\sigma \sigma^{\prime}}_{p q}(\omega) \equiv \langle \Delta I^{\sigma}_{p}(\omega) \Delta I^{\sigma^{\prime}}_{q}(\bar{\omega}) + \Delta I^{\sigma^{\prime}}_{q}(\bar{\omega}) \Delta I^{\sigma}_{p}(\omega) \rangle $. Zero frequency spin-polarised quantum noise $Q^{\sigma \sigma^{\prime}}_{p q}(\omega=\bar{\omega} =0$) \cite{PhysRevB.53.16390}, in a N-sf-N-I-S junction is,
\begin{eqnarray}
Q^{\sigma \sigma^{\prime}}_{pq} && = \frac{2e^2}{h} \int_{-\infty}^{\infty} \sum_{\rho, \rho^{\prime} \in \{\uparrow, \downarrow\}} \sum_{ \substack{k,l \in \{1, 2\} ,\\
x^{\prime}, y^{\prime},\Gamma,\eta \in \{e,h\} } } sgn(x^{\prime}) sgn(y^{\prime}) A^{\rho \rho^{\prime}}_{k,\Gamma;l,\eta}(p x^{\prime},\sigma) A^{\rho^{\prime} \rho}_{l,\eta;k,\Gamma}(q y^{\prime},\sigma^{\prime}) \textit{f}_{k \Gamma}(E) [1-\textit{f}_{l \eta}(E)] dE,
\label{eqn:sn}
\end{eqnarray}{}
where $A^{\rho \rho^{\prime}}_{k \Gamma;l \eta}(p x^{\prime},\sigma) = \delta_{p k} \delta_{p l} \delta_{x^{\prime} \Gamma} \delta_{x^{\prime} \eta}  \delta_{\sigma \rho} \delta_{\sigma \rho^{\prime}} - s^{x^{\prime} \Gamma,\sigma \rho \dagger}_{p k} s^{x^{\prime} \eta,\sigma \rho^{\prime}}_{p l}$, and $sgn(x^{\prime})=sgn(y^{\prime})=+(-)1$ for electron (hole). {$s_{p k}^{x' \Gamma; \sigma \rho}$ are the scattering amplitude for a particle of type $x' \in \{e, h \}$ to scatter from terminal $k \in \{1,2 \}$ with spin $\sigma \in \{\uparrow, \downarrow\}$ to terminal $p \in \{1,2 \}$ as a particle of type $\Gamma \in \{e, h \}$ with spin $\rho \in \{\uparrow, \downarrow \}$.} Spin-polarised quantum noise auto-correlation ($Q^{\sigma \sigma^{\prime}}_{11}$) in a N-sf-N-I-S junction, is as follows,
\begin{eqnarray}
Q^{\sigma \sigma^{\prime}}_{11} && = \frac{2e^2}{h} \int_{-\infty}^{\infty} \sum_{ \rho, \rho^{\prime} \in \{\uparrow, \downarrow\} } \sum_{ \substack{k,l \in \{1, 2\} ,\\ x^{\prime}, y^{\prime},\Gamma,\eta \in \{e,h\}} } sgn(y^{\prime}) sgn(x^{\prime}) A^{\sigma^{\prime} \sigma}_{k,\Gamma;l,\eta}(1 x^{\prime},\sigma) A^{\rho^{\prime} \rho}_{l,\eta;k,\Gamma}(1 y^{\prime},\sigma^{\prime}) \textit{f}_{k \Gamma}(E) [1-\textit{f}_{l \eta}(E)] dE.
\label{B_Snoise}
\end{eqnarray}

The total charge quantum noise ($Q_{11}^{ch}$) is given as $Q_{11}^{ch} = Q_{11}^{\uparrow \uparrow} + Q_{11}^{\uparrow \downarrow} + Q_{11}^{\downarrow \uparrow} + Q_{11}^{\downarrow \downarrow}$, where as spin quantum noise ($Q_{11}^{sp}$) is $Q_{11}^{ch} = Q_{11}^{\uparrow \uparrow} - Q_{11}^{\uparrow \downarrow} - Q_{11}^{\downarrow \uparrow} + Q_{11}^{\downarrow \downarrow}$. One can calculate each spin polarized correlations from Eq. (\ref{B_Snoise}) following the method of Ref. \cite{PhysRevB.53.16390} and calculate both $Q_{11}^{ch}$ and $Q_{11}^{sp}$. 

{In a N-sf-N-I-S junction, one can construct a $s$-matrix, which can relate the incoming and outgoing states and can describe the scattering processes such as normal reflection with and without spin-flip, Andreev reflection with and without spin-flip, transmission of electron and hole-like quasiparticles with and without spin-flip. We denote the $s$-matrix, which relates the outgoing state, which is given as $c_{out} =  \begin{pmatrix}
        c_{1e}^{- \uparrow} &
        c_{1e}^{- \downarrow} &
        c_{2e}^{+ \uparrow} &
        c_{2e}^{+ \downarrow} &
        c_{1h}^{+ \uparrow} &
        c_{1h}^{+ \downarrow}&
        c_{2h}^{- \uparrow} &
        c_{2h}^{- \downarrow}
        \end{pmatrix}^T$ and incoming state $c_{in} = \begin{pmatrix}  
        c_{1e}^{+ \uparrow} &
        c_{1e}^{+ \downarrow} &
        c_{2e}^{- \uparrow} &
        c_{2e}^{- \downarrow} &
        c_{1h}^{- \uparrow} &
        c_{1h}^{- \downarrow}&
        c_{2h}^{+ \uparrow} &
        c_{2h}^{+ \downarrow} 
        \end{pmatrix}^T$ in a N-sf-N-I-S junction by $S_{N-sf-N-I-S}$ and is given as,}

        {
        \begin{equation}
            S_{N-sf-N-I-S} = \begin{pmatrix}
                s_{11}^{ee; \uparrow \uparrow} & s_{11}^{ee; \uparrow \downarrow} & s_{12}^{ee; \uparrow \uparrow} &s_{12}^{ee; \uparrow \downarrow} & s_{11}^{eh; \uparrow \uparrow} & s_{11}^{eh; \uparrow \downarrow} & s_{12}^{eh; \uparrow \uparrow} & s_{12}^{eh; \uparrow \downarrow} \\
                s_{11}^{ee; \downarrow \uparrow} & s_{11}^{ee; \downarrow \downarrow} & s_{12}^{ee; \downarrow \uparrow} &s_{12}^{ee; \downarrow \downarrow} & s_{11}^{eh; \downarrow \uparrow} & s_{11}^{eh; \downarrow \downarrow} & s_{12}^{eh; \downarrow \uparrow} & s_{12}^{eh; \downarrow \downarrow}\\
                s_{21}^{ee; \uparrow \uparrow} & s_{21}^{ee; \uparrow \downarrow} & s_{22}^{ee; \uparrow \uparrow} &s_{22}^{ee; \uparrow \downarrow} & s_{21}^{eh; \uparrow \uparrow} & s_{21}^{eh; \uparrow \downarrow} & s_{22}^{eh; \uparrow \uparrow} & s_{22}^{eh; \uparrow \downarrow} \\
                s_{21}^{ee; \downarrow \uparrow} & s_{21}^{ee; \downarrow \downarrow} & s_{22}^{ee; \downarrow \uparrow} &s_{22}^{ee; \downarrow \downarrow} & s_{21}^{eh; \downarrow \uparrow} & s_{21}^{eh; \downarrow \downarrow} & s_{22}^{eh; \downarrow \uparrow} & s_{22}^{eh; \downarrow \downarrow}\\
                s_{11}^{he; \uparrow \uparrow} & s_{11}^{he; \uparrow \downarrow} & s_{12}^{he; \uparrow \uparrow} &s_{12}^{he; \uparrow \downarrow} & s_{11}^{hh; \uparrow \uparrow} & s_{11}^{hh; \uparrow \downarrow} & s_{12}^{hh; \uparrow \uparrow} & s_{12}^{hh; \uparrow \downarrow} \\
                s_{11}^{he; \downarrow \uparrow} & s_{11}^{he; \downarrow \downarrow} & s_{12}^{he; \downarrow \uparrow} &s_{12}^{he; \downarrow \downarrow} & s_{11}^{hh; \downarrow \uparrow} & s_{11}^{hh; \downarrow \downarrow} & s_{12}^{hh; \downarrow \uparrow} & s_{12}^{hh; \downarrow \downarrow}\\
                s_{21}^{he; \uparrow \uparrow} & s_{21}^{he; \uparrow \downarrow} & s_{22}^{he; \uparrow \uparrow} &s_{22}^{he; \uparrow \downarrow} & s_{21}^{hh; \uparrow \uparrow} & s_{21}^{hh; \uparrow \downarrow} & s_{22}^{hh; \uparrow \uparrow} & s_{22}^{hh; \uparrow \downarrow} \\
                s_{21}^{he; \downarrow \uparrow} & s_{21}^{he; \downarrow \downarrow} & s_{22}^{he; \downarrow \uparrow} &s_{22}^{he; \downarrow \downarrow} & s_{21}^{hh; \downarrow \uparrow} & s_{21}^{hh; \downarrow \downarrow} & s_{22}^{hh; \downarrow \uparrow} & s_{22}^{hh; \downarrow \downarrow}
            \end{pmatrix}.
            \label{eq:1000}
        \end{equation}
        }

{Here, $T$ denotes the transpose of the matrix. The operators $c_{1e}^{+\uparrow(\downarrow)}$ and $c_{1e}^{-\uparrow(\downarrow)}$ represent the incoming and outgoing up (down) spin electron states in the normal metal, respectively. Similarly, $c_{2e}^{-\uparrow(\downarrow)}$ and $c_{2e}^{+\uparrow(\downarrow)}$ correspond to the incoming and outgoing electron states in the superconductor. For holes, $c_{1h}^{-\uparrow(\downarrow)}$ and $c_{1h}^{+\uparrow(\downarrow)}$ represent the incoming and outgoing hole states in the normal metal, while $c_{2h}^{+\uparrow(\downarrow)}$ and $c_{2h}^{-\uparrow(\downarrow)}$ denote the incoming and outgoing hole states in the superconductor.}

{The scattering amplitudes are defined as follows: $s_{11}^{ee; \uparrow\uparrow}$ and $s_{11}^{ee; \downarrow\downarrow}$ are the normal reflection amplitudes without spin-flip, while $s_{11}^{ee; \uparrow\downarrow}$ and $s_{11}^{ee; \downarrow\uparrow}$ describe spin-flip normal reflections. $s_{11}^{he; \uparrow\uparrow}$ and $s_{11}^{he; \downarrow\downarrow}$ are the Andreev reflection amplitudes without spin-flip, and $s_{11}^{he; \uparrow\downarrow}$ and $s_{11}^{he; \downarrow\uparrow}$ correspond to spin-flip Andreev reflections. Transmission amplitudes from the normal metal to the superconductor are given by $s_{21}^{ee; \uparrow\uparrow}$ and $s_{21}^{ee; \downarrow\downarrow}$ for electron-like quasiparticles without spin-flip, and by $s_{21}^{ee; \uparrow\downarrow}$, $s_{21}^{ee; \downarrow\uparrow}$ for those with spin-flip. Similarly, $s_{21}^{he; \uparrow\uparrow}$, $s_{21}^{he; \downarrow\downarrow}$ denote the transmission amplitudes for hole-like quasiparticles without spin-flip, while $s_{21}^{he; \uparrow\downarrow}$, $s_{21}^{he; \downarrow\uparrow}$ represent the spin-flip counterparts. Additional scattering amplitudes such as $s_{11}^{hh}$, $s_{11}^{eh}$, and $s_{21}^{hh}$, $s_{21}^{eh}$, with appropriate spin indices, describe normal and Andreev reflections as well as transmission processes for hole-like and electron-like quasiparticles with or without spin-flip for an incident hole. Corresponding amplitudes for quasiparticles incident from the superconducting side can be defined analogously; see Eq.~(\ref{eq:1000}) for details.}

{For the derivation of both charge and spin quantum noise, we only require the amplitudes from the first, second, fifth, and sixth row elements of the $s$-matrix as in Eq. (\ref{eq:1000}). Therefore, we ignore the $s$-matrix amplitudes from the third, fourth, seventh, and eighth rows of $S_{N\text{-}sf\text{-}N\text{-}I\text{-}S}$ as in Eq. (\ref{eq:1000}). We denote $s_{11}^{ee; \uparrow \uparrow} = r_N^{\uparrow \uparrow}$, $s_{11}^{ee; \downarrow \uparrow} = r_N^{\downarrow \uparrow}$, $s_{11}^{he; \uparrow \uparrow} = r_{Na}^{\uparrow \uparrow}$, $s_{11}^{he; \downarrow \uparrow} = r_{Na}^{\downarrow \uparrow}$, $s_{12}^{ee; \uparrow \uparrow} = \sqrt{|u|^2 - |v|^2} \, c_S^{\uparrow \uparrow}$, $s_{12}^{ee; \downarrow \uparrow} = \sqrt{|u|^2 - |v|^2} \, c_S^{\downarrow \uparrow}$, $s_{12}^{he; \uparrow \uparrow} = \sqrt{|u|^2 - |v|^2} \, d_S^{\uparrow \uparrow}$, $s_{12}^{he; \downarrow \uparrow} = \sqrt{|u|^2 - |v|^2} \, d_S^{\downarrow \uparrow}$. Here, $\mathcal{A}^{\uparrow \uparrow} = |r_{aN}^{\uparrow \uparrow}|^2$, $\mathcal{A}^{\downarrow \uparrow} = |r_{aN}^{\downarrow \uparrow}|^2$ are the Andreev reflection probabilities without and with spin-flip, $\mathcal{B}^{\uparrow \uparrow} = |r_{N}^{\uparrow \uparrow}|^2$, $\mathcal{B}^{\downarrow \uparrow} = |r_{N}^{\downarrow \uparrow}|^2$ are the normal reflection probabilities without and with spin-flip, $\mathcal{C}_S^{\uparrow \uparrow} = (|u|^2 - |v|^2) |c_S^{\uparrow \uparrow}|^2$ and $\mathcal{C}_S^{\downarrow \uparrow} = (|u|^2 - |v|^2) |c_S^{\downarrow \uparrow}|^2$ are the transmission probabilities of electron-like quasiparticles without and with spin-flip, and $\mathcal{D}_S^{\uparrow \uparrow} = (|u|^2 - |v|^2) |d_S^{\uparrow \uparrow}|^2$ and $\mathcal{D}_S^{\downarrow \uparrow} = (|u|^2 - |v|^2) |d_S^{\downarrow \uparrow}|^2$ are the transmission probabilities of hole-like quasiparticles without and with spin-flip from superconductor to the normal metal.}

{From the unitarity of $S_{N\text{-}sf\text{-}N\text{-}I\text{-}S}$, we obtain: $s_{11}^{ee; \downarrow \downarrow} = s_{11}^{ee; \uparrow \uparrow}$, $s_{11}^{hh; \uparrow \uparrow} = s_{11}^{hh; \downarrow \downarrow} = s_{11}^{ee; \uparrow \uparrow ^*}$, $s_{11}^{ee; \uparrow \downarrow} = s_{11}^{ee; \downarrow \uparrow}$, $s_{11}^{hh; \downarrow \uparrow} = s_{11}^{hh; \uparrow \downarrow} = s_{11}^{ee; \downarrow \uparrow ^*}$, $s_{11}^{he; \downarrow \downarrow} = -s_{11}^{he; \uparrow \uparrow}$, $- s_{11}^{eh; \uparrow \uparrow} =  s_{11}^{eh; \downarrow \downarrow} =  s_{11}^{he; \uparrow \uparrow ^*}$, $s_{11}^{he; \uparrow \downarrow} = -s_{11}^{he; \downarrow \uparrow}$, $-s_{11}^{eh; \downarrow \uparrow} = s_{11}^{eh; \uparrow \downarrow} = s_{11}^{he; \uparrow \uparrow ^*}$, $s_{12}^{ee; \downarrow \downarrow} = s_{12}^{ee; \uparrow \uparrow}$, $s_{12}^{hh; \uparrow \uparrow} = s_{12}^{hh; \downarrow \downarrow} = s_{12}^{ee; \uparrow \uparrow ^*}$, $s_{12}^{ee; \uparrow \downarrow} = s_{12}^{ee; \downarrow \uparrow}$, $s_{12}^{hh; \downarrow \uparrow} = s_{12}^{hh; \uparrow \downarrow} = s_{11}^{ee; \downarrow \uparrow ^*}$, $s_{12}^{he; \downarrow \downarrow} = -s_{12}^{he; \uparrow \uparrow}$, $- s_{12}^{eh; \uparrow \uparrow} =  s_{12}^{eh; \downarrow \downarrow} =  s_{12}^{he; \uparrow \uparrow ^*}$, $s_{12}^{he; \uparrow \downarrow} = -s_{12}^{he; \downarrow \uparrow}$, $-s_{12}^{eh; \downarrow \uparrow} = s_{12}^{eh; \uparrow \downarrow} = s_{12}^{he; \uparrow \uparrow ^*}$. Furthermore, unitarity of $S_{N\text{-}sf\text{-}N\text{-}I\text{-}S}$ gives the additional constraint: $(|u|^2 - |v|^2) (c_S^{\uparrow \uparrow} d_S^{\downarrow \uparrow ^*} - c_S^{\downarrow \uparrow} d_S^{\uparrow \uparrow ^*}) = r_N^{\downarrow \uparrow} r_{Na}^{\uparrow \uparrow ^*} - r_N^{\uparrow \uparrow} r_{Na}^{\downarrow \uparrow ^*}$ and $(|u|^2 - |v|^2) (c_S^{\uparrow \uparrow ^*} d_S^{\downarrow \uparrow} - c_S^{\downarrow \uparrow ^*} d_S^{\uparrow \uparrow}) = r_N^{\downarrow \uparrow ^*} r_{Na}^{\uparrow \uparrow} - r_N^{\uparrow \uparrow ^*} r_{Na}^{\downarrow \uparrow}$.}
{These relations further imply: $\mathcal{C}_S^{\uparrow \uparrow} \mathcal{D}_S^{\downarrow \uparrow} + \mathcal{C}_S^{\downarrow \uparrow} \mathcal{D}_S^{\uparrow \uparrow} - c_S^{\uparrow \uparrow} d_S^{\downarrow \uparrow ^*} c_S^{\downarrow \uparrow ^*} d_S^{\uparrow \uparrow} - c_S^{\uparrow \uparrow ^*} d_S^{\downarrow \uparrow} c_S^{\downarrow \uparrow} d_S^{\uparrow \uparrow ^*} = \mathcal{A}^{\uparrow \uparrow} \mathcal{B}^{\downarrow \uparrow} + \mathcal{A}^{\downarrow \uparrow} \mathcal{B}^{\uparrow \uparrow} - r_N^{\downarrow \uparrow} r_{Na}^{\uparrow \uparrow ^*} r_N^{\uparrow \uparrow ^*} r_{Na}^{\downarrow \uparrow} - r_N^{\downarrow \uparrow ^*} r_{Na}^{\uparrow \uparrow} r_N^{\uparrow \uparrow} r_{Na}^{\downarrow \uparrow ^*}$. We will use these relations further.}

{The charge quantum noise ($Q_{11}^{ch}$) can be derived now from Eq. (\ref{B_Snoise}) directly by using the $s$-matrix elements given above by finding $Q_{11}^{\sigma \sigma'}$ for $\sigma, \sigma \in \{\uparrow, \downarrow\}$. The expression of $Q_{11}^{ch}$ is given as}

{
\begin{equation}
\begin{split}
    Q_{11}^{ch} = &\frac{4e^2}{h} \int_0^{\infty} dE \bigg[ \bigg(\mathcal{A}^2 + (1 - \mathcal{B})^2 + 2 \mathcal{A} (1 - \mathcal{B}) + 2\mathcal{A}^{\uparrow \uparrow} \mathcal{A}^{\downarrow \uparrow} + 2 \mathcal{B}^{\uparrow \uparrow} \mathcal{B}^{\downarrow \uparrow} + 2 Re(r_{Na}^{\downarrow \uparrow^2} r_{Na}^{\uparrow \uparrow^{*2}}) +2 Re(r_{N}^{\downarrow \uparrow^2} r_{N}^{\uparrow \uparrow^{*2}})\\ & \quad \quad \quad \quad \quad+ 8 Re(r_{Na}^{\uparrow \uparrow} r_{Na}^{\downarrow \uparrow^*}) Re(r_{N}^{\uparrow \uparrow} r_{N}^{\downarrow \uparrow^*})\bigg) \bigg(f_{1e} (1 - f_{1e}) + f_{1h} (1-f_{1h})\bigg) + \bigg( 4 \mathcal{A}^{\downarrow \uparrow} \mathcal{B}^{\downarrow \uparrow} + 4 \mathcal{A}^{\uparrow \uparrow} \mathcal{B}^{\uparrow \uparrow} + 8 Re(r_{Na}^{\uparrow \uparrow} r_{N}^{\uparrow \uparrow} r_{Na}^{\downarrow \uparrow^{*}} r_{N}^{\downarrow \uparrow^*})\bigg)\bigg]\\
    & \quad \quad \quad \quad \quad \bigg(f_{1e}(1 - f_{1h}) + f_{1h} (1 - f_{1e})\bigg) + \bigg(2 (\mathcal{C}_S + \mathcal{D}_S)^2 - 4 \mathcal{C}_S \mathcal{D}_S + 4 \mathcal{C}_S^{\downarrow \uparrow} \mathcal{C}_S^{\uparrow \uparrow} + 4 \mathcal{D}_S^{\downarrow \uparrow} \mathcal{D}_S^{\uparrow \uparrow} + (|u|^2 - |v|^2)^2 (c_{S}^{\downarrow \uparrow^{2}} c_{S}^{\uparrow \uparrow^{*^2}} \\
    & \quad \quad \quad \quad \quad + c_{S}^{\uparrow \uparrow^{2}} c_{S}^{\downarrow \uparrow^{*^2}}  + d_{S}^{\downarrow \uparrow^{2}} d_{S}^{\uparrow \uparrow^{*^2}} + d_{S}^{\uparrow \uparrow^{2}} d_{S}^{\downarrow \uparrow^{*^2}}) + (|u|^2 - |v|^2)^2 16 Re (c_S^{\uparrow \uparrow} c_S^{\downarrow \uparrow^*}) Re (d_S^{\uparrow \uparrow} d_S^{\downarrow \uparrow^*}) + 8 \mathcal{C}_S^{\downarrow \uparrow} \mathcal{D}_S^{\downarrow \uparrow} + 8 \mathcal{C}_S^{\uparrow \uparrow} \mathcal{D}_S^{\uparrow \uparrow} \\
    &\quad \quad \quad \quad \quad + (|u|^2 - |v|^2)^2 16 Re(c_S^{\downarrow \uparrow} d_S^{\downarrow \uparrow} c_S^{\uparrow \uparrow^*} d_S^{\uparrow \uparrow^*})\bigg) f_{2e}(1-f_{2e}) + \bigg( 4 (|u|^2 - |v|^2) (Re(c_S^{\uparrow \uparrow} c_S^{\downarrow \uparrow^{*}}) +  Re(d_S^{\uparrow \uparrow} d_S^{\downarrow \uparrow^{*}}))(Re(r_{N}^{\uparrow \uparrow} r_{N}^{\downarrow \uparrow^{*}}) \\
    & \quad \quad \quad \quad \quad +  Re(r_{Na}^{\uparrow \uparrow} r_{Na}^{\downarrow \uparrow^{*}})) + 2 (|u|^2 - |v|^2) (c_{S}^{\uparrow \uparrow} d_S^{\downarrow \uparrow^*} - c_S^{\downarrow \uparrow}d_S^{\uparrow \uparrow^*}) (r_{Na}^{\uparrow \uparrow} r_N^{\downarrow \uparrow^*} - r_{Na}^{\downarrow \uparrow}r_{N}^{\uparrow \uparrow^*}) + 2 (|u|^2 - |v|^2)(c_{S}^{\uparrow \uparrow^*} d_S^{\downarrow \uparrow} - c_S^{\downarrow \uparrow^*}d_S^{\uparrow \uparrow}) \\
    &\quad \quad \quad \quad \quad(r_{Na}^{\uparrow \uparrow^*} r_N^{\downarrow \uparrow} - r_{Na}^{\downarrow \uparrow^*}r_{N}^{\uparrow \uparrow})  \bigg) \bigg(f_{1e}(1-f_{2e}) + f_{2e} (1-f_{1e}) + f_{1h}(1-f_{2e}) + f_{2e}(1-f_{1h})\bigg)\bigg).
    \end{split}
    \label{eq:1001}
\end{equation}
}

{Here, $\mathcal{A} = \mathcal{A}^{\uparrow \uparrow} + \mathcal{A}^{\downarrow \uparrow}, \mathcal{B} = \mathcal{B}^{\uparrow \uparrow} + \mathcal{B}^{\downarrow \uparrow}, \mathcal{C}_S = \mathcal{C}_S^{\uparrow \uparrow} + \mathcal{C}_S^{\downarrow \uparrow}$ and $\mathcal{D}_S = \mathcal{D}_S^{\uparrow \uparrow} + \mathcal{D}_S^{\downarrow \uparrow}$. Now, we utilize the property $(|u|^2 - |v|^2)(c_S^{\uparrow \uparrow} d_S^{\downarrow \uparrow^*} - c_S^{\downarrow \uparrow^*}d_S^{\uparrow \uparrow}) (r_{Na}^{\uparrow \uparrow^*} r_{N}^{\downarrow \uparrow} - r_{Na}^{\downarrow \uparrow^*} r_N^{\uparrow \uparrow}) = \mathcal{A}^{\uparrow \uparrow} \mathcal{B}^{\downarrow \uparrow} + \mathcal{A}^{\downarrow \uparrow} \mathcal{B}^{\uparrow \uparrow} - 2 * \text{Re}(r_{N}^{\downarrow \uparrow} r_{Na}^{\uparrow \uparrow^*} r_{Na}^{\downarrow \uparrow} r_{N}^{\downarrow \uparrow^*})$ and also using $f_{1e}(1 - f_{1h}) + f_{1h} (1 - f_{1e}) = f_{1e} (1 - f_{1e}) + f_{1h} (1-f_{1h}) + (f_{1e} - f_{1h})^2$ and $f_{1e}(1 - f_{2e}) + f_{2e} (1 - f_{1e}) + f_{1h}(1 - f_{2e}) + f_{2e} (1 - f_{1h}) = f_{1e} (1 - f_{1e}) + f_{2e} (1-f_{2e}) + (f_{1e} - f_{2e})^2 + f_{1h} (1 - f_{1h}) + f_{2e} (1-f_{2e}) + (f_{1h} - f_{2e})^2$.  We also utilize the property $f_{1h}(-E) = 1 - f_{1e} (E), f_{2e}(-E) = 1- f_{2e}(E)$ and also the symmetric nature of all the $s$-matrix elements and probabilities with $E$, we get the final expression for charge quantum noise to be}

{
\begin{equation}
\begin{split}
    Q_{11}^{ch} = &\frac{4e^2}{h} \int_{-\infty}^{\infty} dE \bigg[ \bigg((1-\mathcal{B} + \mathcal{A})^2 + 4 \mathcal{A}(1 - \mathcal{B}) + 2\mathcal{A}^{\uparrow \uparrow} \mathcal{A}^{\downarrow \uparrow} + 2 \mathcal{B}^{\uparrow \uparrow} \mathcal{B}^{\downarrow \uparrow} + 8 Re(r_{Na}^{\uparrow \uparrow} r_{Na}^{\downarrow \uparrow^*}) Re(r_{N}^{\uparrow \uparrow} r_{N}^{\downarrow \uparrow^*})\bigg) \bigg(f_{1e} (1 - f_{1e})\bigg) \\& + \bigg( 2 \mathcal{A}^{\downarrow \uparrow} \mathcal{B}^{\downarrow \uparrow} + 2 \mathcal{A}^{\uparrow \uparrow} \mathcal{B}^{\uparrow \uparrow} + 4 Re(r_{Na}^{\uparrow \uparrow} r_{N}^{\uparrow \uparrow} r_{Na}^{\downarrow \uparrow^{*}} r_{N}^{\downarrow \uparrow^*})\bigg) \bigg(f_{1e}-f_{1h}\bigg)^2 + \bigg( (\mathcal{C}_S + \mathcal{D}_S)^2 + (\mathcal{A} + \mathcal{B})(\mathcal{C}_S + \mathcal{D}_S)  + 2 \mathcal{C}_S^{\downarrow \uparrow} \mathcal{C}_S^{\uparrow \uparrow} \\& + 2 \mathcal{D}_S^{\downarrow \uparrow} \mathcal{D}_S^{\uparrow \uparrow} + 2 (|u|^2 - |v|^2)^2 Re(c_{S}^{\downarrow \uparrow^{2}} c_{S}^{\uparrow \uparrow^{*^2}})  +2 (|u|^2 - |v|^2)^2 Re (d_{S}^{\downarrow \uparrow^{2}} d_{S}^{\uparrow \uparrow^{*^2}}) + 8 (|u|^2 - |v|^2)^2  Re (c_S^{\uparrow \uparrow} c_S^{\downarrow \uparrow^*}) Re (d_S^{\uparrow \uparrow} d_S^{\downarrow \uparrow^*})\\& + 4(|u|^2 - |v|^2)^2 (Re(c_S^{\uparrow \uparrow} c_S^{\downarrow \uparrow *}) + Re(d_S^{\uparrow \uparrow} d_S^{\downarrow \uparrow *})) (Re(r_N^{\uparrow \uparrow} r_N^{\downarrow \uparrow *}) + Re(r_{Na}^{\uparrow \uparrow} r_{Na}^{\downarrow \uparrow *})) \bigg) f_{2e}(1-f_{2e}) + ((\mathcal{A} + \mathcal{B})(\mathcal{C}_S + \mathcal{D}_S)\\& + 4 (|u|^2 - |v|^2)\left(\text{Re}(c_S^{\uparrow \uparrow} c_S^{\downarrow \uparrow *}) + \text{Re}(d_S^{\uparrow \uparrow} d_S^{\downarrow \uparrow *})\right) \times \left(\text{Re}(r_N^{\uparrow \uparrow} r_N^{\downarrow \uparrow *}) + \text{Re}(r_{Na}^{\uparrow \uparrow} r_{Na}^{\downarrow \uparrow *})\right) + 4 \mathcal{A}^{\uparrow \uparrow} \mathcal{B}^{\downarrow \uparrow} + 4 \mathcal{A}^{\downarrow \uparrow} \mathcal{B}^{\uparrow \uparrow} \\& - 8 \text{Re} \left(r_N^{\downarrow \uparrow} r_{Na}^{\uparrow \uparrow *} r_{Na}^{\downarrow \uparrow} r_N^{\uparrow \uparrow *} \right)\bigg)\bigg(f_{1e} - f_{2e}\bigg)^2\bigg].
    \end{split}
    \label{eq:1002}
\end{equation}
}

{In our work, we consider $k_B T \ll \Delta_0$, where $\Delta_0$ is the superconducting gap, therefore, only the scattering for energies below this gap matters. This implies that the amplitudes $s_{12}^{ee; \uparrow \uparrow}, s_{12}^{ee; \downarrow \uparrow}, s_{12}^{he; \uparrow \uparrow}, s_{12}^{he; \downarrow \uparrow}$ vanishes and therefore the probabilities $\mathcal{C}_S^{\uparrow \uparrow}$, $\mathcal{C}_S^{\downarrow \uparrow}$, $\mathcal{D}_S^{\uparrow \uparrow}$ and $\mathcal{D}_S^{\downarrow \uparrow}$ also vanish.}
{Therefore, the final expression for $Q_{11}^{ch}$ is given as}
{
\begin{equation}
\begin{split}
    Q_{11}^{ch} &= \frac{4e^2}{h} \int_{-\infty}^{\infty} dE ( 1 - \mathcal{B} + 3\mathcal{A} + 2 \mathcal{A}^{\uparrow \uparrow } \mathcal{A}^{\downarrow \uparrow} + 2 \mathcal{B}^{\uparrow \uparrow} \mathcal{B}^{\downarrow \uparrow} + 2 \text{Re}(r_{Na}^{\downarrow \uparrow^2} r_{Na}^{\uparrow \uparrow^{*2}}) + 2 \text{Re}(r_{N}^{\downarrow \uparrow^2} r_{N}^{\uparrow \uparrow^{*2}}) + 8 \text{Re}(r_{Na}^{\uparrow \uparrow} r_{Na}^{\downarrow \uparrow^*}) \text{Re}(r_{N}^{\uparrow \uparrow} r_{N}^{\downarrow \uparrow^*}) ) f_{1e} (1-f_{1e}) \\
    &\quad + \frac{4e^2}{h} \int_{-\infty}^{\infty} dE (2 \mathcal{A}^{\downarrow \uparrow} \mathcal{B}^{\downarrow \uparrow} + 2 \mathcal{A}^{\uparrow \uparrow} \mathcal{B}^{\uparrow \uparrow} + 4 \text{Re}(r_{Na}^{\uparrow \uparrow} r_N^{\uparrow \uparrow} r_{Na}^{\downarrow \uparrow^*} r_{N}^{\downarrow \uparrow^*})) (f_{1e} - f_{1h})^2 \\
    &\quad + \frac{4e^2}{h} \int_{-\infty}^{\infty} dE (4 \mathcal{A}^{\uparrow \downarrow} \mathcal{B}^{\downarrow \uparrow} + 4 \mathcal{A}^{\downarrow \uparrow} \mathcal{B}^{\uparrow \uparrow} - 8 \text{Re} (r_N^{\downarrow \uparrow} r_{Na}^{\uparrow \uparrow^*} r_{N}^{\uparrow \uparrow^*} r_{Na}^{\downarrow \uparrow})) (f_{1e} - f_{2e})^2 \\
    &= Q_{11}^{ch; th} + Q_{11}^{ch; sh}
\end{split}
\label{eq:B3}
\end{equation}
}
{where $Q_{11}^{ch; th} = \frac{4e^2}{h} \int_{-\infty}^{\infty} dE (1 - \mathcal{B} + 3\mathcal{A} + 2 \mathcal{A}^{\uparrow \uparrow } \mathcal{A}^{\downarrow \uparrow} + 2 \mathcal{B}^{\uparrow \uparrow} \mathcal{B}^{\downarrow \uparrow} + 2 \text{Re}(r_{Na}^{\downarrow \uparrow^2} r_{Na}^{\uparrow \uparrow^{*2}}) + 2 \text{Re}(r_{N}^{\downarrow \uparrow^2} r_{N}^{\uparrow \uparrow^{*2}}) + 8 \text{Re}(r_{Na}^{\uparrow \uparrow} r_{Na}^{\downarrow \uparrow^*}) \text{Re}(r_{N}^{\uparrow \uparrow} r_{N}^{\downarrow \uparrow^*}) ) f_{1e} (1-f_{1e})$ is the thermal noise-like contribution and $Q_{11}^{ch; sh} =  \frac{4e^2}{h} \int_{-\infty}^{\infty} dE (2 \mathcal{A}^{\downarrow \uparrow} \mathcal{B}^{\downarrow \uparrow} + 2 \mathcal{A}^{\uparrow \uparrow} \mathcal{B}^{\uparrow \uparrow} + 4 \text{Re}(r_{Na}^{\uparrow \uparrow} r_N^{\uparrow \uparrow} r_{Na}^{\downarrow \uparrow^*} r_{N}^{\downarrow \uparrow^*})) (f_{1e} - f_{1h})^2  + \frac{4e^2}{h} \int_{-\infty}^{\infty} dE (4 \mathcal{A}^{\uparrow \downarrow} \mathcal{B}^{\downarrow \uparrow} + 4 \mathcal{A}^{\downarrow \uparrow} \mathcal{B}^{\uparrow \uparrow} - 8 \text{Re} (r_N^{\downarrow \uparrow} r_{Na}^{\uparrow \uparrow^*} r_{N}^{\uparrow \uparrow^*} r_{Na}^{\downarrow \uparrow})) (f_{1e} - f_{2e})^2$ is the shot noise-like contribution to the total charge quantum noise $Q_{11}^{ch}$. However, at zero voltage bias, $f_{1e} = f_{1h}$ and therefore the shot noise contribution proportional to $(f_{1e} - f_{1h})^2$ vanishes. Therefore, at zero voltage bias and finite temperature bias, the term proportional to $(f_{1e} - f_{2e})^2$ only remains finite, which is exactly the charge $\Delta_T$ noise, which we denote as $\Delta_T^{ch}$.}

{Similarly, the expression for spin quantum noise ($Q_{11}^{sp}$) can also be derived directly from Eq. (\ref{B_Snoise}) and utilizing the same properties explained below Eq. (\ref{eq:1001}), it is given as, }

{
\begin{equation}
\begin{split}
    Q_{11}^{sp} = &\frac{4e^2}{h} \int_{-\infty}^{\infty} dE \bigg[ \mathcal{A}^{\downarrow \uparrow 2}  + \mathcal{A}^{\uparrow \uparrow 2} + (1 - \mathcal{B}^{\uparrow \uparrow} + \mathcal{B}^{\downarrow \uparrow})^2 + 2 (\mathcal{A}^{\uparrow \uparrow} - \mathcal{A}^{\downarrow \uparrow}) (1 - \mathcal{B}^{\uparrow \uparrow} + \mathcal{B}^{\downarrow \uparrow}) - 2 \text{Re}(r_{Na}^{\downarrow \uparrow^2} r_{Na}^{\uparrow \uparrow^{*2}}) - 2 \text{Re}(r_{N}^{\downarrow \uparrow^2} r_{N}^{\uparrow \uparrow^{*2}}) \\ &  - 8 \text{Re}(r_{Na}^{\uparrow \uparrow} r_{Na}^{\downarrow \uparrow^*}) \text{Re}(r_{N}^{\uparrow \uparrow} r_{N}^{\downarrow \uparrow^*}) - 8 \text{Re}(r_{Na}^{\uparrow \uparrow} r_{Na}^{\downarrow \uparrow *} r_{N}^{\uparrow \uparrow} r_{N}^{\downarrow \uparrow * })\bigg) \bigg(f_{1e} (1 - f_{1e})\bigg) + \bigg( 2 \mathcal{A} \mathcal{B} -8 Re(r_{Na}^{\uparrow \uparrow} r_{N}^{\uparrow \uparrow}) Re(r_{Na}^{\downarrow \uparrow^{*}} r_{N}^{\downarrow \uparrow^*})\bigg) \\& \bigg(f_{1e}-f_{1h}\bigg)^2 + \bigg( \mathcal{C}_S^{\downarrow \uparrow 2} + \mathcal{D}_S^{\downarrow \uparrow 2} + 2 \mathcal{C}_S \mathcal{D}_S +  (\mathcal{A} + \mathcal{B})(\mathcal{C}_S + \mathcal{D}_S)  + 2 \mathcal{C}_S^{\downarrow \uparrow} \mathcal{C}_S^{\uparrow \uparrow}  + 2 \mathcal{D}_S^{\downarrow \uparrow} \mathcal{D}_S^{\uparrow \uparrow} - 2 (|u|^2 - |v|^2)^2 Re(c_{S}^{\downarrow \uparrow^{2}} c_{S}^{\uparrow \uparrow^{*^2}}) \\& - 2 (|u|^2 - |v|^2)^2 Re (d_{S}^{\downarrow \uparrow^{2}} d_{S}^{\uparrow \uparrow^{*^2}}) - 8 (|u|^2 - |v|^2)^2  Re (c_S^{\uparrow \uparrow} c_S^{\downarrow \uparrow^*}) Re (d_S^{\uparrow \uparrow} d_S^{\downarrow \uparrow^*}) - 4(|u|^2 - |v|^2)^2 (Re(c_S^{\uparrow \uparrow} c_S^{\downarrow \uparrow *})  + Re(d_S^{\uparrow \uparrow} d_S^{\downarrow \uparrow *})) \\&(Re(r_N^{\uparrow \uparrow} r_N^{\downarrow \uparrow *}) + Re(r_{Na}^{\uparrow \uparrow} r_{Na}^{\downarrow \uparrow *})) \bigg) f_{2e}(1-f_{2e}) + ((\mathcal{A} + \mathcal{B})(\mathcal{C}_S + \mathcal{D}_S) - 4 (|u|^2 - |v|^2)\left(\text{Re}(c_S^{\uparrow \uparrow} c_S^{\downarrow \uparrow *}) + \text{Re}(d_S^{\uparrow \uparrow} d_S^{\downarrow \uparrow *})\right) \\& \left(\text{Re}(r_N^{\uparrow \uparrow} r_N^{\downarrow \uparrow *}) + \text{Re}(r_{Na}^{\uparrow \uparrow} r_{Na}^{\downarrow \uparrow *})\right) - 4 \mathcal{A}^{\uparrow \uparrow} \mathcal{B}^{\downarrow \uparrow} - 4 \mathcal{A}^{\downarrow \uparrow} \mathcal{B}^{\uparrow \uparrow}  + 8 \text{Re} \left(r_N^{\downarrow \uparrow} r_{Na}^{\uparrow \uparrow *} r_{Na}^{\downarrow \uparrow} r_N^{\uparrow \uparrow *} \right)\bigg)\bigg(f_{1e} - f_{2e}\bigg)^2\bigg].
    \end{split}
    \label{eq:1003}
\end{equation}
}

{Here too, considering only excitations below the superconducting gap $\Delta_0$, the general expression for $Q_{11}^{sp}$ is simplified to}

{
\begin{align}
    \begin{split}
        Q_{11}^{sp} &= \frac{4e^2}{h} \int_{-\infty}^{\infty} dE (\mathcal{A}^{\uparrow \uparrow^2} + \mathcal{A}^{\downarrow \uparrow^2} + (1 - \mathcal{B}^{\uparrow \uparrow} + \mathcal{B}^{\downarrow \uparrow})^2 + 2 \mathcal{A}^{\uparrow \uparrow} - 2 \mathcal{A}^{\downarrow \uparrow} -2 \mathcal{A} \mathcal{B}  - 2 \text{Re}(r_{Na}^{\downarrow \uparrow^2} r_{Na}^{\uparrow \uparrow^{*2}}) - 2 \text{Re}(r_{N}^{\downarrow \uparrow^2} r_{N}^{\uparrow \uparrow^{*2}}) \\ & \quad \quad \quad - 8 \text{Re}(r_{Na}^{\uparrow \uparrow} r_{Na}^{\downarrow \uparrow^*}) \text{Re}(r_{N}^{\uparrow \uparrow} r_{N}^{\downarrow \uparrow^*})) f_{1e}(1-f_{1e})  + \frac{4e^2}{h} \int_{-\infty}^{\infty} dE (2 \mathcal{A} \mathcal{B} - 8 \text{Re} (r_{Na}^{\uparrow \uparrow} r_{Na}^{\uparrow \downarrow^*}) \text{Re}(r_{N}^{\uparrow \uparrow} r_{N}^{\uparrow \downarrow^*})) (f_{1e} - f_{1h})^2 \\ &
        \quad \quad \quad + \frac{4e^2}{h} \int_{-\infty}^{\infty} dE (-4 \mathcal{A}^{\uparrow \downarrow} \mathcal{B}^{\downarrow \uparrow} - 4 \mathcal{A}^{\downarrow \uparrow} \mathcal{B}^{\uparrow \uparrow} + 8 \text{Re} (r_N^{\downarrow \uparrow} r_{Na}^{\uparrow \uparrow^*} r_{N}^{\uparrow \uparrow^*} r_{Na}^{\downarrow \uparrow})) (f_{1e} - f_{2e})^2  \\
        & \quad \quad \quad= Q_{11}^{sp; th} + Q_{11}^{sp; sh}.
    \end{split}
    \label{eq:B4}
\end{align}}
{where $Q_{11}^{sp; th} = \frac{4e^2}{h} \int_{-\infty}^{\infty} dE (\mathcal{A}^{\uparrow \uparrow^2} + \mathcal{A}^{\downarrow \uparrow^2} + (1 - \mathcal{B}^{\uparrow \uparrow} + \mathcal{B}^{\downarrow \uparrow})^2 + 2 \mathcal{A}^{\uparrow \uparrow} - 2 \mathcal{A}^{\downarrow \uparrow} -2 \mathcal{A} \mathcal{B}  - 2 \text{Re}(r_{Na}^{\downarrow \uparrow^2} r_{Na}^{\uparrow \uparrow^{*2}}) - 2 \text{Re}(r_{N}^{\downarrow \uparrow^2} r_{N}^{\uparrow \uparrow^{*2}})  - 8 \text{Re}(r_{Na}^{\uparrow \uparrow} r_{Na}^{\downarrow \uparrow^*}) \text{Re}(r_{N}^{\uparrow \uparrow} r_{N}^{\downarrow \uparrow^*}) f_{1e}(1-f_{1e}) $ is the thermal noise-like contribution and $Q_{11}^{ch; sh} = \frac{4e^2}{h} \int_{-\infty}^{\infty} dE (2 \mathcal{A} \mathcal{B} - 8 \text{Re} (r_{Na}^{\uparrow \uparrow} r_{Na}^{\uparrow \downarrow^*}) \text{Re}(r_{N}^{\uparrow \uparrow} r_{N}^{\uparrow \downarrow^*})) (f_{1e} - f_{1h})^2 + \frac{4e^2}{h} \int_{-\infty}^{\infty} (-4 \mathcal{A}^{\uparrow \downarrow} \mathcal{B}^{\downarrow \uparrow} - 4 \mathcal{A}^{\downarrow \uparrow} \mathcal{B}^{\uparrow \uparrow} + 8 \text{Re} (r_N^{\downarrow \uparrow} r_{Na}^{\uparrow \uparrow^*} r_{N}^{\uparrow \uparrow^*} r_{Na}^{\downarrow \uparrow})) (f_{1e} - f_{2e})^2$ is the shot noise-like contribution to the total spin quantum noise $Q_{11}^{sp}$. At zero voltage bias and finite temperature bias, the term proportional to $(f_{1e} - f_{2e})^2$, which is same as spin $\Delta_T$ noise, which we denote as $\Delta_T^{sp}$.}

\section{Andreev and normal reflection probabilities for YSR states and MBS}
\label{App:prob}
{In this section, we provide the plots of Andreev and normal reflection probabilities for YSR and MBS states in various configurations, see Sec. \ref{theory} D. Also, we provide the probability values at zero energy in Table \ref{Table170}-\ref{Table173} for configurations: 1 and 2 at $J = 4.5$ and -4.9 with $Z = 0.781$. Note: Configuration 3, 4 for electron-down incident will be exactly similar to configuration 2 and 1 respectively.}

 \begin{figure}[H]
\centering
\includegraphics[width=1.0\linewidth]{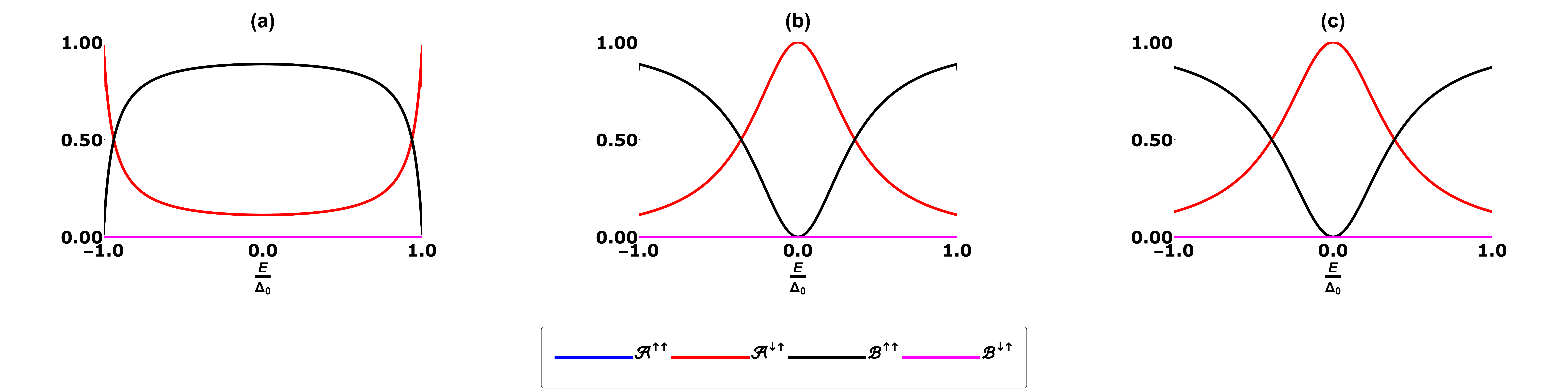}
\caption{Reflection probabilities vs. the excitation energy of incident electron $\left(\frac{E}{\Delta_0}\right)$ in configuration-1 for (a) YSR ($s$-wave), (b) MBS (chiral $p$-wave) and (c) MBS (spinless $p$-wave). The parameters are: $Z = 0.781, k_F a = 0.85 \pi$, $J = 4.5$, $T = 0.0K$. }
\label{fig:200}
\end{figure} 

\begin{table}[H]
\caption{Reflection probabilities at zero excitation energy of incident electron $\left(\frac{E}{\Delta_0}\right)$ for configuration 1 for $Z = 0.781, k_F a = 0.85 \pi, J = 4.5$.}
\centering
\begin{tabular}{|l|llll|lll|lll|lll|}
\hline
States & \multicolumn{4}{l|}{$\mathcal{A}^{\uparrow \uparrow}$} & \multicolumn{3}{l|}{$\mathcal{A}^{\downarrow \uparrow}$} & \multicolumn{3}{l|}{$\mathcal{B}^{\uparrow \uparrow}$} & \multicolumn{3}{l|}{$\mathcal{B}^{\downarrow \uparrow}$} \\ \hline
YSR ($s$-wave) & \multicolumn{4}{l|}{0.00} & \multicolumn{3}{l|}{0.12} & \multicolumn{3}{l|}{0.88} & \multicolumn{3}{l|}{0.00} \\ \hline
MBS (chiral $p$-wave) & \multicolumn{4}{l|}{0.00} & \multicolumn{3}{l|}{1.00} & \multicolumn{3}{l|}{0.00} & \multicolumn{3}{l|}{0.00} \\ \hline
MBS (spinless $p$-wave) & \multicolumn{4}{l|}{0.00} & \multicolumn{3}{l|}{1.00} & \multicolumn{3}{l|}{0.00} & \multicolumn{3}{l|}{0.00} \\ \hline
\end{tabular}
\label{Table170}
\end{table}

 \begin{figure}[H]
\centering
\includegraphics[width=1.0\linewidth]{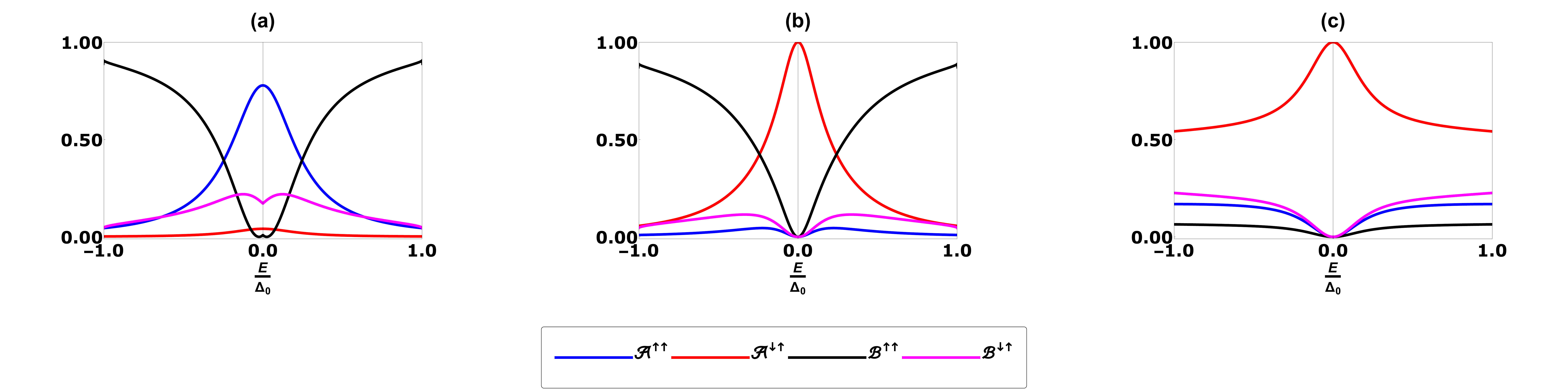}
\caption{Reflection probabilities vs. the excitation energy of incident electron $\left(\frac{E}{\Delta_0}\right)$ in configuration 2 for (a) YSR ($s$-wave), (b) MBS (chiral $p$-wave) and (c) MBS (spinless $p$-wave). The parameters are: $Z = 0.781, k_F a = 0.85 \pi$, $J = 4.5$, $T = 0.0K$.}
\label{fig:201}
\end{figure}

\begin{table}[H]
\caption{Reflection probabilities at zero excitation energy of incident electron $\left(\frac{E}{\Delta_0}\right)$ for configuration 2 for $Z = 0.781, k_F a = 0.85 \pi, J = 4.5, T = 0.0K$}
\centering
\begin{tabular}{|l|llll|lll|lll|lll|}
\hline
States & \multicolumn{4}{l|}{$\mathcal{A}^{\uparrow \uparrow}$} & \multicolumn{3}{l|}{$\mathcal{A}^{\downarrow \uparrow}$} & \multicolumn{3}{l|}{$\mathcal{B}^{\uparrow \uparrow}$} & \multicolumn{3}{l|}{$\mathcal{B}^{\downarrow \uparrow}$} \\ \hline
YSR ($s$-wave) & \multicolumn{4}{l|}{0.78} & \multicolumn{3}{l|}{0.04} & \multicolumn{3}{l|}{0.009} & \multicolumn{3}{l|}{0.17} \\ \hline
MBS (chiral $p$-wave) & \multicolumn{4}{l|}{0.00} & \multicolumn{3}{l|}{1.00} & \multicolumn{3}{l|}{0.00} & \multicolumn{3}{l|}{0.00} \\ \hline
MBS (spinless $p$-wave) & \multicolumn{4}{l|}{0.00} & \multicolumn{3}{l|}{1.00} & \multicolumn{3}{l|}{0.00} & \multicolumn{3}{l|}{0.00} \\ \hline
\end{tabular}
\label{Table171}
\end{table}

 \begin{figure}[H]
\centering
\includegraphics[width=1.0\linewidth]{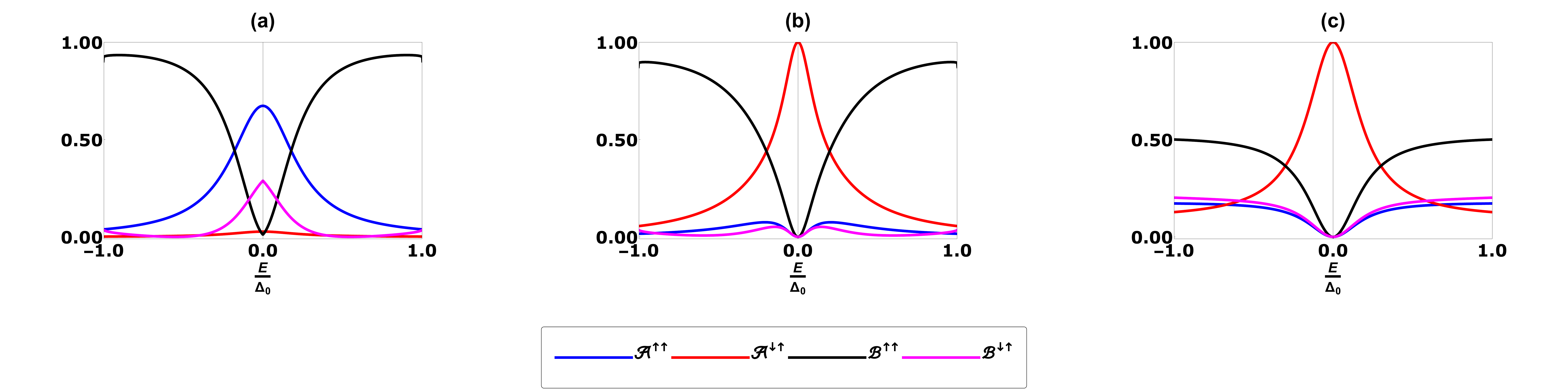}
\caption{Reflection probabilities vs. the excitation energy of incident electron $\left(\frac{E}{\Delta_0}\right)$ in configuration 1 for (a) YSR ($s$-wave), (b) MBS (chiral $p$-wave) and (c) MBS (spinless $p$-wave). The parameters are: $Z = 0.781, k_F a = 0.85 \pi$, $J = -4.9$, $T = 0.0K$.}
\label{fig:202}
\end{figure}

\begin{table}[H]
\caption{Reflection probabilities at zero excitation energy of incident electron $\left(\frac{E}{\Delta_0}\right)$ for configuration 1 for $Z = 0.781, k_F a = 0.85 \pi, J = -4.9, T = 0.0K$}
\centering
\begin{tabular}{|l|llll|lll|lll|lll|}
\hline
States & \multicolumn{4}{l|}{$\mathcal{A}^{\uparrow \uparrow}$} & \multicolumn{3}{l|}{$\mathcal{A}^{\downarrow \uparrow}$} & \multicolumn{3}{l|}{$\mathcal{B}^{\uparrow \uparrow}$} & \multicolumn{3}{l|}{$\mathcal{B}^{\downarrow \uparrow}$} \\ \hline
YSR ($s$-wave) & \multicolumn{4}{l|}{0.00} & \multicolumn{3}{l|}{0.13} & \multicolumn{3}{l|}{0.87} & \multicolumn{3}{l|}{0.00} \\ \hline
MBS (chiral $p$-wave) & \multicolumn{4}{l|}{0.00} & \multicolumn{3}{l|}{1.00} & \multicolumn{3}{l|}{0.00} & \multicolumn{3}{l|}{0.00} \\ \hline
MBS (spinless $p$-wave) & \multicolumn{4}{l|}{0.00} & \multicolumn{3}{l|}{1.00} & \multicolumn{3}{l|}{0.00} & \multicolumn{3}{l|}{0.00} \\ \hline
\end{tabular}
\label{Table172}
\end{table}

 \begin{figure}[H]
\centering
\includegraphics[width=1.0\linewidth]{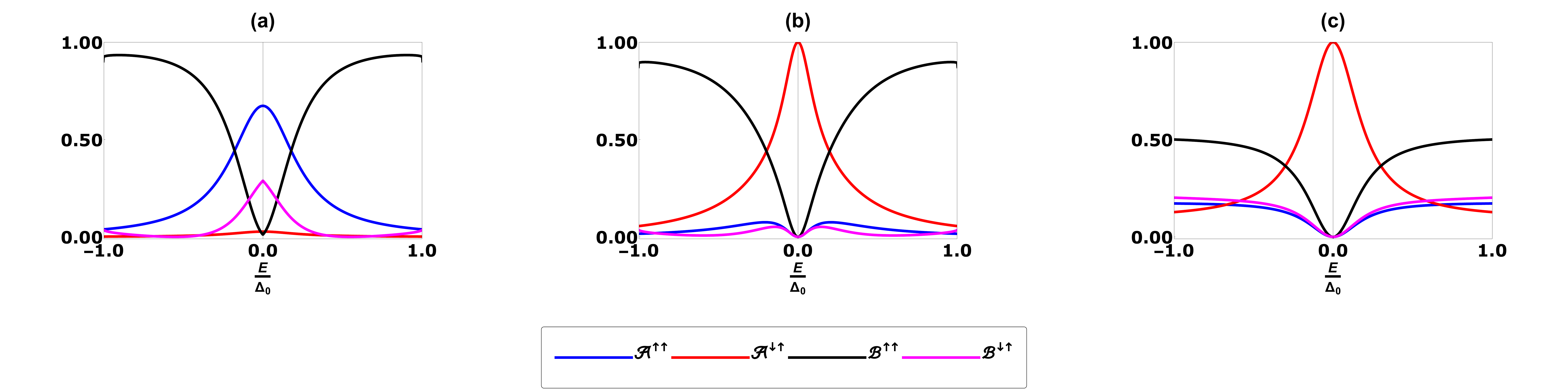}
\caption{Reflection probabilities vs. the excitation energy of incident electron $\left(\frac{E}{\Delta_0}\right)$ in configuration-1 for (a) YSR ($s$-wave), (b) MBS (chiral $p$-wave) and (c) MBS (spinless $p$-wave). The parameters are: $Z = 0.781, k_F a = 0.85 \pi$, $J = -4.9$, $T = 0.0K$.}
\label{fig:203}
\end{figure}

\begin{table}[H]
\caption{Reflection probabilities at zero excitation energy of incident electron $\left(\frac{E}{\Delta_0}\right)$ for configuration 2 for $Z = 0.781, k_F a = 0.85 \pi, J = -4.9, T = 0.0K$}
\centering
\begin{tabular}{|l|llll|lll|lll|lll|}
\hline
States & \multicolumn{4}{l|}{$\mathcal{A}^{\uparrow \uparrow}$} & \multicolumn{3}{l|}{$\mathcal{A}^{\downarrow \uparrow}$} & \multicolumn{3}{l|}{$\mathcal{B}^{\uparrow \uparrow}$} & \multicolumn{3}{l|}{$\mathcal{B}^{\downarrow \uparrow}$} \\ \hline
YSR ($s$-wave) & \multicolumn{4}{l|}{0.68} & \multicolumn{3}{l|}{0.027} & \multicolumn{3}{l|}{0.012} & \multicolumn{3}{l|}{0.28} \\ \hline
MBS (chiral $p$-wave) & \multicolumn{4}{l|}{0.00} & \multicolumn{3}{l|}{1.00} & \multicolumn{3}{l|}{0.00} & \multicolumn{3}{l|}{0.00} \\ \hline
MBS (spinless $p$-wave) & \multicolumn{4}{l|}{0.00} & \multicolumn{3}{l|}{1.00} & \multicolumn{3}{l|}{0.00} & \multicolumn{3}{l|}{0.00} \\ \hline
\end{tabular}
\label{Table173}
\end{table}

\end{widetext}

\bibliography{apssamp5}

\end{document}